\documentclass[aps,amsmath,amssymb,onecolumn,nofootinbib,pra,superscriptaddress,longbibliography]{revtex4-2} 

\usepackage[colorlinks,citecolor=blue,urlcolor=blue,bookmarks=false,hypertexnames=true]{hyperref} 

\usepackage[utf8]{inputenc}
\usepackage[T1]{fontenc}
\usepackage{graphicx}
\usepackage{bm}
\usepackage{epsfig,amsthm}
\usepackage{mathtools}
\usepackage{relsize}
\usepackage{accents}
\usepackage{wasysym}  
\usepackage{times}
\usepackage[mathscr]{euscript}
\usepackage[english]{babel} 

\usepackage[normalem]{ulem} 

\usepackage{amsmath,amsfonts,bm,thmtools}
\usepackage{mathptmx}
\usepackage{mathrsfs}
\DeclareMathAlphabet{\mathcal}{OMS}{cmsy}{m}{n}

\usepackage{multirow}

\newcommand{\fig}[1]{Fig.~\ref{#1}}
\newcommand{\tr}{\textrm{Tr}}
\newcommand{\Symbol}[2]{{#1}_{\textrm{#2}}}
\newcommand{\bfj}{\textbf{j}}
\newcommand{\bfq}{\textbf{q}}
\newcommand{\HEOMLS}{\hat{\mathcal{M}}}

\newcommand{\code}[1]{\texttt{#1}}

\begin{document}

\title{HierarchicalEOM.jl: An efficient Julia framework for hierarchical equations of motion \\in open quantum systems}

\author{Yi-Te Huang}
\thanks{These authors contributed equally.}
\affiliation{Department of Physics, National Cheng Kung University, 701 Tainan, Taiwan}
\affiliation{Center for Quantum Frontiers of Research and Technology, NCKU, 701 Tainan, Taiwan}

\author{Po-Chen Kuo}
\thanks{These authors contributed equally.}
\affiliation{Department of Physics, National Cheng Kung University, 701 Tainan, Taiwan}
\affiliation{Center for Quantum Frontiers of Research and Technology, NCKU, 701 Tainan, Taiwan}
\affiliation{Theoretical Quantum Physics Laboratory, Cluster for Pioneering Research, RIKEN, Wakoshi, Saitama 351-0198, Japan}

\author{Neill Lambert}
\email{nwlambert@gmail.com}
\affiliation{Theoretical Quantum Physics Laboratory, Cluster for Pioneering Research, RIKEN, Wakoshi, Saitama 351-0198, Japan}

\author{Mauro Cirio}
\email{cirio.mauro@gmail.com}
\affiliation{Graduate School of China Academy of Engineering Physics, Haidian District, Beijing, 100193, China}

\author{Simon Cross}
\affiliation{Theoretical Quantum Physics Laboratory, Cluster for Pioneering Research, RIKEN, Wakoshi, Saitama 351-0198, Japan}

\author{Shen-Liang Yang}
\affiliation{Department of Physics, National Cheng Kung University, 701 Tainan, Taiwan}
\affiliation{Center for Quantum Frontiers of Research and Technology, NCKU, 701 Tainan, Taiwan}

\author{Franco Nori}
\affiliation{Theoretical Quantum Physics Laboratory, Cluster for Pioneering Research, RIKEN, Wakoshi, Saitama 351-0198, Japan}
\affiliation{Center for Quantum Computing (RQC), RIKEN, Wakoshi, Saitama 351-0198, Japan}
\affiliation{Physics Department, The University of Michigan, Ann Arbor, Michigan 48109-1040, USA.}

\author{Yueh-Nan Chen}
\email{yuehnan@mail.ncku.edu.tw}
\affiliation{Department of Physics, National Cheng Kung University, 701 Tainan, Taiwan}
\affiliation{Center for Quantum Frontiers of Research and Technology, NCKU, 701 Tainan, Taiwan}

\begin{abstract}
The hierarchical equations of motion (HEOM) approach can describe the reduced dynamics of a system simultaneously coupled to multiple bosonic and fermionic environments. The complexity of exactly describing the system-environment interaction with the HEOM method usually results in time-consuming calculations and a large memory cost. Here, we introduce an open-source software package called HierarchicalEOM.jl: a Julia framework integrating the HEOM approach. HierarchicalEOM.jl features a collection of methods to compute bosonic and fermionic spectra, stationary states, and the full dynamics in the extended space of all auxiliary density operators (ADOs). The required handling of the ADOs multi-indexes is achieved through a user-friendly interface. We exemplify the functionalities of the package by analyzing a single impurity Anderson model, and an ultra-strongly coupled charge-cavity system interacting with bosonic and fermionic reservoirs. HierarchicalEOM.jl achieves a significant speedup with respect to the corresponding method in the Quantum Toolbox in Python (QuTiP), upon which this package is founded.
\end{abstract}

\maketitle
\newpage
\section{Introduction}
The time evolution of a closed quantum system can offer important insights about its nature and properties. However, the dynamics are inevitable affected by interactions with external environments~\cite{Robert1960,Vernon1963}, which can involve the exchange of energy or particles, and the suppression of quantum coherence. Due to the effective continuum of degrees of freedom present in these external baths~\cite{Caldeira1983,Caldeira1987}, modeling the dynamics of an open quantum system can be challenging. This is especially the case when perturbative approaches~\cite{HongBin2015} are no longer valid due to non-Markovian effects emerging in the presence of strong interaction with the bath~\cite{Tanimura_2_1989,Tanimura_1_1990}. In this regime, standard Markovian master equations are no longer applicable, and non-perturbative techniques are required~\cite{Ralf2008,WeiMin_PRL2012,Strasberg2016,Brenes2020,Gauger2020}.

In particular, here we consider the hierarchical equations of motion (HEOM) approach, which offers a non-perturbative~\cite{Yan_2_PRL} characterization of all the environmental effects on the system. This is achieved by using a hierarchy of auxiliary density operators (ADOs) to model system-bath correlations and entanglement~\cite{Tanimura_3_2020,QuTiP-BoFiN}. This makes the HEOM suitable for studying complex systems strongly coupled to either bosonic (with applications in quantum biology~\cite{Lambert2019,Limmer2022}, quantum optics~\cite{Ma2012}, quantum thermodynamics~\cite{Tanimura_4_2016}, and quantum information~\cite{GuiLu2022}), or fermionic environments (for the analysis of quantum transport~\cite{Yan_1,Ishizaki_1_2009,Hartle2013} and condensed matter physics~\cite{Yan_5_2016}). Additionally, quantum systems interacting simultaneously with both bosonic and fermionic environments can be found in the study of electron transport through both natural and artificial molecules~\cite{Schinabeck2018,Thoss_2_2021}.  Naturally, such an increase in the complexity of the environment leads to an increase in computational resources which, in the case of the HEOM method, corresponds to an increase in the size of the HEOM Liouvillian superoperator (HEOMLS) matrix. To deal with this issue, it is  beneficial to explore the numerical efficiency of programming languages designed to optimize different computational resources.

In this work, we use \code{Julia}~\cite{Julia2012, Julia2017}: a dynamic general-purpose, high-level programming language capable of high-performance in scientific applications. It was created in 2012 with the goal to build a language with the speed of \code{C}~\cite{C2006}, the dynamism of \code{Ruby}~\cite{Ruby2007}, the practicality of \code{Python}~\cite{Python2009}, and able to solve statistics and linear algebra tasks like \code{R}~\cite{R2021} and \code{Matlab}~\cite{Matlab2016}, respectively. \code{Julia} is fast because it uses a just-in-time compiler~\cite{LLVM2004} to convert source code into machine code before running it. Recently, many researchers are turning to \code{Julia} to build high-performance simulations and models for open quantum system dynamics~\cite{HOQST2022}, quantum optics~\cite{QuantumOptics.jl-2018}, quantum algorithms~\cite{Yao2020} and quantum information theories~\cite{QuantumInformation2018}. Moreover, the variable name in \code{Julia} supports UTF-8 encoding (e.g., Greek symbols and mathematical symbols) which allows to write equations in code more elegantly.

\code{HierarchicalEOM.jl} is developed following the \code{Julia} design philosophy: one can have machine performance without sacrificing human convenience~\cite{Julia2012}. While integrating many of the features presented in other open-source HEOM packages, \code{HierarchicalEOM.jl} also includes other functionalities, such as the estimation of importance values for all ADOs, the calculation of spectra for both bosonic and fermionic systems, the construction of HEOMLS matrices for even- or odd-parity auxiliary density operators, and a user-friendly interface (which interrogates the ADOs multi-indexes) for gaining access to bath properties. By wrapping some functions from other \code{Julia} packages~\cite{DifferentialEquations.jl-2017,LinearSolve.jl,FastExpm.jl-2011}, we could further optimize the computations of the dynamics and stationary states for all ADOs. The entire ecosystem of \code{HierarchicalEOM.jl} package is summarized in \fig{fig:HierarchicalEOM.jl}. 

\section{Results}
\subsection{Preliminaries for hierarchical equations of motion}
Throughout this work, we consider an open quantum system (s) interacting with \textit{fermionic} (f) and \textit{bosonic} (b) \textit{environments} described by the following total (T) \textit{Hamiltonian} ($\hbar$ is set to unity throughout this work): 
\begin{equation}
\begin{aligned}
H_{\textrm{T}} = H_{\textrm{s}}(t) + H_{\textrm{f}} + H_{\textrm{b}} + H_{\textrm{sf}} + H_{\textrm{sb}},
\label{eq:H_total}
\end{aligned}
\end{equation}
where $\Symbol{H}{s}(t)$ is the (possibly time-dependent) system Hamiltonian containing boson and fermion particles. Here, we allow the fermionic environment to be composed by multiple baths of non-interacting fermionic degrees of freedom described by the Hamiltonian
\begin{equation}
\begin{aligned}
\Symbol{H}{f} = \sum_{\alpha}\sum_{k}\epsilon_{\alpha,k}c_{\alpha,k}^{\dagger}c_{\alpha,k},
\label{eq:H_f}
\end{aligned}
\end{equation}
where $c_{\alpha,k}$ $(c_{\alpha,k}^{\dagger})$ annihilates (creates) a fermion (f) in the state $k$ (with energy $\epsilon_{\alpha,k}$) of the $\alpha$th fermionic bath. Analogously,
\begin{equation}
\begin{aligned}
\Symbol{H}{b} = \sum_{\beta}\sum_{k}\omega_{\beta,k}b_{\beta,k}^{\dagger}b_{\beta,k},
\label{eq:H_b}
\end{aligned}
\end{equation}
describes a generic bosonic environment which can accommodate multiple non-interacting bosonic baths in which $b_{\beta,k}$ $(b_{\beta,k}^{\dagger})$ is the bosonic annihilation (creation) operator associated to the $k$th mode (with frequency $\omega_{\beta,k}$) in the $\beta$th bosonic bath.
This bosonic environment can represent a broad range of physical environments, e.g., electromagnetic environments~\cite{Mauro2016,Stockklauser2017}, phonon environments~\cite{Gustafsson2014,Manenti2017,Iorsh2020}, surface plasmon modes~\cite{Benz2016,Po2020} and vibronic environments in molecules (e.g., nuclear motion in a photosynthetic complex~\cite{Ishizaki_1_2009}). The interaction Hamiltonian between a fermionic system and the fermionic environments can be written as
\begin{equation}
\begin{aligned}
 \Symbol{H}{sf} = \sum_{\alpha,k}\left(g_{\alpha,k}c_{\alpha,k}^{\dagger}\Symbol{d}{s}+g_{\alpha,k}^*\Symbol{d}{s}^\dagger c_{\alpha,k}\right),
\label{eq:H_sf}
\end{aligned}
\end{equation}
in terms of the \textit{coupling strengths} $g_{\alpha,k}$. Analogously, the interaction between a bosonic or fermionic system and the exterior bosonic environments can be modeled by 
\begin{equation}
\begin{aligned}
\Symbol{H}{sb} = \Symbol{V}{s}\sum_{\beta,k}g_{\beta,k}(b_{\beta,k}+b_{\beta,k}^{\dagger}),
\label{eq:H_sb}
\end{aligned}
\end{equation}
in terms of the coupling strengths $g_{\beta,k}$. 
Here, $\Symbol{d}{s}$ and $\Symbol{V}{s}$ refer to the coupling operators acting on the system's degrees of freedom. In particular, $\Symbol{d}{s}$ is a odd-parity operator destroying a fermion in the system, while $\Symbol{V}{s}$ is in general a Hermitian operator which can act on both fermionic and bosonic systems. When $\Symbol{V}{s}$ is acting on the fermionic system, it must have even-parity to be compatible with charge conservation. Furthermore, one can easily generalize to the case where the system contains multiple bosonic and fermionic quantum numbers (such as frequency, energy, or spin) interacting with either individual or shared environment(s) [see Example 1 and Example 2].

To derive the HEOM, we assume the following three conditions: (1) The system and the environments (baths) are initialized in a separable state. (2) Each of the fermionic (bosonic) baths is initially in thermal equilibrium characterized by a Fermi-Dirac (Bose-Einstein) distribution. (3) The bath operator within the system-bath interaction Hamiltonian should be linear in the bath annihilation and creation operators, as shown in Eq.~(\ref{eq:H_sf}) and Eq.~(\ref{eq:H_sb}).

Without resorting to any approximation, the reduced density matrix for the system at time $t$ can be written in terms of the following Dyson series
\begin{equation}\label{eq:G_series}
\begin{aligned}
\rho_{\textrm{s}}^{p=+}(t)=\hat{\mathcal{G}}(t)\left[\rho_{\textrm{s}}^{p=+}(0)\right],
\end{aligned}
\end{equation}
where, the propagator $\hat{\mathcal{G}}(t)[\cdot]$ is a \textit{canonical superoperator}~\cite{Mauro2022} which propagates the even-parity ($p=+$) or odd-parity ($p=-$) operator initially in $\rho_{\textrm{s}}^{p}(0)$. Furthermore, the reduced density matrix of the system has even-parity and can be written as $\rho_{\textrm{s}}^{+}(0)$. The propagator takes the form
\begin{equation}
\begin{aligned}
\hat{\mathcal{G}}(t)[\cdot]
=\hat{\mathcal{T}}\exp\Big\{-\int_{0}^{t}d t_{1}\int_{0}^{t_{1}}d t_{2}
\Big[\hat{\mathcal{W}}_{\textrm{f}}(t_1,t_2)[\cdot]+\hat{\mathcal{W}}_{\textrm{b}}(t_1,t_2)[\cdot]\Big]\Big\}
,\label{G_op}
\end{aligned}
\end{equation}
in terms of the time-ordering operator $\hat{\mathcal{T}}$, together with fermionic~\cite{Mauro2022} and bosonic~\cite{Lambert2019,QuTiP-BoFiN} operators explicitly written as
\begin{equation}
\begin{aligned}
\hat{\mathcal{W}}_{\textrm{f}}(t_1,t_2)[\cdot]
=\sum_{\alpha}~\sum_{p=\pm}~\sum_{\nu=\pm 1}
\Big\{
C^{\nu}_{\alpha}(t_1,t_2)
\Big[\Symbol{d}{s}^{\bar{\nu}}(t_2),\Symbol{d}{s}^{\nu}(t_1)\cdot\Big]_{-p}
+C^{\nu}_{\alpha}(t_2,t_1)
\Big[\cdot \Symbol{d}{s}^{\bar{\nu}}(t_2),\Symbol{d}{s}^{\nu}(t_1)\Big]_{-p}
\Big\}
,\label{eq:W_f}
\end{aligned}
\end{equation}
and 
\begin{equation}
\begin{aligned}
\hat{\mathcal{W}}_{\textrm{b}}(t_1,t_2)[\cdot]
=\sum_{\beta}\Big[\Symbol{V}{s}(t_1),
\left(
C^{\mathbb{R}}_{\beta}(t_1,t_2)[\Symbol{V}{s}(t_2),\cdot]_{-}
+i C^{\mathbb{I}}_{\beta}(t_1,t_2)[\Symbol{V}{s}(t_2),\cdot]_{+}
\right)\Big]_{-}
,\label{eq:W_b}
\end{aligned}
\end{equation}
respectively. For simplicity, we define $\nu$ to denote the presence ($\nu = +1$) or absence ($\nu = -1$) of a Hermitian conjugation and $\bar{\nu}:=-\nu$. Here, $[\cdot,\cdot]_{-}$ and $[\cdot,\cdot]_{+}$ represent the commutator and the anti-commutator introduced to allow the propagator to keep track of the parity sector (labeled by $p=\pm$) it is acting on. This dependence intuitively originates from the properties of partial traces over composite fermionic spaces, where operators do not necessarily commute. 

As an example, for an environment made out of a single fermion, the reduced matrix elements $\langle{i}|\rho_\textrm{s}^p|{j}\rangle$ (in a basis  labeled by $\langle i|$ and $|{j}\rangle$) involve the perturbative sum  of expressions of the form $\langle i| (c \tilde{\rho}_\textrm{e} \tilde{\rho}_\textrm{s}^p c^\dagger+\tilde{\rho}_\textrm{e} \tilde{\rho}_\textrm{s}^p)|{j}\rangle$ (in terms of environmental operators $\tilde{\rho}_{\textrm{e}}$, system operators $\tilde{\rho}_\textrm{s}^p$ with parity $p$, and the environment-annihilation operator $c$). These quantities depend on the commutator between $\tilde{\rho}_\textrm{s}^p$ and $c$, which is trivial only for even-parity ($p=+$). In the odd-parity ($p=-$) case, the partial trace over the environment requires further anti-commutations, ultimately resulting in extra minus signs in the expression for the effective propagator describing the reduced dynamics~\cite{Mauro2022}. It is important to explicitly note that, here, by parity we do not refer to the presence of an odd or even number of fermions in the system but, rather, to the number of fermionic operators needed to represent $\rho_\textrm{s}^p$. In summary, while commutators ($-p=-$) appear for the propagation of even-parity operators $\rho_{\textrm{s}}^{p=+}(0)$ (which include the reduced density operator of the system), anti-commutators ($-p=+$) appear for the propagation of odd-parity operators $\rho_{\textrm{s}}^{p=-}(0)$ [e.g., $\Symbol{d}{s}\rho_{\textrm{s}}^{+}(0)$ or $\Symbol{d}{s}^\dagger\rho_{\textrm{s}}^{+}(0)$], which appear in the  computation of observables such as the density of states (see the Methods section: Numerical computation of the spectrum).

As it can be seen in Eq.~(\ref{eq:W_f}) and Eq.~(\ref{eq:W_b}), the effects of fermionic and bosonic environments [initially in thermal equilibrium (eq) and linearly coupled to the system] are completely encoded in the \textit{two-time correlation functions} $C(t_1,t_2)$. In the fermionic case, they depend on the spectral density $J_{\alpha}(\omega)=2\pi\sum_{k} |g_{\alpha,k}|^{2}\delta(\omega-\omega_{k})$ and the Fermi–Dirac distribution $n_{\alpha}^{\textrm{eq}}(\omega)=\{\exp[(\omega-\mu_{\alpha})/k_{\textrm{B}}T_{\alpha}]+1\}^{-1}$ as
\begin{equation}
\begin{aligned}
C^{\nu}_{\alpha}(t_{1},t_{2})
=\frac{1}{2\pi}\int_{-\infty}^{\infty} d\omega 
J_{\alpha}(\omega)\Big[\frac{1-\nu}{2}+\nu n_{\alpha}^{\textrm{eq}}(\omega)
\Big]e^{\nu i\omega (t_{1}-t_{2})}.
\label{eq:C_f}
\end{aligned}
\end{equation}
Analogously, in the bosonic case, they depend on the spectral density $J_{\beta}(\omega)=2\pi\sum_{k} g_{\beta,k}^{2}\delta(\omega-\omega_{k})$ and the Bose–Einstein distribution $n_{\beta}^{\textrm{eq}}(\omega)=\{\exp[\omega/k_{\textrm{B}}T_{\beta}]-1\}^{-1}$ as
\begin{equation}
\begin{aligned}
C_{\beta}(t_{1},t_{2})
=\frac{1}{2\pi}\int_{0}^{\infty} d\omega 
J_{\beta}(\omega)\Big[n_{\beta}^{\textrm{eq}}(\omega)e^{i\omega (t_{1}-t_{2})}
+(n_{\beta}^{\textrm{eq}}(\omega)+1)e^{-i\omega (t_{1}-t_{2})}
\Big].
\label{eq:C_b}
\end{aligned}
\end{equation}
Here, $k_{\textrm{B}}$ is the Boltzmann constant and $T_{\alpha}$ $(T_{\beta})$ represents the absolute temperature of the $\alpha$-fermionic ($\beta$-bosonic) bath. A non-zero chemical potential ($\mu_{\alpha}\neq 0$) in the $\alpha$-fermionic bath can account for non-equilibrium physics.

A more practical representation of Eq.~(\ref{eq:G_series}) can be found by expressing the bath correlation functions as a sum of exponential terms (exponents). This allows to define an iterative procedure which leads to the celebrated hierarchical equations of motion (HEOM).
Specifically, different approaches, such as the Matsubara~\cite{Shi2009} or the Pad\'{e}~\cite{Jie2011} spectral decompositions, can be used (see Methods) to write the fermionic and bosonic correlation functions as, respectively,
\begin{equation}
\begin{aligned}
C^{\nu}_{\alpha}(\tau)=\sum_{h=1}^{N_\alpha}\eta_{\alpha,h}^{\nu}\exp(-\gamma_{\alpha,h}^\nu \tau), 
\label{eq:C_f_exp}
\end{aligned}
\end{equation}%
and
\begin{equation}
\begin{aligned}
C_{\beta}(\tau)=\sum_{l=1}^{N_\beta}\xi_{\beta,l}\exp(-\chi_{\beta,l}\tau), 
\label{eq:C_b_exp}
\end{aligned}
\end{equation}%
where $\tau=t_{1}-t_{2}$, and $N_\alpha$  ($N_\beta$) is the total number of exponentials for the $\alpha$-fermionic ($\beta$-bosonic) bath. When $\chi_{\beta,l}\neq\chi_{\beta,l}^*$, a closed form for the HEOM can be obtained by further decomposing the bosonic correlation function into its real ($\mathbb{R}$) and imaginary ($\mathbb{I}$) parts as
\begin{equation}
\begin{aligned}
C_{\beta}(\tau) = \sum_{u=\mathbb{R},\mathbb{I}} 
(\delta_{u,\mathbb{R}}+i\delta_{u,\mathbb{I}})C_{\beta}^{u}(\tau)
\label{eq:C_beta_RI}
\end{aligned}
\end{equation}%
where $\delta$ is the Kronecker delta function and where
\begin{equation}
\begin{aligned}
C_{\beta}^{u}(\tau) = \sum_{l=1}^{N_\beta^u}
\xi_{\beta,l}^{u}\exp(-\chi_{\beta,l}^{u}\tau).
\label{eq:C_beta_u}
\end{aligned}
\end{equation}

The expressions for the fermionic [Eq.~(\ref{eq:C_f_exp})] and bosonic [Eq.~(\ref{eq:C_beta_RI})] correlation functions can now be used in Eq.~(\ref{G_op}). The resulting expression can be recursively differentiated in time~\cite{Ma2012,Mauro2022,Pochen2023} to define a local master equation in an enlarged space made out of \textit{auxiliary density operators} (ADOs) $\rho^{(m,n,p)}_{\bfj \vert \bfq}(t)$. Intuitively, these variables encode environmental effects related to different  exponential terms present in the correlation function and provide an iterative description of high-order system-baths memory effects. In $\rho^{(m,n,p)}_{\bfj \vert \bfq}(t)$, the tuple $(m, n, p)$ represents the $m$th-level-bosonic-and-$n$th-level-fermionic ADO with parity $p$, and $\bfj$ ($\bfq$) denotes a vector [$j_m,\cdots,j_1$] ([$q_n,\cdots,q_1$]), where each $j$ ($q$) represents a specific multi-index ensemble $\{\beta,u,l,\Symbol{\sigma}{b}\}$ ($\{\alpha,\nu,h,\Symbol{\sigma}{f}\}$). The indexes  $\Symbol{\sigma}{b}$ and $\Symbol{\sigma}{f}$ are used to specify other quantum numbers (such as energy or spin) of the system interacting with a bosonic or fermionic bath, respectively. In this way, the hierarchical equations of motion in the Schr\"{o}dinger picture can be expressed as
\begin{equation}
\begin{aligned}
\partial_{t}\rho^{(m,n,p)}_{\bfj \vert \bfq}(t)
\equiv&\HEOMLS\rho^{(m,n,p)}_{\bfj \vert \bfq}(t)\\
=&-\Big(
i\hat{\mathcal{L}}_{\textrm{s}}+\sum_{r=1}^{m}\chi_{j_{r}}+\sum_{w=1}^{n}\gamma_{q_{w}}
\Big)
\rho^{(m,n,p)}_{\bfj \vert \bfq}(t)
\\
&-i\sum_{q'\notin\bfq}\hat{\mathcal{A}}_{q'}
\rho^{(m,n+1,p)}_{\bfj\vert \bfq^+}(t)
-i\sum_{j'} \hat{\mathcal{B}}_{j'}
\rho^{(m+1,n,p)}_{\bfj^+ \vert \bfq}(t)
\\&
-i\sum_{w=1}^{n}(-1)^{n-w}\hat{\mathcal{C}}_{q_{w}}
\rho^{(m,n-1,p)}_{\bfj\vert \bfq_{w}^{-}}(t)
-i\sum_{r=1}^{m}\hat{\mathcal{D}}_{j_{r}}
\rho^{(m-1,n,p)}_{\bfj_{r}^{-} \vert \bfq}(t),
\label{eq:HEOM}
\end{aligned}
\end{equation}%
where we used the multi-index notations:
\begin{equation}
\begin{aligned}
\bfj^{+}&=[j',j_{m},\cdots,j_1],\\
\bfq^{+}&=[q',q_n,\cdots,q_{1}],\\
\bfj_{r}^{-}&=[j_m,\cdots,j_{r+1},j_{r-1},\cdots,j_{1}],\\
\bfq_{w}^{-}&=[q_n,\cdots,q_{w+1},q_{w-1},\cdots,q_{1}].
\end{aligned}
\end{equation}
Here, $j'$ can be chosen from any one of the elements in $\bfj$ or any other bosonic multi-index ensemble that has not yet be considered in $\bfj$, whereas $q'\notin\bfq$ due to the Pauli exclusion principle. 

Furthermore, we defined $\hat{\mathcal{L}}_{\textrm{s}}[\cdot]=[H_{\textrm{s}}(t),\cdot]_{-}$ characterizing the bare system dynamics. System-bath interactions are encoded by the fermionic superoperators $\hat{\mathcal{A}}_{q}$, $\hat{\mathcal{C}}_{q}$ (to couple the $n$th-level-fermionic ADOs to the $(n+1)$th-level- and $(n-1)$th-level- fermionic ADOs, respectively), and the bosonic superoperators  $\hat{\mathcal{B}}_{j}$, $\hat{\mathcal{D}}_{j}$ (to couple the $m$th-level-bosonic ADOs to the $(m+1)$th-level- and $(m-1)$th-level- bosonic ADOs, respectively). Their explicit forms are given by 
\begin{equation}
\begin{aligned}
\hat{\mathcal{A}}_{q}&[\cdot] = 
(-1)^{\delta_{p,-}}\Big\{
d^{\bar{\nu}}_{\Symbol{\sigma}{f}}[\cdot]-\hat{\mathcal{P}}_{\textrm{s}}\left[[\cdot]d^{\bar{\nu}}_{\Symbol{\sigma}{f}}\right]
\Big\},\\
\hat{\mathcal{C}}_{q}&[\cdot] = 
(-1)^{\delta_{p,-}}\Big\{
\eta^{\nu}_{\alpha,h}d^{\nu}_{\Symbol{\sigma}{f}}[\cdot]+(\eta^{\bar{\nu}}_{\alpha,h})^*\hat{\mathcal{P}}_{\textrm{s}}\left[[\cdot]d^{\nu}_{\Symbol{\sigma}{f}}\right]
\Big\},\\
\hat{\mathcal{B}}_{j}&[\cdot] = \Big[V_{\Symbol{\sigma}{b}},\cdot\Big]_{-},\\
\hat{\mathcal{D}}_{j}&[\cdot] = \delta_{u,\mathbb{R}}\xi^{\mathbb{R}}_{\beta,l}\Big[V_{\Symbol{\sigma}{b}},\cdot\Big]_{-}+i\delta_{u,\mathbb{I}}\xi^{\mathbb{I}}_{\beta,l}\Big[V_{\Symbol{\sigma}{b}},\cdot\Big]_{+},\\
\hat{\mathcal{P}_{\textrm{s}}}&\left[\rho^{(m,n,\pm)}_{\bfj \vert \bfq}(t)d^{\nu}_{\Symbol{\sigma}{f}}\right]=\mp (-1)^{n} \rho^{(m,n,\pm)}_{\bfj \vert \bfq}(t)d^{\nu}_{\Symbol{\sigma}{f}}.
\end{aligned}
\end{equation}
The form of the HEOM in Eq.~(\ref{eq:HEOM}) can be applied to auxiliary density operators with arbitrary parity symmetry. For example,  $\rho_{|}^{(0,0,+)}(t)$ is the system reduced density operator, and the odd-parity operator $\rho_{|}^{(0,0,-)}(t)$ can take the form $d^{\nu}_{\textrm{s}}\rho_{|}^{(0,0,+)}(t)$ as mentioned after Eq.~(\ref{eq:W_b}). To project in the different parity system sub-spaces, we introduced the superoperator 
\begin{equation}
    \hat{\mathcal{P}}_{\textrm{s}}[\cdot]=\left(\prod_{\Symbol{\sigma}{f}}\exp[i\pi d^{\dagger}_{\Symbol{\sigma}{f}}d_{\Symbol{\sigma}{f}}]\right)[\cdot]\left(\prod_{\Symbol{\sigma}{f}}\exp[i\pi d^{\dagger}_{\Symbol{\sigma}{f}}d_{\Symbol{\sigma}{f}}]\right).
\end{equation}
Overall, the HEOM \textit{Liouvillian superoperator} (HEOMLS) $\HEOMLS$ characterizes the dynamics in the full ADOs space and it can, numerically, be expressed as a matrix.

In practice, the dimensions of this matrix must be finite and
the hierarchical equations must be truncated at a suitable bosonic-level ($m_\textrm{max}$) and fermionic-level ($n_\textrm{max}$). Previous works~\cite{Hartle2013,Wenderoth2016} showed that the computational effort can be optimized by associating an \textit{importance value} to each ADO and then discarding all ADOs (in the second and higher levels) whose importance value is smaller than a threshold value $\mathcal{I}_\textrm{th}$. The importance value $\mathcal{I}$ for a given ADO $\rho^{(m,n,p)}_{\bfj \vert \bfq}$ is defined as
\begin{equation}\label{eq:importance}
\mathcal{I}\left(\rho^{(m,n,p)}_{\bfj \vert \bfq}\right)=\left|\prod_{r=1}^{m}\left(\frac{\xi_{j_r}}{\textrm{Re}[\chi_{j_r}]}\frac{1}{\sum_{x=1}^r\textrm{Re}[\chi_{j_x}]}\right)\right| \times \left|\prod_{w=1}^{n}\left(\frac{\eta_{q_w}}{\textrm{Re}[\gamma_{q_w}]}\frac{1}{\sum_{x=1}^w\textrm{Re}[\gamma_{q_x}]}\right)\right|,
\end{equation}
where the first and second absolute value represents the importance of the bosonic part~\cite{Thoss_2_2021} and fermionic part~\cite{Hartle2015} of the hierarchy, respectively. Note that for hybrid (consider bosonic and fermionic baths at the same time) cases, we retain all the ADOs where their hierarchy levels $(m, n)\in\{(0, 0), (0, 1), (1, 0), (1, 1)\}$, and all the other high-level ADOs may be neglected if their importance value is less than $\Symbol{\mathcal{I}}{th}$. Moreover, the full hierarchical equations can be recovered in the limiting case $\Symbol{\mathcal{I}}{th}\rightarrow0$. We will provide a concrete example and detailed discussion later.

\begin{figure}[tb]
\centering \includegraphics[width=18cm]{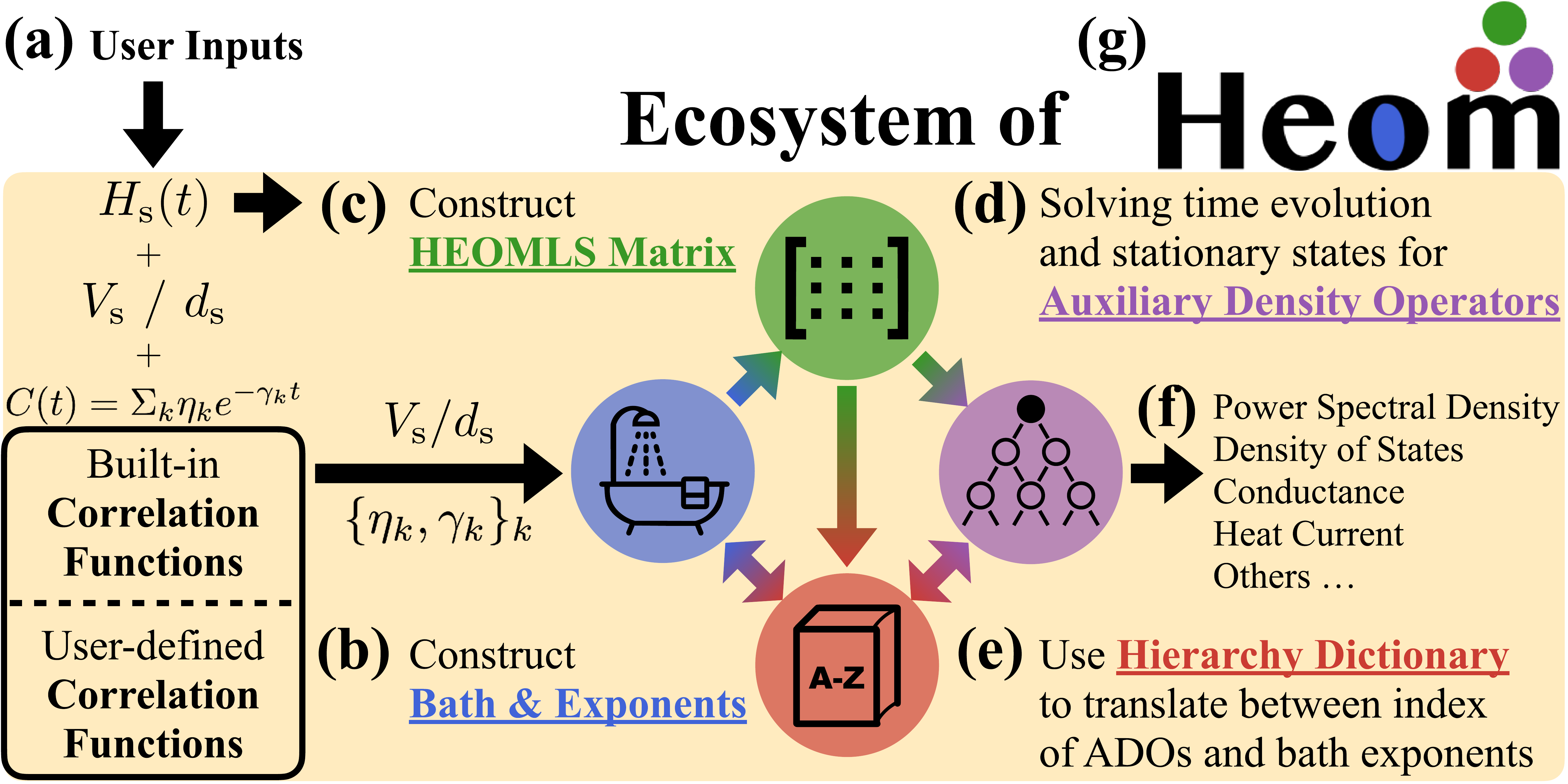}
\caption{
\textbf{The ecosystem of the HierarchicalEOM.jl package.}
\textbf{(a)} Users should specify the system Hamiltonian $H_\textrm{s}(t)$, coupling operators ($V_\textrm{s}$ or $d_\textrm{s}$), and the bath correlation function $C(t)$. For the exponent $\{\eta_k,\gamma_k\}$, users can either specify the physical parameters characterizing the spectral density of the bath by built-in functions, or directly providing a list of exponents.
\textbf{(b)} Construction of the bath-object which includes the system coupling operator and a list of exponents characterizing the bath correlation function.
\textbf{(c)} Construction of the HEOM Liouvillian superoperator (HEOMLS) matrix $\hat{\mathcal{M}}$ which defines the hierarchical equations of motion from the system Hamiltonian and the bath-objects.
\textbf{(d)} Computation of the dynamics and stationary states for all auxiliary density operators using $\hat{\mathcal{M}}$.
\textbf{(e)} The hierarchy dictionary translates the index of each ADO into the corresponding multi-index ensembles together with the exponents of the bath, and vice-versa.
\textbf{(f)} The hierarchy dictionary allows a high-level interpretation of the ADOs to compute some physical properties.
\textbf{(g)} Logo of \code{HierarchicalEOM.jl} package.}
\label{fig:HierarchicalEOM.jl}
\end{figure}

\subsection{Package architecture and design philosophy}
The package \code{HierarchicalEOM.jl} is designed to integrate the efficiency of \code{Julia} with the functionalities provided by other existing HEOM packages~\cite{PHI,Yan_5_2016,DM-HEOM,PyHEOM,QuTiP-BoFiN}. This leads to an intuitive interface to construct arbitrary Hamiltonians and initial states (which can be accomplished directly with the built-in library for linear algebra in \code{Julia} or by taking advantage of the package \code{QuantumOptics.jl}~\cite{QuantumOptics.jl-2018}). We now introduce the ecosystem of the package as summarized in Fig.~\ref{fig:HierarchicalEOM.jl}.

Following Fig.~\ref{fig:HierarchicalEOM.jl}(a) and Fig.~\ref{fig:HierarchicalEOM.jl}(b), users should specify the system Hamiltonian $H_\textrm{s}(t)$, system coupling operators $V_\textrm{s}$ ($d_\textrm{s}$) describing the interaction with bosonic (fermionic) baths, and the bath correlation functions. For the bath correlation functions in Eq.~(\ref{eq:C_f}) and Eq.~(\ref{eq:C_b}), users can specify them by an exponential series as shown in Eqs.~(\ref{eq:C_f_exp}-\ref{eq:C_beta_u}). \code{HierarchicalEOM.jl} provides built-in functions  to construct these series from physical parameters (e.g., coupling strength, temperature, etc.). For example, it is possible to choose a Drude-Lorentz spectral density for bosonic baths and a Lorentzian spectral density for fermionic baths. For each of these cases, the correlations are computed using either the Matsubara~\cite{Shi2009} or Pad\'{e} spectral decomposition~\cite{Jie2011}. Alternatively, \code{HierarchicalEOM.jl} offers the possibility to manually define the correlation functions (simply supplying the list of exponents). Additional spectral densities could be incorporated into the built-in functions in future releases. We explicitly note that the package allows for any combination of fermionic and bosonic baths.

After the definition of the bath, users can further construct the HEOMLS matrix $\HEOMLS$ in Eq.~(\ref{eq:HEOM}) together with the system Hamiltonian $\Symbol{H}{s}(t)$, see Fig.~\ref{fig:HierarchicalEOM.jl}(c). At this stage, \code{HierarchicalEOM.jl} offers the possibility to optimize this matrix by neglecting all the ADOs (at second or higher level) whose importance values in Eq.~(\ref{eq:importance}) are below the user-specified threshold $\Symbol{\mathcal{I}}{th}$. It is also possible to construct $\HEOMLS$ within the even- or odd-parity sector depending on the parity of the auxiliary density operator upon which the HEOMLS matrix is acting on (see the Methods section: Numerical computation of the spectrum). Moreover, our package allows users to further add a \textit{Lindbladian} (superoperator) to $\HEOMLS$. The Lindbladian describes the dissipative interaction between the system and extra environment while its explicit form is given by
\begin{equation}\label{eq:Lindbladian}
    \hat{\mathcal{J}}(F)\Big[\cdot\Big]=F\left[\cdot\right]F^\dagger-\frac{1}{2}\Big[F^\dagger F, \cdot\Big]_+,
\end{equation}
where $F$ is the jump operator and describes the dissipation dynamics of the system. Including such Lindbladians when some of the baths are weakly coupled to the system  is more efficient than solving the full HEOM since it does not require any ADOs for such baths and hence reduces the size of $\HEOMLS$ (see Example 2 for more details).

As shown in Fig.~\ref{fig:HierarchicalEOM.jl}(d), with this information it is possible to proceed with solving for the dynamics of all the ADOs as described in Eq.~(\ref{eq:HEOM}). \code{HierarchicalEOM.jl} provides two distinct methods to compute the dynamics. The first one relies on \code{DifferentialEquations.jl}~\cite{DifferentialEquations.jl-2017} which provides a set of low-level solvers for ordinary differential equations. The second method directly builds the propagator when the system Hamiltonian is time-independent: $\hat{\mathcal{G}}(t)=\exp(\HEOMLS t)$ using \code{FastExpm.jl}~\cite{FastExpm.jl-2011} which is optimized for the exponentiation of either large-dense or sparse matrices. In addition, \code{HierarchicalEOM.jl} can also directly solve for the stationary states of all the ADOs using \code{LinearSolve.jl}~\cite{LinearSolve.jl}, which offers a unified interface to solve linear equations in \code{Julia}.

To allow further analysis of specific physical properties, \code{HierarchicalEOM.jl} provides a hierarchy dictionary, as depicted in Fig.~\ref{fig:HierarchicalEOM.jl}(e) and Fig.~\ref{fig:HierarchicalEOM.jl}(f). This dictionary translates the index of each ADO in terms of the corresponding exponential terms of the bath, and vice-versa. This feature is designed to allow a high-level description of the ADOs, which can be useful in the analysis of electronic currents~\cite{Yan_1, Hartle2013}, heat currents~\cite{Velizhanin2008, Kato2015, Tanimura_4_2016}, and higher-order moments of heat currents~\cite{Song2017}. Moreover, \code{HierarchicalEOM.jl} can calculate the spectrum for both bosonic and fermionic systems using \code{LinearSolve.jl} (see the Methods section: Numerical computation of the spectrum). 

We now demonstrate the use of \code{HierarchicalEOM.jl} with two examples: a single impurity strongly coupled to fermionic reservoirs and an  ultra-strongly coupled charge–cavity system interacting with both bosonic and fermionic reservoirs.

\begin{figure}[b]
\centering \includegraphics[width=8cm]{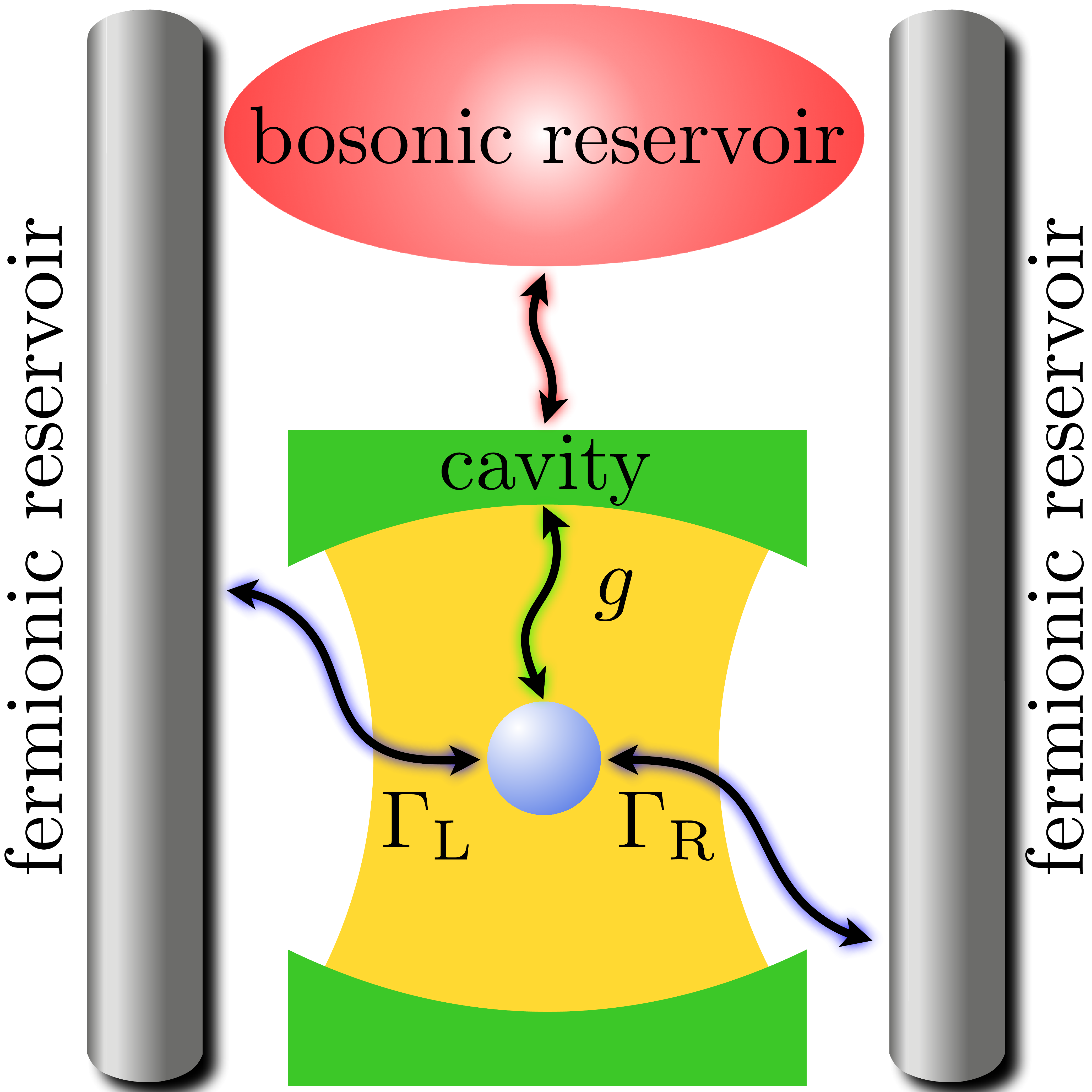}
\caption{
\textbf{Schematic illustration of a light-matter system (an electron coupled to a bosonic mode in a cavity) interacting with bosonic and fermionic reservoirs.} Here, $g$ is the coupling strength between the electron and cavity, $\Delta$ is the coupling strength between the cavity and the bosonic reservoir, and $\Gamma_\textrm{L}=\Gamma_\textrm{R}=\Gamma$ are the coupling strengths between the electron and its fermionic reservoirs.
}
\label{fig:System}
\end{figure}

\begin{table}[b]
    \centering
    \begin{tabular}{ccccccccccccccccc}
        \hline
        \hline
        Example ~&~ $\epsilon$ ~&~ $U$ ~&~ $\omega_\textrm{c}$ ~&~ $g$ ~&~ $\Gamma$ ~&~ $\Delta$ ~&~ $W_\beta$ ~&~ $W_\alpha$ ~&~ $e\Phi$ ~&~ $k_{\textrm{B}} T$ ~&~ $\Symbol{N}{photon}$ ~&~ $N_\beta$ ~&~ $N_\alpha$ ~&~ $\Symbol{m}{max}$ ~&~ $\Symbol{n}{max}$ ~&~ $\Symbol{\mathcal{I}}{th}$\\
        \hline
        (1) & -5 & 10 & $\times$ & $\times$ & 1 & $\times$ & $\times$ & 10 & [0, 4] & 0.5 & $\times$ & $\times$ & 7 & $\times$ & 4 & $10^{-7}$ \\
        (2) & -3 & $\times$ & 1 & [0, 0.5] & 1 & 0.01 & 0.2 & 10 & [0, 10] & 0.5 & 6 & 5 & 7 & 4 & 3 & $10^{-6}$ \\ 
        \hline
        \hline
    \end{tabular}
    \caption{\textbf{Parameters used in the examples.} Here, $\epsilon$ and $\omega_\textrm{c}$ are, respectively, the energies of the electron state and the frequency of the cavity mode, while $U$ represents the Coulomb repulsion energy. The electron-cavity, electron-(fermionic reservoir), and cavity-(bosonic reservoir) couplings strengths are specified by $g$, $\Gamma$, and $\Delta$, respectively. $W_\beta$ and $W_\alpha$ are the band-widths of the bosonic and fermionic reservoir, respectively. The bias voltage $\Phi$ is used to characterize the chemical potential of the fermionic reservoirs as $\mu_\textrm{L}=-\mu_\textrm{R}=e\Phi/2$ (in terms of the elementary charge $e$). All the reservoirs are initially at equilibrium at the same value of $k_{\textrm{B}} T$, where $k_{\textrm{B}}$ is the Boltzmann constant and $T$ is the absolute temperature. Note that all the values are considered in units of milli-electronvolt (meV). The truncation parameters are: $\Symbol{N}{photon}$ for the Fock space dimension, $N_\beta$ for the number of exponential terms for the bosonic reservoirs, $N_\alpha$ for the number of exponential terms for the fermionic reservoirs,  $\Symbol{m}{max}$ for the tier of the bosonic hierarchy, $\Symbol{n}{max}$ for the tier of the fermionic hierarchy, and $\Symbol{\mathcal{I}}{th}$ for the importance threshold.
    }
    \label{tab:parameters}
\end{table}

\subsection{Example 1: Electronic system strongly interacting with fermionic reservoirs}
The investigation of the Kondo effect in electronic systems is crucial as it serves both as a valuable testing ground for the theories of the Kondo effect and has the potential to lead to a better understanding of this intrinsic many-body phenomena~\cite{Sprinzak2002,Keller2013,Hur2015}, with applications in building new generation nanoscale electronic devices~\cite{Park2002,Wingreen2004,Natelson2004} and quantum computation~\cite{Smith2019}. In this sense, we here consider a single-level electronic system [which can be populated by a spin up, $\Symbol{\sigma}{f}=\uparrow$, or spin down electron, $\Symbol{\sigma}{f}=\downarrow$] coupled to two [left (L) and right (R)] fermionic reservoirs, as illustrated in Fig.~(\ref{fig:System}). In this case, the different Hamiltonian terms in Eq.~(\ref{eq:H_total}) take the form
\begin{equation}
\begin{aligned}
H_\textrm{s} &= \epsilon \left(d^\dagger_\uparrow d_\uparrow + d^\dagger_\downarrow d_\downarrow \right) + U\left(d^\dagger_\uparrow d_\uparrow d^\dagger_\downarrow d_\downarrow\right),\\
\Symbol{H}{f} &=\sum_{\alpha=\textrm{L},\textrm{R}}\sum_{\Symbol{\sigma}{f}=\uparrow,\downarrow}\sum_{k}\epsilon_{\alpha,\Symbol{\sigma}{f},k}c_{\alpha,\Symbol{\sigma}{f},k}^{\dagger}c_{\alpha,\Symbol{\sigma}{f},k},\\
\Symbol{H}{sf} &=\sum_{\alpha=\textrm{L},\textrm{R}}\sum_{\Symbol{\sigma}{f}=\uparrow,\downarrow}\sum_{k}g_{\alpha,k}c_{\alpha,\Symbol{\sigma}{f},k}^{\dagger}d_{\Symbol{\sigma}{f}} + g_{\alpha,k}^*d_{\Symbol{\sigma}{f}}^{\dagger}c_{\alpha,\Symbol{\sigma}{f},k}, \\
\end{aligned}
\end{equation}
where $\epsilon$ is the energy of the impurity and $U$ is the Coulomb repulsion energy for double occupation. We further assume both fermionic reservoirs to have a Lorentzian-shaped spectral density. As we discussed above, the correlation function in Eq.~(\ref{eq:C_f}) can be expressed as a sum of exponential terms, see Eq.~(\ref{eq:C_f_exp}), using the Pad\'{e} decomposition (see Methods). These spectral densities depend on the following physical parameters: the coupling strength $\Symbol{\Gamma}{L}=\Symbol{\Gamma}{R}=\Gamma$ between system and reservoirs, the band-width $W_\alpha$, the temperature $T$, and the chemical potential $\Symbol{\mu}{L}=-\Symbol{\mu}{R}=e\Phi/2$, where $e$ is the elementary charge, and $\Phi$ is the bias voltage. We summarize the values of these parameters in Table.~\ref{tab:parameters}. With these choices, it is now possible to use \code{HierarchicalEOM.jl} to construct the HEOMLS matrix $\HEOMLS$ and solve for the stationary states of all the auxiliary density operators (ADOs).

\begin{figure}[tb]
\centering \includegraphics[width=18cm]{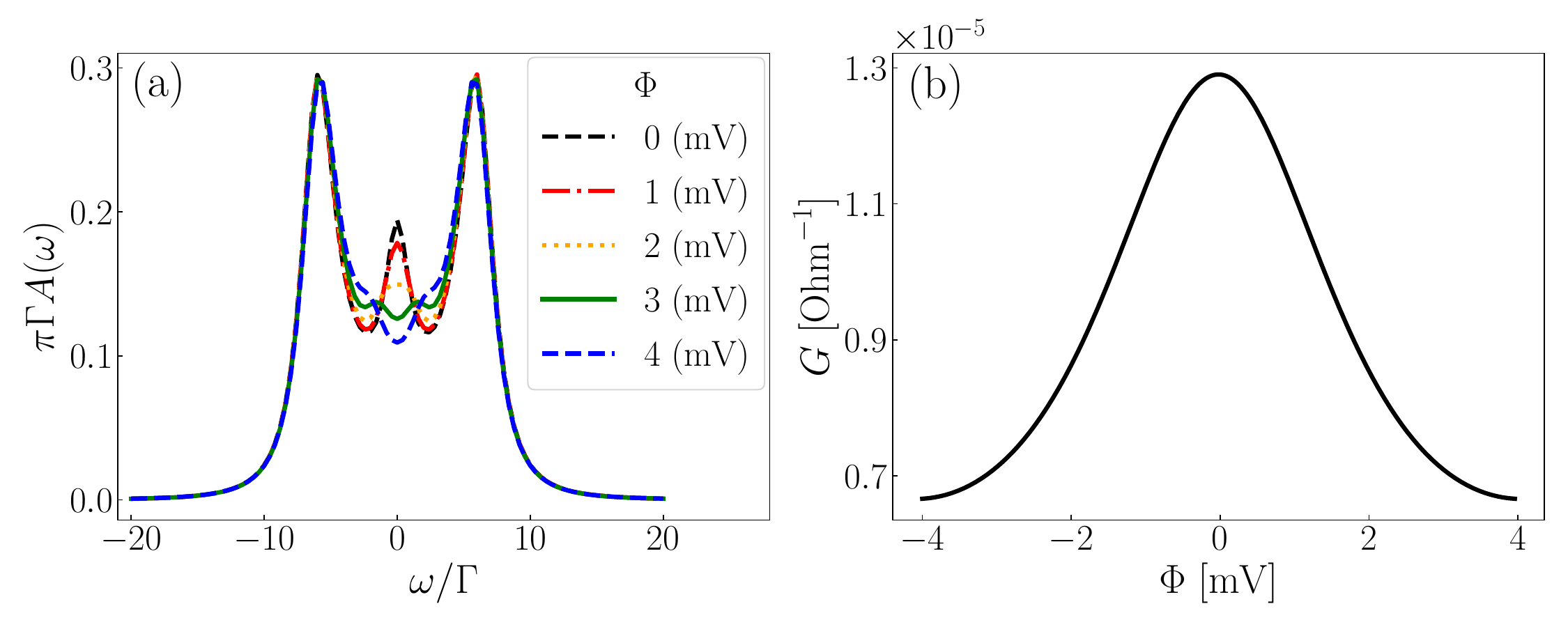}
\caption{
\textbf{The Kondo resonance from equilibrium to non-equilibrium.} In ($\textbf{a}$), the black dashed, red dash-dotted, orange dotted, green solid, and blue dashed curves represent the density of states $A(\omega)$ for bias voltage $\Phi$ at $0$, $1$, $2$, $3$, and $4~\textrm{mV}$, respectively. One can observe that the density of states of the electronic system exhibits two Hubbard peaks and an additional central peak at equilibrium (i.e., bias voltage $\Phi=0$). By increasing the bias voltage $\Phi$ to non-equilibrium, the central peak is suppressed and splits into two. In ($\textbf{b}$), the black curve shows the effects of the Kondo resonance on the differential conductance as a function of bias voltage $\Phi$. The conductance $G$ has a peak at $\Phi=0$ as the Kondo resonance acts as a transport channel for the electron.
}
\label{fig:Spin_result}
\end{figure}

By using the built-in function in \code{HierarchicalEOM.jl}, we first compute the spectrum of the system (also known as density of states (DOS); see Methods) in the stationary state and plot the results in Fig.~\ref{fig:Spin_result}(a). The DOS of the system exhibits two Hubbard peaks and an additional central peak. The negative frequency Hubbard peak is a consequence of the singly occupied electron state with either spin-up or spin-down. Similarly, double occupation of the state contributes to the resonance at energy $\epsilon + U$. When the system is in equilibrium, the Kondo-resonance central peak is a consequence of the many-body entanglement between the single electron populating the system and the electrons in the fermionic reservoirs. As the bias voltage $\Phi$ is increased, the central peak splits into two~\cite{Yan_7_2013}. This ultimately results in two distinct energies for the Kondo cloud which depend on the Fermi-level of the fermionic reservoirs~\cite{Kouwenhoven2001,VBorzenets2020}.

Alternatively, the Kondo effect can also be probed by calculating the differential conductance $G=dI/d\Phi$ between the system and fermionic reservoirs. To compute this quantity, we first compute the electronic current in the stationary states of the ADOs (see Methods) as a function of the bias voltage $\Phi$ and then take the derivative numerically. The zero-bias conductance peak shown in Fig.~\ref{fig:Spin_result}(b) is another manifestation of the existence of a single-channel spin-1/2 Kondo resonance. In fact, in this regime, the system effectively acts as a one-dimensional channel for the moving electrons thereby increasing the conductance~\cite{Luke2022}.

\subsection{Example 2: The ultra-strongly coupled charge–cavity system interacting with bosonic and fermionic reservoirs}
A crucial aspect of quantum transport research involves investigating impurities, coupled to normal metal contacts, whose charge degree of freedom is strongly coupled to cavity photons~\cite{vanderWiel2002,Kontos2018,Wallraff2018}. By integrating quantum dots with microwave cavities, researchers can investigate such novel cavity quantum electrodynamics systems~\cite{Scarlino2019}, yielding insights into a diverse range of quantum phenomena such as out-of-equilibrium state occupation in double quantum dots~\cite{Kontos2014}, reservoir-induced charge relaxation in single dots~\cite{Cottet2016}, and photo-assisted tunneling processes~\cite{vanderWiel2002,Souquet2014}. Going beyond the strong coupling regime, we here consider a single-level electronic system ultra-strongly coupled to a single-mode cavity~\cite{Kockum2019}. The electronic impurity is in contact with two distinct, left (L) and right (R), fermionic reservoirs, while the cavity is in contact with a bosonic reservoir, as illustrated in Fig.~(\ref{fig:System}). In order to focus on the charge-photon interaction, this model neglects the electronic spin degree of freedom. As a consequence, the different contributions to the total Hamiltonian, as given in Eq.~(\ref{eq:H_total}), can be expressed as
\begin{equation}
\begin{aligned}
H_\textrm{s}&=\epsilon d^\dagger d + \omega_c a^\dagger a + g d^\dagger d \left(a + a^\dagger\right),\\
H_\textrm{f} &= \sum_{\alpha=\textrm{L},\textrm{R}}\sum_{k}\epsilon_{\alpha,k}c_{\alpha,k}^{\dagger}c_{\alpha,k},\\
H_\textrm{b} &= \sum_{k}\omega_{k}b_{k}^{\dagger}b_{k},\\
H_\textrm{sf} &= \sum_{\alpha=\textrm{L},\textrm{R}}\sum_{k}g_{\alpha,k}c_{\alpha,k}^{\dagger}d + g_{\alpha,k}^* d^{\dagger}c_{\alpha,k}, \\
H_\textrm{sb} &= \sum_{k}g_{k}(a+a^\dagger)(b_{k}+b_{k}^{\dagger}),\\
\end{aligned}
\end{equation}
in terms of the energy $\epsilon$  of the electronic state, the frequency $\Symbol{\omega}{c}$  of the single-mode cavity, the electron-cavity coupling strength $g$, and the annihilation (creation) operator $a$ ($a^\dagger$) of the single-mode cavity. We assume a Drude-Lorentz (Lorentzian) spectral density for the bosonic (fermionic) reservoir(s). As previously discussed, the correlation function in Eq.~(\ref{eq:C_f}) and Eq.~(\ref{eq:C_b}) can be expressed as a sum of exponential terms using the Pad\'{e} decomposition, as shown in Eq.~(\ref{eq:C_f_exp}) and Eq.~(\ref{eq:C_b_exp}), respectively. 

The environmental spectral densities depend on the following physical parameters: the coupling strength $\Symbol{\Gamma}{L}=\Symbol{\Gamma}{R}=\Gamma$ between the electron and the fermionic reservoirs, the coupling strength $\Delta$ between the single-mode cavity and the bosonic reservoir, the band-width $W_\alpha$ ($W_\beta$) of the fermionic (bosonic) reservoirs, the chemical potentials $\Symbol{\mu}{L}=-\Symbol{\mu}{R}=e\Phi/2$ for the left and right fermionic reservoirs, and the temperature $T$ common to all reservoirs. All values of the parameters are summarized in Table.~\ref{tab:parameters}. Using these parameters as input in \code{HierarchicalEOM.jl}, we can construct the HEOMLS matrix $\HEOMLS$ and solve for the stationary states of all the ADOs.

\begin{figure}[tb]
\centering \includegraphics[width=18cm]{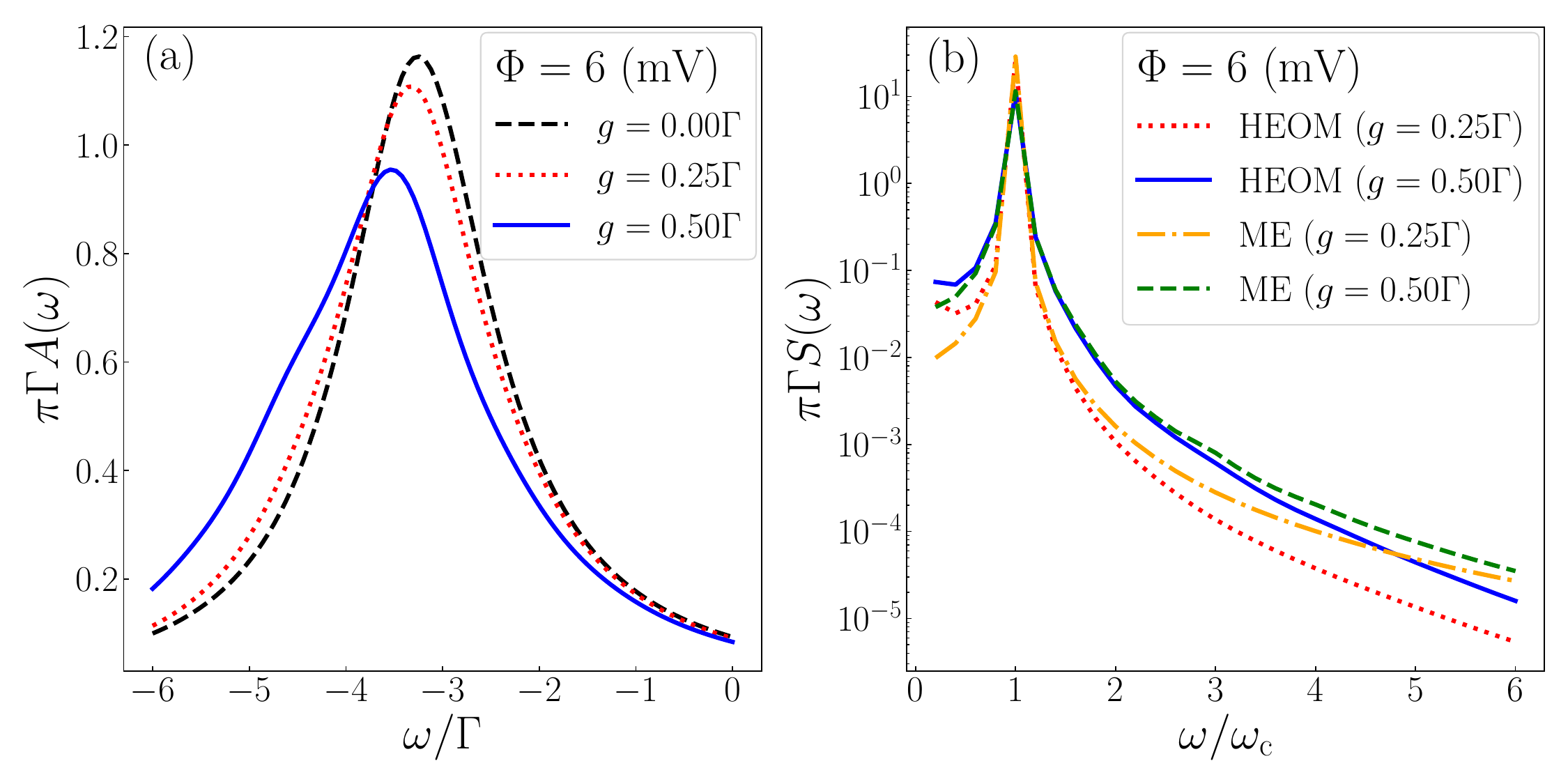}
\caption{
\textbf{Effects of the electron-cavity coupling on the spectra of the charge and cavity system.} In \textbf{(a)}, the black dashed, red dotted, and blue solid curves represent the density of states $A(\omega)$ for the charge-cavity coupling strength $g$ at $0$, $0.25\Gamma$, and $0.5\Gamma$, respectively. The resonance in the density of states of the charge system is reduced and shifts towards lower energies as $g$ increases. In \textbf{(b)}, we plot the power spectral density $S(\omega)$ of the cavity and uses two different methods [HEOM (red dotted and blue solid curves for $g$ at $0.25\Gamma$ and $0.5\Gamma$, respectively) and perturbative master equation (abbreviated as ME, which is represented by orange dash-dotted and green dashed curves for $g$ at $0.25\Gamma$ and $0.5\Gamma$, respectively)] to describe the interaction between the cavity and the bosonic reservoir. Both approaches show a single peak at $\omega=\omega_{\textrm{c}}$ for $g\in(0, 0.5]$. The Born-Markov approximation used to derive the ME may result in an inaccurate  estimation of the magnitude of the $S(\omega)$ compared to the HEOM, especially in the high-frequency range.
}
\label{fig:Charge_spectrum}
\end{figure} 

We now compute the spectra (see Methods) of both the charge (density of states; DOS) and cavity (power spectral density; PSD) using the built-in function in \code{HierarchicalEOM.jl}, and plot the results in Fig.~\ref{fig:Charge_spectrum}. Here, we consider the bias voltage $\Phi=6$, such that the site energy of the charge system aligns with the Fermi-level of one of the fermionic reservoirs, increasing the likelihood of electron tunneling to the fermionic reservoirs. We first discuss the DOS of the charge system, as shown in Fig.~\ref{fig:Charge_spectrum}(a). 
As the coupling strength $g$ between the charge and the cavity increases, the DOS of the charge system decreases and its peak shifts towards lower energies. This can be attributed to the energy shift induced by the cavity field on the electronic level. To explore the charge-cavity interaction, we now analyze the PSD of the cavity system, see Fig.~\ref{fig:Charge_spectrum}(b). The PSD of the cavity exhibits a single large peak at $\omega=\Gamma=\Symbol{\omega}{c}$ for $g\in(0, 0.5]$, indicating that the cavity resonance is largely unaffected by the electronic system, but that it does induce additional broadening. 

It is interesting to further compare the differences between the HEOM and a perturbative master equation (ME) approach. We do this by reducing the interaction strength  $\Delta=0.01\Symbol{\omega}{c}$ between the cavity and its bosonic reservoir to a weak-coupling regime. We model the interaction between the system and bosonic reservoir with a Born-Markov master equation~\cite{Breuer2007} using the Lindbladian given in Eq.~(\ref{eq:Lindbladian}). More specifically, we replace the bosonic part of HEOM with the following term 
\begin{equation}
    J_\beta(\Symbol{\omega}{c}) \left[(n^{\textrm{eq}}_\beta(\Symbol{\omega}{c})+1)\hat{\mathcal{J}}(a)+n^{\textrm{eq}}_\beta(\Symbol{\omega}{c})\hat{\mathcal{J}}(a^\dagger)\right]\rho^{(0, n, p)}_{~|\bfq}(t)
\end{equation}
on the right-hand side of Eq.~(\ref{eq:HEOM}). Here, $J_\beta$ is the spectral density of the bosonic bath (assumed to be Drude-Lorentz in this case) and $n^{\textrm{eq}}_\beta$ is the Bose-Einstein distribution. In this specific scenario, it is evident that the outcomes derived from the master equation approach may overestimate the magnitude when compared to the HEOM results in the high-frequency range of the PSD. The origin of this inconsistency can be attributed to the non-Markovian effect~\cite{HongBin2015,WeiMin_PRL2012,Yueh-Nan2009,Xiong2015} arising from the bosonic reservoir, characterized by the narrow band-width of the Drude-Lorentz spectral density ($W_\beta=0.2$) we use to describe it; even though the coupling strength is weak, this narrow width implies the environment can still have memory. In addition, the master equation has limited capability in addressing the combined influence of electron-cavity interactions and electron-fermionic-reservoir hybridization. Nevertheless, these challenges can be effectively tackled using the HEOM approach.

\begin{figure}[tb]
\centering \includegraphics[width=9cm]{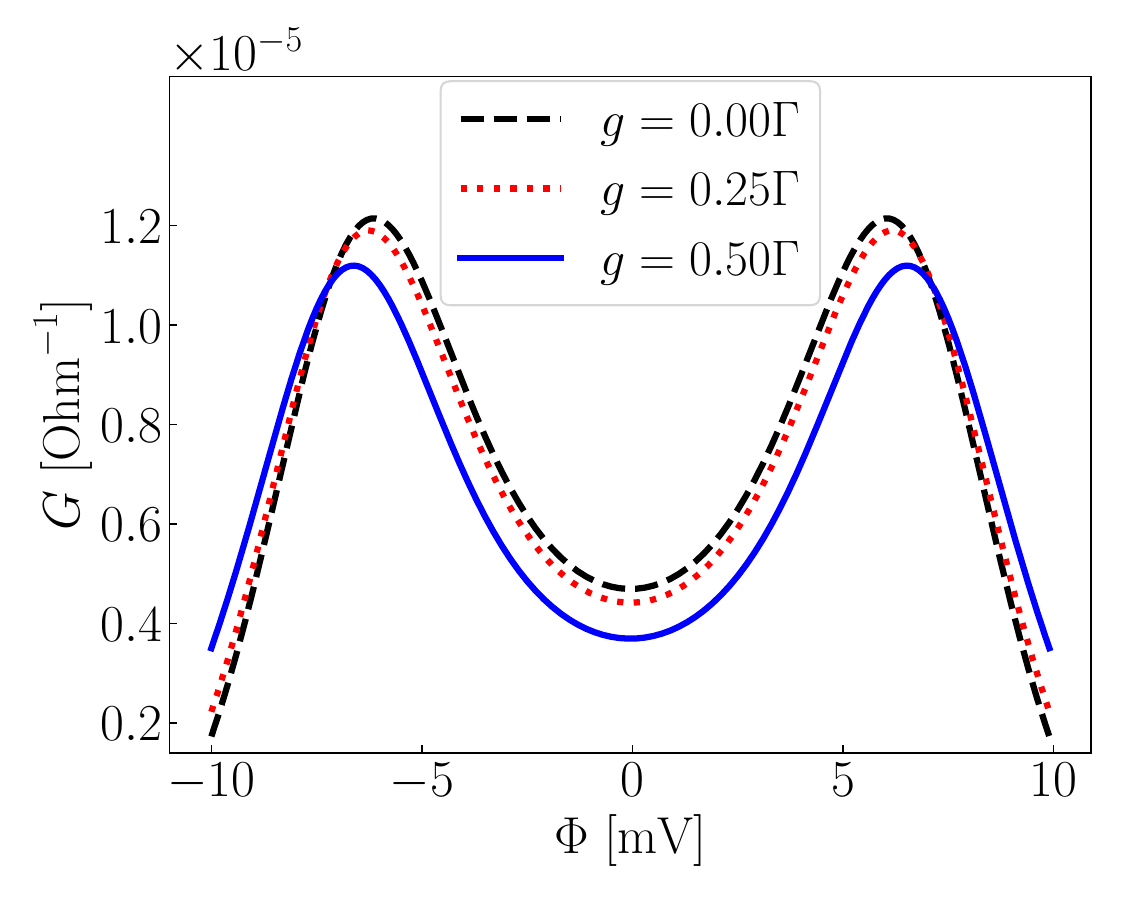}
\caption{
\textbf{Effect of the electron-cavity coupling on the conductance between the charge system and fermionic reservoirs.} The black dashed, red dotted, and blue solid curves represent the conductance $G$ for the charge-cavity coupling strength $g$ at $0$, $0.25\Gamma$, and $0.5\Gamma$, respectively. The conductance depends on the absolute value of the bias voltage $\Phi$ and reaches its maximum when the average energy of the charge system is equal to the Fermi-level of one of the fermionic reservoirs, i.e., when $e\left|\Phi\right|/2=\epsilon$. As the electron-cavity coupling $g$ increases, both the conductance $G$ and the average energy of the charge system  decrease, resulting in a wider splitting of the conductance peaks.
}
\label{fig:Charge_Conductance}
\end{figure}

Applying the same methodology as in the previous example, one can also calculate the conductance between the charge system and the fermionic reservoirs, as shown in Fig.~\ref{fig:Charge_Conductance}. We begin by examining the case when the charge system is not coupled to the cavity ($g=0$). In this scenario, due to the finite band-width  of the fermionic reservoirs  $W_\alpha$ and a site energy of the charge system $\epsilon=-3$  being far away from the Fermi-level, a local minimum of the conductance is observed at $\Phi=0$. Additionally, the conductance increases with increasing $|\Phi|$, reaching its maximum value when the site energy $\epsilon$ coincides with the Fermi-level of one of the fermionic reservoirs, resulting in the splitting of the peaks in the conductance. 

Conversely, when the charge is coupled to the cavity ($g>0$), the average energy of the charge in the electronic system decreases away from the Fermi-levels of the fermionic reservoirs, resulting in a decrease in conductance. It should be noted that the maximum conductance will be achieved with larger $|\Phi|$, leading to a wider splitting of the conductance peak.

\begin{table}[tb]
    \centering
    \begin{tabular}{lcccccc}
        \hline
        \hline
        Package Name ~&~ PHI~\cite{PHI} ~&~ HEOM-QUICK~\cite{Yan_5_2016} ~&~ DM-HEOM~\cite{DM-HEOM} ~&~ PyHEOM~\cite{PyHEOM} ~&~ QuTiP-BoFiN~\cite{QuTiP-BoFiN} ~&~ HierarchicalEOM.jl\\
        \hline
        Language & Fortran & Fortran & C/C++ & C++/Python & Python & Julia\\
        Documentation & $\checkmark$ & $\checkmark$ & $\checkmark$ & $\times$ & $\checkmark$ & $\checkmark$\\
        RunTests & $\times$ & $\times$ & $\times$ & $\times$ & $\checkmark$ & $\checkmark$\\
        Bosonic Bath & $\checkmark$ & $\times$ & $\checkmark$ & $\checkmark$ & $\checkmark$ & $\checkmark$\\
        Fermionic Bath & $\times$ & $\checkmark$ & $\times$ & $\times$ & $\checkmark$ & $\checkmark$\\
        Hybrid Bath & $\times$ & $\times$ & $\times$ & $\times$ & $\checkmark$ & $\checkmark$\\
        T-D system support & $\times$ & $\checkmark$ & $\times$ & $\times$ & $\checkmark$ & $\checkmark$\\
        Parity support & $\times$ & $\checkmark$ & $\times$ & $\times$ & $\times$ & $\checkmark$\\
        Importance support & $\times$ & $\times$ & $\times$ & $\times$ & $\times$ & $\checkmark$\\
        \hline
        \hline
    \end{tabular}
     \caption{\textbf{Comparison between different open source HEOM packages.} For each package, the check (cross) mark represents the presence (absence) of support for the listed functionalities. Language specifies the programming language. Documentation  labels the availability of instructions for installing and running the code. RunTests describes the availability of a suite of tests to ensure the correct functionality of the package. T-D system describes the availability of the system Hamiltonian $\Symbol{H}{s}$ to be time-dependent (T-D). Parity describes the availability of constructing the HEOMLS matrices for even- or odd-parity auxiliary density operators.}
    \label{tab:compare_pkg}
\end{table}

\subsection{Capabilities of HierarchicalEOM.jl and comparison with other packages}
Recent developments in the field of HEOM have lead to an explosion of open source software, such as \code{PHI}~\cite{PHI}, \code{HEOM-QUICK}~\cite{Yan_5_2016}, \code{DM-HEOM}~\cite{DM-HEOM}, \code{PyHEOM}~\cite{PyHEOM}, and \code{QuTiP-BoFiN}~\cite{QuTiP-BoFiN,QuTiP,QuTiP2}. These packages are written in different programming languages which have different capabilities. Thus, in Table~\ref{tab:compare_pkg}, we briefly compare their differences and similarities with respect to \code{HierarchicalEOM.jl}.

\begin{table}[tb]
    \centering
    \begin{tabular}{c|r r r r r r r r r}
        \hline
        \hline\\
        $n_\textrm{max} \backslash \mathcal{I}_\textrm{th}$ & \multicolumn{1}{c}{$10^{-3}$} & \multicolumn{1}{c}{$10^{-4}$} & \multicolumn{1}{c}{$10^{-5}$} & \multicolumn{1}{c}{$10^{-6}$} & \multicolumn{1}{c}{$10^{-7}$} & \multicolumn{1}{c}{$10^{-8}$} & \multicolumn{1}{c}{$10^{-9}$} & \multicolumn{1}{c}{$10^{-10}$} & \multicolumn{1}{c}{$0.0$}\\
        \hline
        1 & 57 & 57 & 57 & 57 & 57 & 57 & 57 & 57 & 57\\
        2 & 249 & 873 & 1,193 & 1,421 & 1,569 & 1,597 & 1,597 & 1,597 & 1,597\\
        3 & 305 & 1,489 & 4,241 & 12,713 & 18,933 & 23,693 & 27,161 & 28,645 & 29,317\\
        4 & 305 & 1,489 & 5,011 & 18,901 & 49,713 & 126,715 & 205,803 & 274,249 & 396,607\\
        5 & 305 & 1,489 & 5,011 & 19,013 & 55,173 & 170,297 & 418,589 & 931,551 & 4,216,423\\
        6 & 305 & 1,489 & 5,011 & 19,013 & 55,201 & 171,557 & 444,979 & 1,137,239 & 36,684,859\\
        \hline
        \hline
    \end{tabular}
        \caption{\textbf{Number of auxiliary density operators for different truncation values of the fermionic hierarchy level and different thresholds of the importance value.} Here, we use the parameters from example 1 (see Table~\ref{tab:parameters}), where $n_{\textrm{max}}$ is the truncation tier of the fermionic hierarchy and $\mathcal{I}_{\textrm{th}}$ is the importance threshold. Note that we only neglect the ADOs in the second and higher levels ($n\geq2$) whose importance value is smaller than $\mathcal{I}_\textrm{th}$.}
    \label{tab:ADOs_num}
\end{table}

\begin{figure}[tb]
    \centering
    \includegraphics[width=9cm]{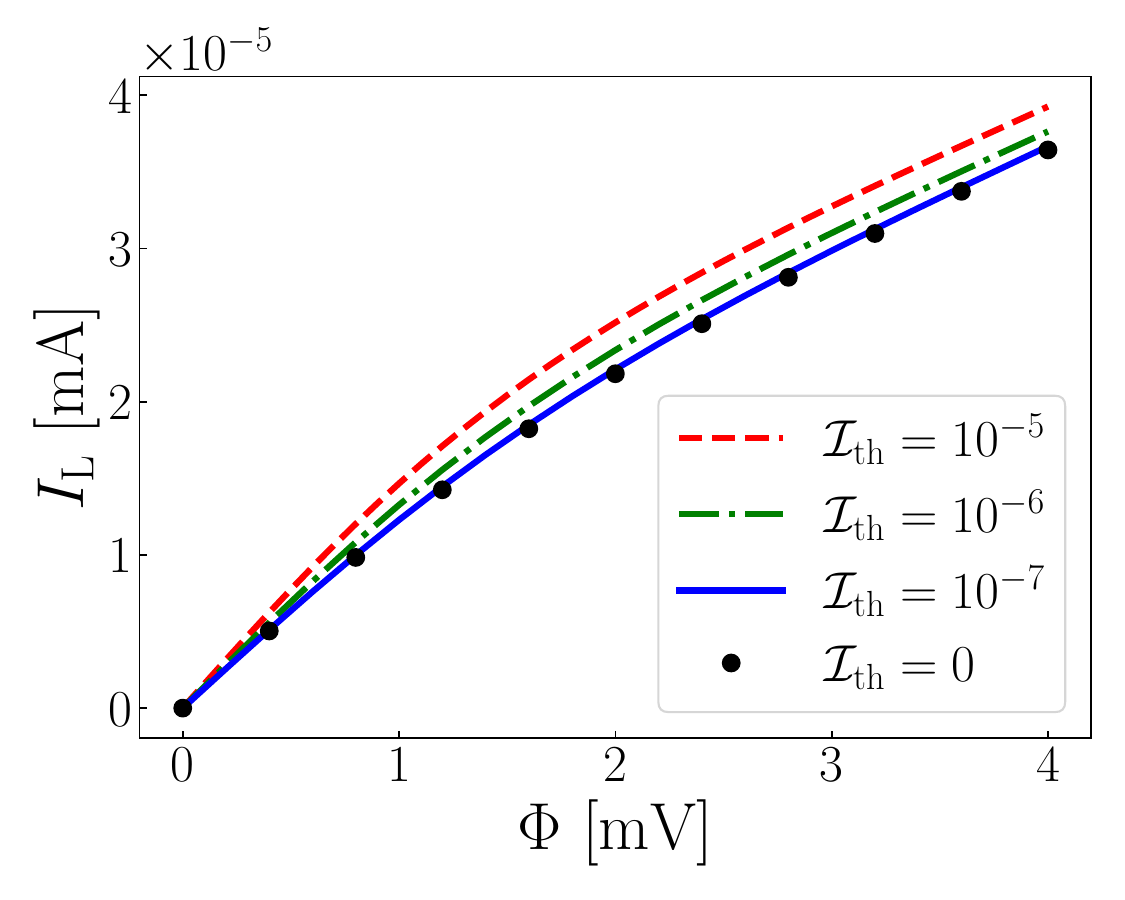}
    \caption{\textbf{Convergence of the electronic current with respect to the importance threshold.} Here, we consider the parameters setting in example 1 (see Table~\ref{tab:parameters}) and show the electronic current $
    \Symbol{I}{L}$ (from the left-hand side fermionic reservoir into the system) as a function of bias voltage $\Phi$ where the red dashed, green dash-dotted, blue solid curves, and black dots correspond to the importance threshold $\Symbol{\mathcal{I}}{th}$ at $10^{-5}$, $10^{-6}$, $10^{-7}$, and $0$, respectively.
    The values of electronic current converge when $\Symbol{\mathcal{I}}{th}=10^{-7}$, which reduces the number of ADOs from 396,607 ($\Symbol{\mathcal{I}}{th}=0$) to 49,713, as shown in Table~\ref{tab:ADOs_num}.
    }
    \label{fig:importance_coverge}
\end{figure}

We first emphasize the advantage of specifying the importance threshold $\Symbol{\mathcal{I}}{th}$ and estimating the importance for each ADOs with Eq.~(\ref{eq:importance}), which are not available for other packages listed in Table~\ref{tab:compare_pkg}. The main computational complexity can be quantified by the total number of ADOs because it directly affects the size of $\HEOMLS$. The importance criterion $\mathcal{I}\left(\rho_{\bfj|\bfq}^{(m,n,p)}\right)\geq\Symbol{\mathcal{I}}{th}$ allows us to only consider the ADOs which affects the dynamics more, and thus, reduce the size of $\HEOMLS$. Table~\ref{tab:ADOs_num} summarizes the number of ADOs with respect to different truncation level $\Symbol{n}{max}$ and importance thresholds $\mathcal{I}_{\textrm{th}}$. We observe a reduction in the number of auxiliary density operators when $\mathcal{I}_{\textrm{th}}$ increases.

We further discuss the consistency of the importance criterion and consider the parameter settings in example 1 (see Table~\ref{tab:parameters}). In this case, we first increase the truncation level until we find out that the values of electronic current converges at $\Symbol{n}{max}=4$, which requires 396,607 ADOs (as shown in Table.~\ref{tab:ADOs_num}) to describe the dynamics. The effects on the convergence of the electronic current while using the importance criterion are shown in Fig.~\ref{fig:importance_coverge}. For the equilibrium case ($\Phi=0$), one can observe that the electronic current already converged at $\Symbol{\mathcal{I}}{th}=10^{-5}$. However, by increasing the bias voltage to non-equilibrium $\Phi>0$, the results obtained with a higher $\Symbol{\mathcal{I}}{th}$ show larger deviation from the converged results. The results converge at $\Symbol{\mathcal{I}}{th}=10^{-7}$ and require only 49,713 ADOs (as shown in Table.~\ref{tab:ADOs_num}). For too large thresholds $\Symbol{\mathcal{I}}{th}\geq 10^{-4}$, we observe unphysical negative electronic currents or even fail to obtain the stationary states of the ADOs.

\begin{figure}[tb]
\centering \includegraphics[width=18cm]{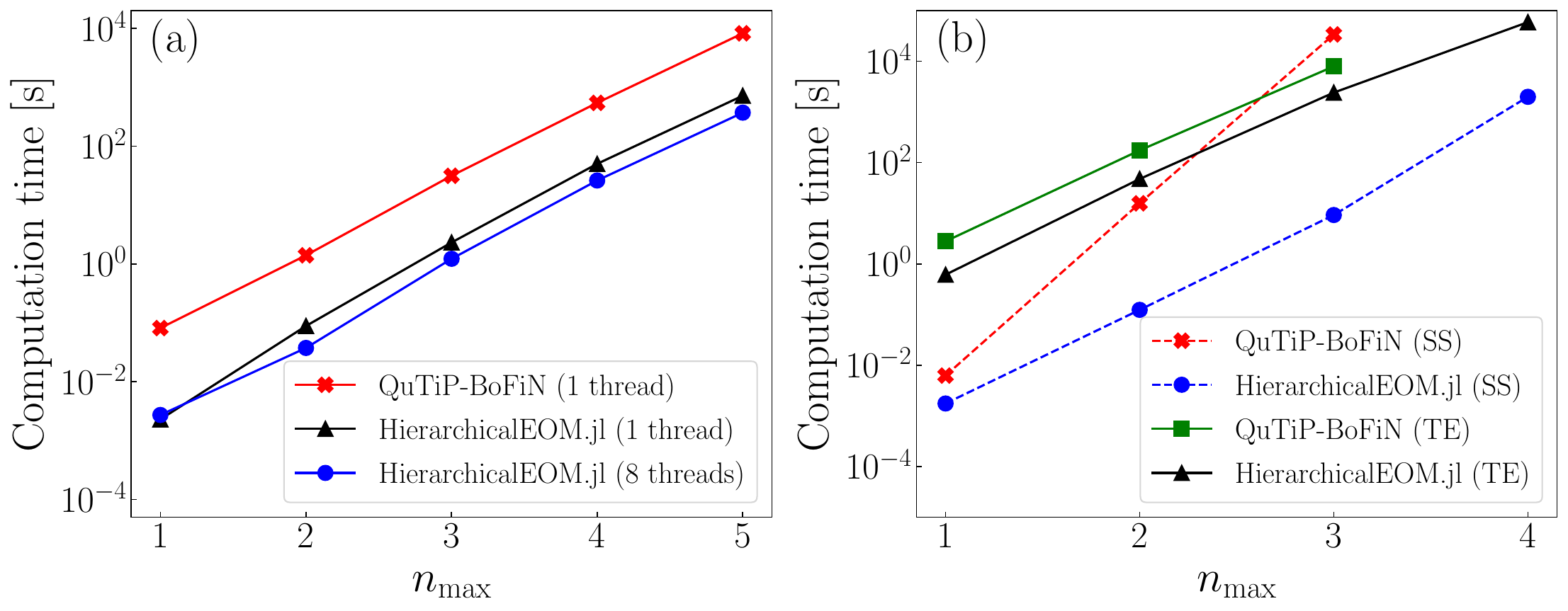}
\caption{
\textbf{Benchmark of the performance between HierarchicalEOM.jl and QuTiP-BoFiN.} We consider the parameters used in example 1 (see Table~\ref{tab:parameters}). Note that \code{QuTiP-BoFiN}~\cite{QuTiP-BoFiN,QuTiP,QuTiP2} currently does not support the estimation of importance for ADOs, and thus, we set $\Symbol{\mathcal{I}}{th}=0$ here. The benchmark was done on a workstation with two Intel(R) Xeon(R) Silver 4110 @2.10GHz CPUs and 256 GB memory. The software versions were \code{QuTiP} 4.7.1, \code{HierarchicalEOM.jl} 0.1.0, \code{Python} 3.7.6, and \code{Julia} 1.8.2. The operating system was CentOS Linux 8.1.1911. In ($\textbf{a}$), we show the total runtime versus truncation level $\Symbol{n}{max}$ for the construction of the HEOMLS matrices $\HEOMLS$ using \code{QuTiP-BoFiN} (red crosses) and \code{HierarchicalEOM.jl} (black tirangles and blue circles represent 1 and 8 threads used during the process of constructing the matrices, respectively). Note that \code{QuTiP-BoFiN} currently only supports single-thread process for the construction of $\HEOMLS$. In ($\textbf{b}$), we show the total runtime versus truncation level $\Symbol{n}{max}$ for calculating the stationary states (SS), represented as dashed lines, and time evolution (TE), represented as solid lines, of the auxiliary density operators using \code{QuTiP-BoFiN} (red crosses and green squares) and \code{HierarchicalEOM.jl} (blue circles and black triangles).
}
\label{fig:compare}
\end{figure}

We now compare the performance in constructing HEOMLS matrix $\HEOMLS$, solving time evolution (TE) of ADOs, and solving the stationary states (SS) of ADOs between \code{HierarchicalEOM.jl} and \code{QuTiP-BoFiN}~\cite{QuTiP-BoFiN,QuTiP,QuTiP2}. This choice is motivated by the fact that \code{QuTiP-BoFiN} served as an inspiration and guide for the development of this package, and thus has similar features, and by the fact that they are both purely written in high-level languages (\code{Python} and \code{Julia}). For this comparison, we consider the same setting in example 1 (see Table~\ref{tab:parameters}). The benchmark result of constructing $\HEOMLS$ are shown in Fig.~\ref{fig:compare}(a). \code{HierarchicalEOM.jl} not only improves the computational time on single-thread processing but also supports multi-threading to make the process of constructing $\HEOMLS$ faster. Note that we only consider the importance threshold $\mathcal{I}_{\textrm{th}}=0.0$ in this benchmark because \code{QuTiP-BoFiN} does not support this functionality at this time. On top of the speed up in constructing $\HEOMLS$, we emphasize that \code{HierarchicalEOM.jl} is also faster in computing the time evolution and stationary states of the ADOs, as shown in Fig.~\ref{fig:compare}(b), largely because of underlying performance beneftis of the numerical libraries in  \code{Julia}~\cite{DifferentialEquations.jl-2017,LinearSolve.jl,FastExpm.jl-2011}. For example, \code{DifferentialEquations.jl}~\cite{DifferentialEquations.jl-2017} provides a unified user interface to solve the differential equations with various choices of solver. \code{HierarchicalEOM.jl} wrapped a subset  of the functions in \code{DifferentialEquations.jl} for solving TE of the ADOs, and users can optimize their solver choice depending on the problem. We have also wrapped some of the functions in \code{LinearSolver.jl}~\cite{LinearSolve.jl} for users to solve the stationary states of ADOs and the spectrum of the system with different solvers. These functionalities improve both runtime and accuracy compared to the libraries currently available in  \code{Python}.

\section{Conclusion}
In conclusion, the \code{HierarchicalEOM.jl} software package provides a user-friendly and efficient tool for simulating complex open quantum systems, including non-Markovian effects due to non-perturbative interaction with one, or multiple, environments. The following characteristics define the package:
\begin{itemize}
    \item It decreases the number of ADOs and the size of the HEOMLS matrices by allowing the users to specify importance thresholds~\cite{Hartle2013,Thoss_2_2021}.
    \item It provides built-in functions to calculate the spectrum for both bosonic and fermionic systems based on \code{LinearSolve.jl}~\cite{LinearSolve.jl}.
    \item It provides a dictionary translating the index of the ADOs to the corresponding bath exponential-expansion series, and vice versa. This allows an easy way to interrogate the ADOs to gain access to bath properties, like electronic and heat currents.
    \item Users can choose to construct the HEOMLS matrices within even- or odd-parity. It depends on the parity of the operator which the HEOMLS matrix is acting on.
    \item It provides built-in functions for the users to further add Lindbladians to the HEOMLS matrices while keeping the size of HEOMLS matrices unchanged (it does not extend the ADO space). The Lindbladian describes the dissipative interactions (under Born-Markov approximation~\cite{Breuer2007}) between the system and extra environments.
\end{itemize}
Moreover, \code{HierarchicalEOM.jl} also takes advantage of other available packages for the following features: 
\begin{itemize}
    \item It is written in \code{Julia}.
    \item It supports different choices of spectral densities and spectral decomposition methods to accurately compute bath correlation functions.
    \item It supports time-dependent system Hamiltonians.
    \item It constructs the HEOMLS matrices for different types (bosonic, fermionic, or hybrid) of baths.
    \item The HEOMLS matrices are constructed using multi-threading.
    \item It provides different methods based on \code{DifferentialEquations.jl}~\cite{DifferentialEquations.jl-2017}, \code{LinearSolve.jl}~\cite{LinearSolve.jl}, and \code{FastExpm.jl}~\cite{FastExpm.jl-2011} to compute the dynamics and stationary states of all ADOs.
\end{itemize}

We exemplified the functionalities of \code{HierarchicalEOM.jl} using two physics examples where we observed order of magnitudes speedups with respect to \code{QuTiP-BoFiN} in constructing HEOMLS matrices and solving dynamics and stationary states for all ADOs. As a result, we believe that \code{HierarchicalEOM.jl} will be a valuable tool for researchers working in different fields such as quantum biology, quantum optics, quantum thermodynamics, quantum information, quantum transport, and condensed matter physics. 

We further plan to extend the software for future releases in two different directions. First, we aim to add fitting routines to the decomposition methods used to compute correlation functions, thereby expanding the domain of spectral densities supported by the HEOM approach. Second, we aim to improve the computational efficiency of the package by reducing the complexity of the HEOM (for example imposing Hermiticity~\cite{Hartle2013, Yan_5_2016}) and by including support for distributed computing on clusters and GPU computing.

\section{Data Availability}
The detailed information for reproducing the numerical data in both examples is available through a public GitHub repo (\href{https://github.com/ytdHuang/HierarchicalEOM.jl-Examples}{https://github.com/ytdHuang/HierarchicalEOM.jl-Examples}). All other data are available upon reasonable request.

\section{Code Availability}
The \code{HierarchicalEOM.jl} package is available through a public GitHub repo (\href{https://github.com/NCKU-QFort/HierarchicalEOM.jl}{https://github.com/NCKU-QFort/HierarchicalEOM.jl}). It is also registered in the \code{Julia} package registry and can be installed by the \code{Julia} package manager. Moreover, detailed information (documentation and other examples) is available through a public website (\href{https://ncku-qfort.github.io/HierarchicalEOM.jl}{https://ncku-qfort.github.io/HierarchicalEOM.jl}).

\section{Acknowledgements}
The authors acknowledge fruitful discussions with Yuan-Ching Lin and Chung-Hsuan Tung. N.L.~acknowledges the Information Systems Division, RIKEN, for the use of their facilities. M.C. acknowledges support from NSFC (Grants No.~12050410264 and No.~11935012) and NSAF (Grant No.~U1930403). F.N. is supported in part by: Nippon Telegraph and Telephone Corporation (NTT) Research, the Japan Science and Technology Agency (JST) [via the Quantum Leap Flagship Program (Q-LEAP), and the Moonshot R\&D Grant Number JPMJMS2061], Office of Naval Research (ONR), and the Asian Office of Aerospace Research and Development (AOARD) (via Grant No. FA2386-20-1-4069). F.N. and N.L. acknowledge the Foundational Questions Institute Fund (FQXi) via Grant No.~FQXi-IAF19-06. YNC acknowledges the support of the National Center for Theoretical Sciences and the National Science and Technology Council, Taiwan (NSTC Grants No. 112-2123-M-006-001). 

\section{Methods}

\subsection{Table of symbols}
Table~\ref{tab:symbols} summarizes the most relevant symbols used throughout the article.

\begin{table}[h]
    \centering
    \begin{tabular}{c|l}
        \hline
        \hline
        Symbol & Description\\
        \hline
        $\sigma_{\textrm{b}}$ & specification of different quantum numbers of the bosonic system\\
        $\sigma_{\textrm{f}}$ & specification of different quantum numbers of the fermionic system\\
        $\beta$ & specification of different bosonic baths\\
        $\alpha$ & specification of different fermionic baths\\
        $[\cdot, \cdot]_-$ & commutator\\
        $[\cdot, \cdot]_+$ & anti-commutator\\
        $\nu\in\{+1,-1\}$ & presence ($+1$) or absence ($-1$) of Hermitian conjugation\\
        $\bar{\nu}$ & opposite of $\nu$: $\bar{\nu}\equiv-\nu$\\
        $p\in\{+,-\}$ & even-parity ($+$) or odd-parity ($-$) of the operator\\
        $u\in\{\mathbb{R},\mathbb{I}\}$ & Real ($\mathbb{R}$) or imaginary ($\mathbb{I}$) part of bosonic correlation function\\
        
        $C(t_1, t_2)$ & two time correlation function of the bath\\
        $n^{\textrm{eq}}(\omega)$ & Fermi-Dirac or Bose-Einstein equilibrium distribution\\
        $J(\omega)$ & spectral density of the bath\\
        $\hat{\mathcal{G}}[\cdot]$ & superoperator which propagates all auxiliary density operators (ADOs)\\
        $\hat{\mathcal{M}}[\cdot]$ & HEOM Liouvillian superoperator (matrix)\\
        $\hat{\mathcal{J}}(F)[\cdot]$ & Lindbladian superoperator with jump operator $F$\\
        $\rho_{\bfj|\bfq}^{(m,n,p)}$ & auxiliary density operator (ADO)\\
        $\rho_{|}^{(0,0,+)}$ & reduced density operator\\
        $m$ & bosonic hierarchy level\\
        $n$ & fermionic hierarchy level\\
        $\bfj$ & vector of bosonic multi-index ensembles: $[j_m, \cdots, j_1]$\\
        $\bfq$ & vector of fermionic multi-index ensembles: $[q_n, \cdots, q_1]$\\
        $\mathcal{I}\left(\rho_{\bfj|\bfq}^{(m,n,p)}\right)$ & importance value of a given auxiliary density operator (ADO)\\
        $\mathcal{I}_\textrm{th}$ & threshold of importance value\\
        $A(\omega)$ & spectrum of fermionic system (density of states, DOS)\\
        $S(\omega)$ & spectrum of bosonic system (power spectral density, PSD)\\
        $I_\alpha$ & electronic current from $\alpha$-fermionic bath into the system\\
        $G$ & conductance\\
        \hline
        \hline
    \end{tabular}
    \caption{\textbf{List of symbols.} We summarize the most relevant symbols used in this article.}
    \label{tab:symbols}
\end{table}

\subsection{Fermion correlation function with Lorentzian spectral density}
In this subsection, we present an example for the decomposition of a $\alpha$-fermionic reservoir correlation function as a sum of $N_\alpha$-exponential terms. In particular, we focus on the following Lorentzian spectral density
\begin{equation}\label{eq:Lorentz}
     J_\alpha(\omega)=\frac{\Gamma_\alpha W_\alpha^2}{(\omega-\mu_\alpha)^2+W_\alpha^2},
\end{equation}
where $\Gamma_\alpha$ represents the coupling strength between system and $\alpha$-fermionic reservoir with band-width $W_\alpha$ and chemical potential $\mu_\alpha$. We now express the Fermi-Dirac distribution $n_{\alpha}^{\textrm{eq}}(x)=\{\exp(x)+1\}^{-1}$ as a series by employing the Pad\'{e} decomposition~\cite{Jie2011} (which has been reported~\cite{QuTiP-BoFiN, Pochen2023} to enjoy a faster convergence than the Matsubara one~\cite{Shi2009}) to obtain
\begin{equation}
    n_{\alpha}^{\textrm{eq}}(x)\approx n_\alpha^\textrm{Pad\'{e}}(x)=\frac{1}{2}-\sum_{h=2}^{N_\alpha}\frac{2\kappa_h x}{x^2+\zeta_h^2},
\end{equation}
where the parameters $\kappa_h$ and $\zeta_h$ are described in previous work~\cite{Jie2011} and depend on the choice of $N_\alpha$. Thus, the integration in Eq.~(\ref{eq:C_f}) can be analytically performed  using the residue theorem and results in Eq.~(\ref{eq:C_f_exp}), which explicitly reads
\begin{equation}
    C_\alpha^\nu(\tau)\approx\sum_{h=1}^{N_\alpha}\eta_{\alpha,h}^{\nu}\exp(-\gamma_{\alpha,h}^\nu \tau)
\end{equation}
with
\begin{equation}
\begin{aligned}
    \eta_{\alpha,1}^\nu&=\frac{\Gamma_\alpha W_\alpha}{2} n_\alpha^\textrm{Pad\'{e}}(\frac{iW_\alpha}{k_{\textrm{B}} T}),\\
    \gamma_{\alpha,1}^\nu&=W_\alpha-\nu i \mu_\alpha,\\
    \eta_{\alpha,h\neq1}^\nu &= -i \kappa_h k_{\textrm{B}} T \cdot \frac{\Gamma_\alpha W_\alpha^2}{-(\zeta_h k_{\textrm{B}} T)^2+W_\alpha^2},\\
    \eta_{\alpha,h\neq1}^\nu &= \zeta_h k_{\textrm{B}} T - \nu i \mu_\alpha.\\
\end{aligned}
\end{equation}

\subsection{Boson correlation function with Drude-Lorentz spectral density}
In this subsection, we present an example for the decomposition of a $\beta$-bosonic reservoir correlation function as a sum of $N_\beta$-exponential terms. In particular, we focus on the following Drude-Lorentz spectral density
\begin{equation}\label{eq:Drude-Lorentz}
     J_\beta(\omega)=\frac{4\Delta_\beta W_\beta\omega}{\omega^2+W_\beta^2},
\end{equation}
in which $\Delta_\beta$ represents the coupling strength between the system and the $\beta$-bosonic reservoir with band-width $W_\beta$. Similarly as in the fermionic case, we can now express the Bose-Einstein distribution $n_{\beta}^{\textrm{eq}}(x)=\{\exp(x)-1\}^{-1}$ as a series by employing the Pad\'{e} decomposition~\cite{Jie2011}, to obtain
\begin{equation}
    n_{\beta}^{\textrm{eq}}(x)\approx n_\beta^\textrm{Pad\'{e}}(x)=\frac{1}{x}-\frac{1}{2}+\sum_{l=2}^{N_\beta}\frac{2\kappa_l x}{x^2+\zeta_l^2},
\end{equation}
where the parameters $\kappa_l$ and $\zeta_l$ are described in previous work~\cite{Jie2011} and depend on the choice of $N_\beta$. Using this decomposition, the integration in Eq.~(\ref{eq:C_b}) can be analytically performed  using the residue theorem and results in Eq.~(\ref{eq:C_b_exp}), which explicitly reads
\begin{equation}
    C_\beta(\tau)\approx\sum_{l=1}^{N_\beta}\xi_{\beta,l}\exp(-\chi_{\beta,l} \tau)
\end{equation}
with
\begin{equation}
\begin{aligned}
    \xi_{\beta,1} &= \Delta_\beta W_\beta \left[-i+\cot\left(\frac{W_\beta}{2 k_{\textrm{B}} T}\right)\right],\\
    \chi_{\beta,1} &=W_\beta,\\
    \xi_{\beta,l\neq1} &= -2 \kappa_l k_{\textrm{B}} T \cdot \frac{2 \Delta_\beta W_\beta \cdot \zeta_l k_{\textrm{B}} T}{-(\zeta_l k_{\textrm{B}} T)^2 + W_\beta^2},\\
    \chi_{\beta,l\neq1} &= \zeta_l k_{\textrm{B}} T.\\
\end{aligned}
\end{equation}
Contrary to the fermionic case, we note that, here, the correlation function does not have to be be further decomposed into its real and imaginary part since $\chi_{\beta, l}=\chi_{\beta, l}^* ~\forall~ l$.

\subsection{Numerical computation of the spectrum}
In this subsection, we briefly summarize how to numerically compute the spectrum associated with the system degree of freedom. Previous work~\cite{Yan_2_PRL} showed that the spectrum can be evaluated either in time or frequency domain. \code{HierarchicalEOM.jl} provides a built-in function which performs the calculation in frequency domain. In the following, we first focus on the bosonic case (to compute the power spectral density) and, next, on fermionic case (to compute the density of states).

In order to compute the power spectral density in the time-domain, we write the system correlation function in terms of the propagator $\hat{\mathcal{G}}(t)=\exp(\HEOMLS t)$, for $t>0$, when the system Hamiltonian is time-independent. The power spectral density associated with $\Symbol{\sigma}{b}$ can be obtained as
\begin{equation}
\begin{aligned}
    \pi S_{\Symbol{\sigma}{b}}(\omega)&=\textrm{Re}\left\{\int_0^\infty dt \langle a_{\Symbol{\sigma}{b}}^\dagger(t)a_{\Symbol{\sigma}{b}}(0)\rangle^* e^{i\omega t}\right\}\\
    &=\textrm{Re}\left\{\int_0^\infty dt \langle a_{\Symbol{\sigma}{b}}^\dagger(t)a_{\Symbol{\sigma}{b}}(0)\rangle e^{-i\omega t}\right\}\\
    &=\textrm{Re}\left\{\int_0^\infty dt \langle a_{\Symbol{\sigma}{b}}^\dagger e^{\HEOMLS t}a_{\Symbol{\sigma}{b}}\rangle e^{-i\omega t}\right\}\\
    &=\textrm{Re}\left\{\int_0^\infty dt \langle a_{\Symbol{\sigma}{b}}^\dagger e^{(\HEOMLS -i\omega)t}a_{\Symbol{\sigma}{b}}\rangle\right\}\\
    &=\textrm{Re}\left\{-\langle a_{\Symbol{\sigma}{b}}^\dagger (\HEOMLS -i\omega)^{-1} a_{\Symbol{\sigma}{b}}\rangle\right\}\\
    &=-\textrm{Re}\left\{\tr\left[ a_{\Symbol{\sigma}{b}}^\dagger (\HEOMLS -i\omega)^{-1} a_{\Symbol{\sigma}{b}}\rho^{(m,n,+)}_{\bfj \vert \bfq}\right]\right\}\\
    &=-\textrm{Re}\left\{\tr\left[ a_{\Symbol{\sigma}{b}}^\dagger \textbf{x}\right]\right\},
\end{aligned}
\end{equation}
where a half-Fourier transform has been introduced in the fifth line. To determine $\textbf{x}$ at a fixed frequency $\omega$, one can express all of the ADOs in a vector form and then solve the linear problem $\textbf{Ax}=\textbf{b}$ (where $\textbf{A}=\HEOMLS-i\omega$ and $\textbf{b}=a_{\Symbol{\sigma}{b}}\rho^{(m,n,+)}_{\bfj \vert \bfq}$) using  the package \code{LinearSolve.jl}~\cite{LinearSolve.jl}. We note that, while in principle all the ADOs are required to solve for $\textbf{x}$, only the reduced density operator ($m=0$ and $n=0$) is considered when taking the final trace.

In order to compute the density of states, we start from the retarded Green's function~\cite{Yan_2_PRL}
\begin{equation}
\begin{array}{lll}
G^{\textrm{R}}_{\Symbol{\sigma}{f}}(t) &=& -i\Theta(t)
                            \Big\{ \langle d_{\Symbol{\sigma}{f}}(t) d_{\Symbol{\sigma}{f}}^\dagger(0)\rangle + 
                            \langle d_{\Symbol{\sigma}{f}}^\dagger(t) d_{\Symbol{\sigma}{f}}(0)\rangle^{\ast} \Big\},
\end{array}
\end{equation}
in which the Heaviside function $\Theta(t)$ is needed to impose causality. Similarly to the bosonic case, the density of states associated with $\Symbol{\sigma}{f}$ can be written in  compact form as
\begin{equation}
\begin{aligned}
    \pi A_{\Symbol{\sigma}{f}}(\omega)&=-\textrm{Im}\int_{-\infty}^{\infty}G^{\textrm{R}}_{\Symbol{\sigma}{f}}(t)e^{i\omega t} dt \\
    &=\textrm{Re}\left\{\int_0^\infty dt \left[\langle d_{\Symbol{\sigma}{f}}(t) d_{\Symbol{\sigma}{f}}^\dagger(0)\rangle + \langle d_{\Symbol{\sigma}{f}}^\dagger(t) d_{\Symbol{\sigma}{f}}(0)\rangle ^* \right] e^{i\omega t}\right\}\\
    &=-\textrm{Re}\left\{\tr\left[ d_{\Symbol{\sigma}{f}} (\HEOMLS + i\omega)^{-1} d_{\Symbol{\sigma}{f}}^\dagger\rho^{(m,n,+)}_{\bfj \vert \bfq}\right]+\tr\left[ d_{\Symbol{\sigma}{f}}^\dagger (\HEOMLS -i\omega)^{-1} d_{\Symbol{\sigma}{f}}\rho^{(m,n,+)}_{\bfj \vert \bfq}\right]\right\}\\
    &=-\textrm{Re}\left\{\tr\left[ d_{\Symbol{\sigma}{f}} \textbf{x}_{+}\right]+\tr\left[ d_{\Symbol{\sigma}{f}}^\dagger \textbf{x}_{-}\right]\right\},
\end{aligned}
\end{equation}
where $\textbf{x}_{+}$ is determined by solving 
\begin{equation}
    (\HEOMLS + i\omega) \textbf{x}_{+} = d_{\Symbol{\sigma}{f}}^\dagger\rho^{(m,n,+)}_{\bfj \vert \bfq},
\end{equation}
and $\textbf{x}_{-}$ is determined by solving
\begin{equation}
    (\HEOMLS - i\omega) \textbf{x}_{-} = d_{\Symbol{\sigma}{f}}\rho^{(m,n,+)}_{\bfj \vert \bfq}.
\end{equation}
Here, the HEOMLS matrix $\HEOMLS$ acts on the odd-parity ($p=-$) space, compatibly with the parity of both the operators $d_{\Symbol{\sigma}{f}}\rho^{(m,n,+)}_{\bfj \vert \bfq}$ and $d_{\Symbol{\sigma}{f}}^\dagger\rho^{(m,n,+)}_{\bfj \vert \bfq}$. As for the computation of the power spectral density, we only need the reduced density operator ($m=0$ and $n=0$) to take the final trace operation.

\subsection{Electronic current}
In this subsection, we discuss how to compute an environmental observable: the electronic current. Within the influence functional approach~\cite{Yan_1,Hartle2013}, the expectation value of the electronic current from the $\alpha$-fermionic bath into the system can be written in terms of the first-level-fermionic ($n=1$) auxiliary density operators, namely
\begin{equation}
\label{eq:last}
\begin{aligned}
\langle I_\alpha(t) \rangle &=(-e) \frac{d\langle \mathcal{N}_\alpha\rangle}{dt}\\
&=i e \sum_{\nu, h, 
\Symbol{\sigma}{f}}(-1)^{\delta_{\nu,-}} ~\textrm{Tr}\left[d_{\Symbol{\sigma}{f}}^{\bar{\nu}}\rho^{(0,1,+)}_{\vert \{\alpha,\nu,h,\Symbol{\sigma}{f}\}}(t)\right],
\end{aligned}
\end{equation}
where $e$ represents the value of the elementary charge, and $\mathcal{N}_\alpha=\sum_k c^\dagger_{\alpha,k}c_{\alpha,k}$ is the occupation number operator for the $\alpha$-fermionic bath. To compute this quantity, it is possible to first use \code{HierarchicalEOM.jl} to obtain the stationary states for the ADOs $\rho^{(m,n,+)}_{\bfj\vert\bfq}(t)$, and then use the hierarchy dictionary [as shown in Fig.~\ref{fig:HierarchicalEOM.jl}(e)] to identify all the zeroth-level-bosonic-and-first-level-fermionic ($m=0$ and $n=1$) ADOs and their corresponding indices $\alpha$, $\nu$, $h$, and $\Symbol{\sigma}{f}$ to be used in Eq.~(\ref{eq:last}).


\begin{thebibliography}{84}%
\makeatletter
\providecommand \@ifxundefined [1]{%
 \@ifx{#1\undefined}
}%
\providecommand \@ifnum [1]{%
 \ifnum #1\expandafter \@firstoftwo
 \else \expandafter \@secondoftwo
 \fi
}%
\providecommand \@ifx [1]{%
 \ifx #1\expandafter \@firstoftwo
 \else \expandafter \@secondoftwo
 \fi
}%
\providecommand \natexlab [1]{#1}%
\providecommand \enquote  [1]{``#1''}%
\providecommand \bibnamefont  [1]{#1}%
\providecommand \bibfnamefont [1]{#1}%
\providecommand \citenamefont [1]{#1}%
\providecommand \href@noop [0]{\@secondoftwo}%
\providecommand \href [0]{\begingroup \@sanitize@url \@href}%
\providecommand \@href[1]{\@@startlink{#1}\@@href}%
\providecommand \@@href[1]{\endgroup#1\@@endlink}%
\providecommand \@sanitize@url [0]{\catcode `\\12\catcode `\$12\catcode `\&12\catcode `\#12\catcode `\^12\catcode `\_12\catcode `\%12\relax}%
\providecommand \@@startlink[1]{}%
\providecommand \@@endlink[0]{}%
\providecommand \url  [0]{\begingroup\@sanitize@url \@url }%
\providecommand \@url [1]{\endgroup\@href {#1}{\urlprefix }}%
\providecommand \urlprefix  [0]{URL }%
\providecommand \Eprint [0]{\href }%
\providecommand \doibase [0]{https://doi.org/}%
\providecommand \selectlanguage [0]{\@gobble}%
\providecommand \bibinfo  [0]{\@secondoftwo}%
\providecommand \bibfield  [0]{\@secondoftwo}%
\providecommand \translation [1]{[#1]}%
\providecommand \BibitemOpen [0]{}%
\providecommand \bibitemStop [0]{}%
\providecommand \bibitemNoStop [0]{.\EOS\space}%
\providecommand \EOS [0]{\spacefactor3000\relax}%
\providecommand \BibitemShut  [1]{\csname bibitem#1\endcsname}%
\let\auto@bib@innerbib\@empty
\bibitem [{\citenamefont {Zwanzig}(1960)}]{Robert1960}%
  \BibitemOpen
  \bibfield  {author} {\bibinfo {author} {\bibfnamefont {R.}~\bibnamefont {Zwanzig}},\ }\bibfield  {title} {\bibinfo {title} {Ensemble method in the theory of irreversibility},\ }\href {https://doi.org/10.1063/1.1731409} {\bibfield  {journal} {\bibinfo  {journal} {J. Chem. Phys.}\ }\textbf {\bibinfo {volume} {33}},\ \bibinfo {pages} {1338} (\bibinfo {year} {1960})}\BibitemShut {NoStop}%
\bibitem [{\citenamefont {Feynman}\ and\ \citenamefont {Vernon}(1963)}]{Vernon1963}%
  \BibitemOpen
  \bibfield  {author} {\bibinfo {author} {\bibfnamefont {R.}~\bibnamefont {Feynman}}\ and\ \bibinfo {author} {\bibfnamefont {F.}~\bibnamefont {Vernon}},\ }\bibfield  {title} {\bibinfo {title} {The theory of a general quantum system interacting with a linear dissipative system},\ }\href {https://doi.org/10.1016/0003-4916(63)90068-X} {\bibfield  {journal} {\bibinfo  {journal} {Ann. Phys.}\ }\textbf {\bibinfo {volume} {24}},\ \bibinfo {pages} {118} (\bibinfo {year} {1963})}\BibitemShut {NoStop}%
\bibitem [{\citenamefont {Caldeira}\ and\ \citenamefont {Leggett}(1983)}]{Caldeira1983}%
  \BibitemOpen
  \bibfield  {author} {\bibinfo {author} {\bibfnamefont {A.}~\bibnamefont {Caldeira}}\ and\ \bibinfo {author} {\bibfnamefont {A.}~\bibnamefont {Leggett}},\ }\bibfield  {title} {\bibinfo {title} {Path integral approach to quantum {B}rownian motion},\ }\href {https://doi.org/10.1016/0378-4371(83)90013-4} {\bibfield  {journal} {\bibinfo  {journal} {Physica A}\ }\textbf {\bibinfo {volume} {121}},\ \bibinfo {pages} {587} (\bibinfo {year} {1983})}\BibitemShut {NoStop}%
\bibitem [{\citenamefont {Hedeg{\aa}rd}\ and\ \citenamefont {Caldeira}(1987)}]{Caldeira1987}%
  \BibitemOpen
  \bibfield  {author} {\bibinfo {author} {\bibfnamefont {P.}~\bibnamefont {Hedeg{\aa}rd}}\ and\ \bibinfo {author} {\bibfnamefont {A.~O.}\ \bibnamefont {Caldeira}},\ }\bibfield  {title} {\bibinfo {title} {Quantum dynamics of a particle in a {F}ermionic environment},\ }\href {https://doi.org/10.1088/0031-8949/35/5/001} {\bibfield  {journal} {\bibinfo  {journal} {Phys. Scripta}\ }\textbf {\bibinfo {volume} {35}},\ \bibinfo {pages} {609} (\bibinfo {year} {1987})}\BibitemShut {NoStop}%
\bibitem [{\citenamefont {Chen}\ \emph {et~al.}(2015)\citenamefont {Chen}, \citenamefont {Lambert}, \citenamefont {Cheng}, \citenamefont {Chen},\ and\ \citenamefont {Nori}}]{HongBin2015}%
  \BibitemOpen
  \bibfield  {author} {\bibinfo {author} {\bibfnamefont {H.-B.}\ \bibnamefont {Chen}}, \bibinfo {author} {\bibfnamefont {N.}~\bibnamefont {Lambert}}, \bibinfo {author} {\bibfnamefont {Y.-C.}\ \bibnamefont {Cheng}}, \bibinfo {author} {\bibfnamefont {Y.-N.}\ \bibnamefont {Chen}},\ and\ \bibinfo {author} {\bibfnamefont {F.}~\bibnamefont {Nori}},\ }\bibfield  {title} {\bibinfo {title} {{Using non-{M}arkovian measures to evaluate quantum master equations for photosynthesis}},\ }\href {https://doi.org/10.1038/srep12753} {\bibfield  {journal} {\bibinfo  {journal} {Sci. Rep.}\ }\textbf {\bibinfo {volume} {5}},\ \bibinfo {pages} {12753} (\bibinfo {year} {2015})}\BibitemShut {NoStop}%
\bibitem [{\citenamefont {Tanimura}\ and\ \citenamefont {Kubo}(1989)}]{Tanimura_2_1989}%
  \BibitemOpen
  \bibfield  {author} {\bibinfo {author} {\bibfnamefont {Y.}~\bibnamefont {Tanimura}}\ and\ \bibinfo {author} {\bibfnamefont {R.}~\bibnamefont {Kubo}},\ }\bibfield  {title} {\bibinfo {title} {Time evolution of a quantum system in contact with a nearly {G}aussian-{M}arkoffian noise bath},\ }\href {https://doi.org/10.1143/JPSJ.58.101} {\bibfield  {journal} {\bibinfo  {journal} {J. Phys. Soc. Jpn.}\ }\textbf {\bibinfo {volume} {58}},\ \bibinfo {pages} {101} (\bibinfo {year} {1989})}\BibitemShut {NoStop}%
\bibitem [{\citenamefont {Tanimura}(1990)}]{Tanimura_1_1990}%
  \BibitemOpen
  \bibfield  {author} {\bibinfo {author} {\bibfnamefont {Y.}~\bibnamefont {Tanimura}},\ }\bibfield  {title} {\bibinfo {title} {Nonperturbative expansion method for a quantum system coupled to a harmonic-oscillator bath},\ }\href {https://doi.org/10.1103/PhysRevA.41.6676} {\bibfield  {journal} {\bibinfo  {journal} {Phys. Rev. A}\ }\textbf {\bibinfo {volume} {41}},\ \bibinfo {pages} {6676} (\bibinfo {year} {1990})}\BibitemShut {NoStop}%
\bibitem [{\citenamefont {Bulla}\ \emph {et~al.}(2008)\citenamefont {Bulla}, \citenamefont {Costi},\ and\ \citenamefont {Pruschke}}]{Ralf2008}%
  \BibitemOpen
  \bibfield  {author} {\bibinfo {author} {\bibfnamefont {R.}~\bibnamefont {Bulla}}, \bibinfo {author} {\bibfnamefont {T.~A.}\ \bibnamefont {Costi}},\ and\ \bibinfo {author} {\bibfnamefont {T.}~\bibnamefont {Pruschke}},\ }\bibfield  {title} {\bibinfo {title} {Numerical renormalization group method for quantum impurity systems},\ }\href {https://doi.org/10.1103/RevModPhys.80.395} {\bibfield  {journal} {\bibinfo  {journal} {Rev. Mod. Phys.}\ }\textbf {\bibinfo {volume} {80}},\ \bibinfo {pages} {395} (\bibinfo {year} {2008})}\BibitemShut {NoStop}%
\bibitem [{\citenamefont {Zhang}\ \emph {et~al.}(2012)\citenamefont {Zhang}, \citenamefont {Lo}, \citenamefont {Xiong}, \citenamefont {Tu},\ and\ \citenamefont {Nori}}]{WeiMin_PRL2012}%
  \BibitemOpen
  \bibfield  {author} {\bibinfo {author} {\bibfnamefont {W.-M.}\ \bibnamefont {Zhang}}, \bibinfo {author} {\bibfnamefont {P.-Y.}\ \bibnamefont {Lo}}, \bibinfo {author} {\bibfnamefont {H.-N.}\ \bibnamefont {Xiong}}, \bibinfo {author} {\bibfnamefont {M.~W.-Y.}\ \bibnamefont {Tu}},\ and\ \bibinfo {author} {\bibfnamefont {F.}~\bibnamefont {Nori}},\ }\bibfield  {title} {\bibinfo {title} {General non-{M}arkovian dynamics of open quantum systems},\ }\href {https://doi.org/10.1103/PhysRevLett.109.170402} {\bibfield  {journal} {\bibinfo  {journal} {Phys. Rev. Lett.}\ }\textbf {\bibinfo {volume} {109}},\ \bibinfo {pages} {170402} (\bibinfo {year} {2012})}\BibitemShut {NoStop}%
\bibitem [{\citenamefont {Strasberg}\ \emph {et~al.}(2016)\citenamefont {Strasberg}, \citenamefont {Schaller}, \citenamefont {Lambert},\ and\ \citenamefont {Brandes}}]{Strasberg2016}%
  \BibitemOpen
  \bibfield  {author} {\bibinfo {author} {\bibfnamefont {P.}~\bibnamefont {Strasberg}}, \bibinfo {author} {\bibfnamefont {G.}~\bibnamefont {Schaller}}, \bibinfo {author} {\bibfnamefont {N.}~\bibnamefont {Lambert}},\ and\ \bibinfo {author} {\bibfnamefont {T.}~\bibnamefont {Brandes}},\ }\bibfield  {title} {\bibinfo {title} {Nonequilibrium thermodynamics in the strong coupling and non-{M}arkovian regime based on a reaction coordinate mapping},\ }\href {https://doi.org/10.1088/1367-2630/18/7/073007} {\bibfield  {journal} {\bibinfo  {journal} {New J. Phys.}\ }\textbf {\bibinfo {volume} {18}},\ \bibinfo {pages} {073007} (\bibinfo {year} {2016})}\BibitemShut {NoStop}%
\bibitem [{\citenamefont {Brenes}\ \emph {et~al.}(2020)\citenamefont {Brenes}, \citenamefont {Mendoza-Arenas}, \citenamefont {Purkayastha}, \citenamefont {Mitchison}, \citenamefont {Clark},\ and\ \citenamefont {Goold}}]{Brenes2020}%
  \BibitemOpen
  \bibfield  {author} {\bibinfo {author} {\bibfnamefont {M.}~\bibnamefont {Brenes}}, \bibinfo {author} {\bibfnamefont {J.~J.}\ \bibnamefont {Mendoza-Arenas}}, \bibinfo {author} {\bibfnamefont {A.}~\bibnamefont {Purkayastha}}, \bibinfo {author} {\bibfnamefont {M.~T.}\ \bibnamefont {Mitchison}}, \bibinfo {author} {\bibfnamefont {S.~R.}\ \bibnamefont {Clark}},\ and\ \bibinfo {author} {\bibfnamefont {J.}~\bibnamefont {Goold}},\ }\bibfield  {title} {\bibinfo {title} {Tensor-network method to simulate strongly interacting quantum thermal machines},\ }\href {https://doi.org/10.1103/PhysRevX.10.031040} {\bibfield  {journal} {\bibinfo  {journal} {Phys. Rev. X}\ }\textbf {\bibinfo {volume} {10}},\ \bibinfo {pages} {031040} (\bibinfo {year} {2020})}\BibitemShut {NoStop}%
\bibitem [{\citenamefont {Sowa}\ \emph {et~al.}(2020)\citenamefont {Sowa}, \citenamefont {Lambert}, \citenamefont {Seideman},\ and\ \citenamefont {Gauger}}]{Gauger2020}%
  \BibitemOpen
  \bibfield  {author} {\bibinfo {author} {\bibfnamefont {J.~K.}\ \bibnamefont {Sowa}}, \bibinfo {author} {\bibfnamefont {N.}~\bibnamefont {Lambert}}, \bibinfo {author} {\bibfnamefont {T.}~\bibnamefont {Seideman}},\ and\ \bibinfo {author} {\bibfnamefont {E.~M.}\ \bibnamefont {Gauger}},\ }\bibfield  {title} {\bibinfo {title} {Beyond {M}arcus theory and the {L}andauer–{B}üttiker approach in molecular junctions. {II}. {A} self-consistent {B}orn approach},\ }\href {https://doi.org/10.1063/1.5143146} {\bibfield  {journal} {\bibinfo  {journal} {J. Chem. Phys.}\ }\textbf {\bibinfo {volume} {152}},\ \bibinfo {pages} {064103} (\bibinfo {year} {2020})}\BibitemShut {NoStop}%
\bibitem [{\citenamefont {Li}\ \emph {et~al.}(2012)\citenamefont {Li}, \citenamefont {Tong}, \citenamefont {Zheng}, \citenamefont {Hou}, \citenamefont {Wei}, \citenamefont {Hu},\ and\ \citenamefont {Yan}}]{Yan_2_PRL}%
  \BibitemOpen
  \bibfield  {author} {\bibinfo {author} {\bibfnamefont {Z.}~\bibnamefont {Li}}, \bibinfo {author} {\bibfnamefont {N.}~\bibnamefont {Tong}}, \bibinfo {author} {\bibfnamefont {X.}~\bibnamefont {Zheng}}, \bibinfo {author} {\bibfnamefont {D.}~\bibnamefont {Hou}}, \bibinfo {author} {\bibfnamefont {J.}~\bibnamefont {Wei}}, \bibinfo {author} {\bibfnamefont {J.}~\bibnamefont {Hu}},\ and\ \bibinfo {author} {\bibfnamefont {Y.}~\bibnamefont {Yan}},\ }\bibfield  {title} {\bibinfo {title} {Hierarchical {L}iouville-space approach for accurate and universal characterization of quantum impurity systems},\ }\href {https://doi.org/10.1103/PhysRevLett.109.266403} {\bibfield  {journal} {\bibinfo  {journal} {Phys. Rev. Lett.}\ }\textbf {\bibinfo {volume} {109}},\ \bibinfo {pages} {266403} (\bibinfo {year} {2012})}\BibitemShut {NoStop}%
\bibitem [{\citenamefont {Tanimura}(2020)}]{Tanimura_3_2020}%
  \BibitemOpen
  \bibfield  {author} {\bibinfo {author} {\bibfnamefont {Y.}~\bibnamefont {Tanimura}},\ }\bibfield  {title} {\bibinfo {title} {Numerically “exact” approach to open quantum dynamics: {T}he hierarchical equations of motion ({HEOM})},\ }\href {https://doi.org/10.1063/5.0011599} {\bibfield  {journal} {\bibinfo  {journal} {J. Chem. Phys.}\ }\textbf {\bibinfo {volume} {153}},\ \bibinfo {pages} {020901} (\bibinfo {year} {2020})}\BibitemShut {NoStop}%
\bibitem [{\citenamefont {Lambert}\ \emph {et~al.}(2023)\citenamefont {Lambert}, \citenamefont {Raheja}, \citenamefont {Cross}, \citenamefont {Menczel}, \citenamefont {Ahmed}, \citenamefont {Pitchford}, \citenamefont {Burgarth},\ and\ \citenamefont {Nori}}]{QuTiP-BoFiN}%
  \BibitemOpen
  \bibfield  {author} {\bibinfo {author} {\bibfnamefont {N.}~\bibnamefont {Lambert}}, \bibinfo {author} {\bibfnamefont {T.}~\bibnamefont {Raheja}}, \bibinfo {author} {\bibfnamefont {S.}~\bibnamefont {Cross}}, \bibinfo {author} {\bibfnamefont {P.}~\bibnamefont {Menczel}}, \bibinfo {author} {\bibfnamefont {S.}~\bibnamefont {Ahmed}}, \bibinfo {author} {\bibfnamefont {A.}~\bibnamefont {Pitchford}}, \bibinfo {author} {\bibfnamefont {D.}~\bibnamefont {Burgarth}},\ and\ \bibinfo {author} {\bibfnamefont {F.}~\bibnamefont {Nori}},\ }\bibfield  {title} {\bibinfo {title} {{QuTiP-BoFiN}: A bosonic and fermionic numerical hierarchical-equations-of-motion library with applications in light-harvesting, quantum control, and single-molecule electronics},\ }\href {https://doi.org/10.1103/PhysRevResearch.5.013181} {\bibfield  {journal} {\bibinfo  {journal} {Phys. Rev. Res.}\ }\textbf {\bibinfo {volume} {5}},\ \bibinfo {pages} {013181} (\bibinfo {year} {2023})}\BibitemShut {NoStop}%
\bibitem [{\citenamefont {Lambert}\ \emph {et~al.}(2019)\citenamefont {Lambert}, \citenamefont {Ahmed}, \citenamefont {Cirio},\ and\ \citenamefont {Nori}}]{Lambert2019}%
  \BibitemOpen
  \bibfield  {author} {\bibinfo {author} {\bibfnamefont {N.}~\bibnamefont {Lambert}}, \bibinfo {author} {\bibfnamefont {S.}~\bibnamefont {Ahmed}}, \bibinfo {author} {\bibfnamefont {M.}~\bibnamefont {Cirio}},\ and\ \bibinfo {author} {\bibfnamefont {F.}~\bibnamefont {Nori}},\ }\bibfield  {title} {\bibinfo {title} {Modelling the ultra-strongly coupled spin-boson model with unphysical modes},\ }\href {https://doi.org/10.1038/s41467-019-11656-1} {\bibfield  {journal} {\bibinfo  {journal} {Nat. Commun.}\ }\textbf {\bibinfo {volume} {10}},\ \bibinfo {pages} {3721} (\bibinfo {year} {2019})}\BibitemShut {NoStop}%
\bibitem [{\citenamefont {Fay}\ and\ \citenamefont {Limmer}(2022)}]{Limmer2022}%
  \BibitemOpen
  \bibfield  {author} {\bibinfo {author} {\bibfnamefont {T.~P.}\ \bibnamefont {Fay}}\ and\ \bibinfo {author} {\bibfnamefont {D.~T.}\ \bibnamefont {Limmer}},\ }\bibfield  {title} {\bibinfo {title} {Coupled charge and energy transfer dynamics in light harvesting complexes from a hybrid hierarchical equations of motion approach},\ }\href {https://doi.org/10.1063/5.0117659} {\bibfield  {journal} {\bibinfo  {journal} {J. Chem. Phys.}\ }\textbf {\bibinfo {volume} {157}},\ \bibinfo {pages} {174104} (\bibinfo {year} {2022})}\BibitemShut {NoStop}%
\bibitem [{\citenamefont {Ma}\ \emph {et~al.}(2012)\citenamefont {Ma}, \citenamefont {Sun}, \citenamefont {Wang},\ and\ \citenamefont {Nori}}]{Ma2012}%
  \BibitemOpen
  \bibfield  {author} {\bibinfo {author} {\bibfnamefont {J.}~\bibnamefont {Ma}}, \bibinfo {author} {\bibfnamefont {Z.}~\bibnamefont {Sun}}, \bibinfo {author} {\bibfnamefont {X.}~\bibnamefont {Wang}},\ and\ \bibinfo {author} {\bibfnamefont {F.}~\bibnamefont {Nori}},\ }\bibfield  {title} {\bibinfo {title} {Entanglement dynamics of two qubits in a common bath},\ }\href {https://doi.org/10.1103/PhysRevA.85.062323} {\bibfield  {journal} {\bibinfo  {journal} {Phys. Rev. A}\ }\textbf {\bibinfo {volume} {85}},\ \bibinfo {pages} {062323} (\bibinfo {year} {2012})}\BibitemShut {NoStop}%
\bibitem [{\citenamefont {Kato}\ and\ \citenamefont {Tanimura}(2016)}]{Tanimura_4_2016}%
  \BibitemOpen
  \bibfield  {author} {\bibinfo {author} {\bibfnamefont {A.}~\bibnamefont {Kato}}\ and\ \bibinfo {author} {\bibfnamefont {Y.}~\bibnamefont {Tanimura}},\ }\bibfield  {title} {\bibinfo {title} {Quantum heat current under non-perturbative and non-{M}arkovian conditions: Applications to heat machines},\ }\href {https://doi.org/10.1063/1.4971370} {\bibfield  {journal} {\bibinfo  {journal} {J. Chem. Phys.}\ }\textbf {\bibinfo {volume} {145}},\ \bibinfo {pages} {224105} (\bibinfo {year} {2016})}\BibitemShut {NoStop}%
\bibitem [{\citenamefont {Chen}\ \emph {et~al.}(2022)\citenamefont {Chen}, \citenamefont {Zhang}, \citenamefont {He}, \citenamefont {Kong}, \citenamefont {Tao}, \citenamefont {Deng}, \citenamefont {Ai},\ and\ \citenamefont {Long}}]{GuiLu2022}%
  \BibitemOpen
  \bibfield  {author} {\bibinfo {author} {\bibfnamefont {X.-Y.}\ \bibnamefont {Chen}}, \bibinfo {author} {\bibfnamefont {N.-N.}\ \bibnamefont {Zhang}}, \bibinfo {author} {\bibfnamefont {W.-T.}\ \bibnamefont {He}}, \bibinfo {author} {\bibfnamefont {X.-Y.}\ \bibnamefont {Kong}}, \bibinfo {author} {\bibfnamefont {M.-J.}\ \bibnamefont {Tao}}, \bibinfo {author} {\bibfnamefont {F.-G.}\ \bibnamefont {Deng}}, \bibinfo {author} {\bibfnamefont {Q.}~\bibnamefont {Ai}},\ and\ \bibinfo {author} {\bibfnamefont {G.-L.}\ \bibnamefont {Long}},\ }\bibfield  {title} {\bibinfo {title} {Global correlation and local information flows in controllable non-{M}arkovian open quantum dynamics},\ }\href {https://doi.org/10.1038/s41534-022-00537-z} {\bibfield  {journal} {\bibinfo  {journal} {npj Quantum Inf.}\ }\textbf {\bibinfo {volume} {8}},\ \bibinfo {pages} {22} (\bibinfo {year} {2022})}\BibitemShut {NoStop}%
\bibitem [{\citenamefont {Jin}\ \emph {et~al.}(2008)\citenamefont {Jin}, \citenamefont {Zheng},\ and\ \citenamefont {Yan}}]{Yan_1}%
  \BibitemOpen
  \bibfield  {author} {\bibinfo {author} {\bibfnamefont {J.}~\bibnamefont {Jin}}, \bibinfo {author} {\bibfnamefont {X.}~\bibnamefont {Zheng}},\ and\ \bibinfo {author} {\bibfnamefont {Y.}~\bibnamefont {Yan}},\ }\bibfield  {title} {\bibinfo {title} {Exact dynamics of dissipative electronic systems and quantum transport: Hierarchical equations of motion approach},\ }\href {https://doi.org/10.1063/1.2938087} {\bibfield  {journal} {\bibinfo  {journal} {J. Chem. Phys.}\ }\textbf {\bibinfo {volume} {128}},\ \bibinfo {pages} {234703} (\bibinfo {year} {2008})}\BibitemShut {NoStop}%
\bibitem [{\citenamefont {Ishizaki}\ and\ \citenamefont {Fleming}(2009)}]{Ishizaki_1_2009}%
  \BibitemOpen
  \bibfield  {author} {\bibinfo {author} {\bibfnamefont {A.}~\bibnamefont {Ishizaki}}\ and\ \bibinfo {author} {\bibfnamefont {G.~R.}\ \bibnamefont {Fleming}},\ }\bibfield  {title} {\bibinfo {title} {Unified treatment of quantum coherent and incoherent hopping dynamics in electronic energy transfer: Reduced hierarchy equation approach},\ }\href {https://doi.org/10.1063/1.3155372} {\bibfield  {journal} {\bibinfo  {journal} {J. Chem. Phys.}\ }\textbf {\bibinfo {volume} {130}},\ \bibinfo {pages} {234111} (\bibinfo {year} {2009})}\BibitemShut {NoStop}%
\bibitem [{\citenamefont {H\"artle}\ \emph {et~al.}(2013)\citenamefont {H\"artle}, \citenamefont {Cohen}, \citenamefont {Reichman},\ and\ \citenamefont {Millis}}]{Hartle2013}%
  \BibitemOpen
  \bibfield  {author} {\bibinfo {author} {\bibfnamefont {R.}~\bibnamefont {H\"artle}}, \bibinfo {author} {\bibfnamefont {G.}~\bibnamefont {Cohen}}, \bibinfo {author} {\bibfnamefont {D.~R.}\ \bibnamefont {Reichman}},\ and\ \bibinfo {author} {\bibfnamefont {A.~J.}\ \bibnamefont {Millis}},\ }\bibfield  {title} {\bibinfo {title} {Decoherence and lead-induced interdot coupling in nonequilibrium electron transport through interacting quantum dots: A hierarchical quantum master equation approach},\ }\href {https://doi.org/10.1103/PhysRevB.88.235426} {\bibfield  {journal} {\bibinfo  {journal} {Phys. Rev. B}\ }\textbf {\bibinfo {volume} {88}},\ \bibinfo {pages} {235426} (\bibinfo {year} {2013})}\BibitemShut {NoStop}%
\bibitem [{\citenamefont {Ye}\ \emph {et~al.}(2016)\citenamefont {Ye}, \citenamefont {Wang}, \citenamefont {Hou}, \citenamefont {Xu}, \citenamefont {Zheng},\ and\ \citenamefont {Yan}}]{Yan_5_2016}%
  \BibitemOpen
  \bibfield  {author} {\bibinfo {author} {\bibfnamefont {L.}~\bibnamefont {Ye}}, \bibinfo {author} {\bibfnamefont {X.}~\bibnamefont {Wang}}, \bibinfo {author} {\bibfnamefont {D.}~\bibnamefont {Hou}}, \bibinfo {author} {\bibfnamefont {R.-X.}\ \bibnamefont {Xu}}, \bibinfo {author} {\bibfnamefont {X.}~\bibnamefont {Zheng}},\ and\ \bibinfo {author} {\bibfnamefont {Y.}~\bibnamefont {Yan}},\ }\bibfield  {title} {\bibinfo {title} {{HEOM}-quick: a program for accurate, efficient, and universal characterization of strongly correlated quantum impurity systems},\ }\href {https://doi.org/10.1002/wcms.1269} {\bibfield  {journal} {\bibinfo  {journal} {WIREs Comput. Mol. Sci.}\ }\textbf {\bibinfo {volume} {6}},\ \bibinfo {pages} {608} (\bibinfo {year} {2016})}\BibitemShut {NoStop}%
\bibitem [{\citenamefont {Schinabeck}\ \emph {et~al.}(2018)\citenamefont {Schinabeck}, \citenamefont {H\"artle},\ and\ \citenamefont {Thoss}}]{Schinabeck2018}%
  \BibitemOpen
  \bibfield  {author} {\bibinfo {author} {\bibfnamefont {C.}~\bibnamefont {Schinabeck}}, \bibinfo {author} {\bibfnamefont {R.}~\bibnamefont {H\"artle}},\ and\ \bibinfo {author} {\bibfnamefont {M.}~\bibnamefont {Thoss}},\ }\bibfield  {title} {\bibinfo {title} {Hierarchical quantum master equation approach to electronic-vibrational coupling in nonequilibrium transport through nanosystems: Reservoir formulation and application to vibrational instabilities},\ }\href {https://doi.org/10.1103/PhysRevB.97.235429} {\bibfield  {journal} {\bibinfo  {journal} {Phys. Rev. B}\ }\textbf {\bibinfo {volume} {97}},\ \bibinfo {pages} {235429} (\bibinfo {year} {2018})}\BibitemShut {NoStop}%
\bibitem [{\citenamefont {B\"atge}\ \emph {et~al.}(2021)\citenamefont {B\"atge}, \citenamefont {Ke}, \citenamefont {Kaspar},\ and\ \citenamefont {Thoss}}]{Thoss_2_2021}%
  \BibitemOpen
  \bibfield  {author} {\bibinfo {author} {\bibfnamefont {J.}~\bibnamefont {B\"atge}}, \bibinfo {author} {\bibfnamefont {Y.}~\bibnamefont {Ke}}, \bibinfo {author} {\bibfnamefont {C.}~\bibnamefont {Kaspar}},\ and\ \bibinfo {author} {\bibfnamefont {M.}~\bibnamefont {Thoss}},\ }\bibfield  {title} {\bibinfo {title} {Nonequilibrium open quantum systems with multiple bosonic and fermionic environments: A hierarchical equations of motion approach},\ }\href {https://doi.org/10.1103/PhysRevB.103.235413} {\bibfield  {journal} {\bibinfo  {journal} {Phys. Rev. B}\ }\textbf {\bibinfo {volume} {103}},\ \bibinfo {pages} {235413} (\bibinfo {year} {2021})}\BibitemShut {NoStop}%
\bibitem [{\citenamefont {Bezanson}\ \emph {et~al.}(2012)\citenamefont {Bezanson}, \citenamefont {Karpinski}, \citenamefont {Shah},\ and\ \citenamefont {Edelman}}]{Julia2012}%
  \BibitemOpen
  \bibfield  {author} {\bibinfo {author} {\bibfnamefont {J.}~\bibnamefont {Bezanson}}, \bibinfo {author} {\bibfnamefont {S.}~\bibnamefont {Karpinski}}, \bibinfo {author} {\bibfnamefont {V.~B.}\ \bibnamefont {Shah}},\ and\ \bibinfo {author} {\bibfnamefont {A.}~\bibnamefont {Edelman}},\ }\bibfield  {title} {\bibinfo {title} {{J}ulia: A fast dynamic language for technical computing},\ }\href {https://doi.org/10.48550/arXiv.1209.5145} {\bibfield  {journal} {\bibinfo  {journal} {arXiv preprint arXiv:1209.5145}\ } (\bibinfo {year} {2012})}\BibitemShut {NoStop}%
\bibitem [{\citenamefont {Bezanson}\ \emph {et~al.}(2017)\citenamefont {Bezanson}, \citenamefont {Edelman}, \citenamefont {Karpinski},\ and\ \citenamefont {Shah}}]{Julia2017}%
  \BibitemOpen
  \bibfield  {author} {\bibinfo {author} {\bibfnamefont {J.}~\bibnamefont {Bezanson}}, \bibinfo {author} {\bibfnamefont {A.}~\bibnamefont {Edelman}}, \bibinfo {author} {\bibfnamefont {S.}~\bibnamefont {Karpinski}},\ and\ \bibinfo {author} {\bibfnamefont {V.~B.}\ \bibnamefont {Shah}},\ }\bibfield  {title} {\bibinfo {title} {{J}ulia: A fresh approach to numerical computing},\ }\href {https://doi.org/10.1137/141000671} {\bibfield  {journal} {\bibinfo  {journal} {{SIAM} Review}\ }\textbf {\bibinfo {volume} {59}},\ \bibinfo {pages} {65} (\bibinfo {year} {2017})}\BibitemShut {NoStop}%
\bibitem [{\citenamefont {Kernighan}\ and\ \citenamefont {Ritchie}(2006)}]{C2006}%
  \BibitemOpen
  \bibfield  {author} {\bibinfo {author} {\bibfnamefont {B.~W.}\ \bibnamefont {Kernighan}}\ and\ \bibinfo {author} {\bibfnamefont {D.~M.}\ \bibnamefont {Ritchie}},\ }\href@noop {} {\emph {\bibinfo {title} {The C programming language}}}\ (\bibinfo  {publisher} {Prentice Hall Professional Technical Reference},\ \bibinfo {year} {2006})\BibitemShut {NoStop}%
\bibitem [{\citenamefont {Flanagan}\ and\ \citenamefont {Matsumoto}(2007)}]{Ruby2007}%
  \BibitemOpen
  \bibfield  {author} {\bibinfo {author} {\bibfnamefont {D.}~\bibnamefont {Flanagan}}\ and\ \bibinfo {author} {\bibfnamefont {Y.}~\bibnamefont {Matsumoto}},\ }\href@noop {} {\emph {\bibinfo {title} {The Ruby Programming Language}}}\ (\bibinfo  {publisher} {O'Reilly Media, Inc.},\ \bibinfo {year} {2007})\BibitemShut {NoStop}%
\bibitem [{\citenamefont {Van~Rossum}\ and\ \citenamefont {Drake}(2009)}]{Python2009}%
  \BibitemOpen
  \bibfield  {author} {\bibinfo {author} {\bibfnamefont {G.}~\bibnamefont {Van~Rossum}}\ and\ \bibinfo {author} {\bibfnamefont {F.~L.}\ \bibnamefont {Drake}},\ }\href@noop {} {\emph {\bibinfo {title} {Python 3 Reference Manual}}}\ (\bibinfo  {publisher} {CreateSpace},\ \bibinfo {address} {Scotts Valley, CA},\ \bibinfo {year} {2009})\BibitemShut {NoStop}%
\bibitem [{\citenamefont {{R Core Team}}(2021)}]{R2021}%
  \BibitemOpen
  \bibfield  {author} {\bibinfo {author} {\bibnamefont {{R Core Team}}},\ }\href {https://www.R-project.org/} {\emph {\bibinfo {title} {R: A Language and Environment for Statistical Computing}}}\ (\bibinfo  {publisher} {R Foundation for Statistical Computing},\ \bibinfo {address} {Vienna, Austria},\ \bibinfo {year} {2021})\BibitemShut {NoStop}%
\bibitem [{\citenamefont {Higham}\ and\ \citenamefont {Higham}(2017)}]{Matlab2016}%
  \BibitemOpen
  \bibfield  {author} {\bibinfo {author} {\bibfnamefont {D.~J.}\ \bibnamefont {Higham}}\ and\ \bibinfo {author} {\bibfnamefont {N.~J.}\ \bibnamefont {Higham}},\ }\href@noop {} {\emph {\bibinfo {title} {{MATLAB} Guide}}}\ (\bibinfo  {publisher} {Society for Industrial and Applied Mathematics},\ \bibinfo {address} {Philadelphia, PA, USA},\ \bibinfo {year} {2017})\ pp.\ \bibinfo {pages} {xxvi+476}\BibitemShut {NoStop}%
\bibitem [{\citenamefont {Lattner}\ and\ \citenamefont {Adve}(2004)}]{LLVM2004}%
  \BibitemOpen
  \bibfield  {author} {\bibinfo {author} {\bibfnamefont {C.}~\bibnamefont {Lattner}}\ and\ \bibinfo {author} {\bibfnamefont {V.}~\bibnamefont {Adve}},\ }\bibfield  {title} {\bibinfo {title} {{LLVM}: a compilation framework for lifelong program analysis \& transformation},\ }in\ \href {https://doi.org/10.1109/CGO.2004.1281665} {\emph {\bibinfo {booktitle} {International Symposium on Code Generation and Optimization, 2004. CGO 2004.}}}\ (\bibinfo {year} {2004})\ p.~\bibinfo {pages} {75}\BibitemShut {NoStop}%
\bibitem [{\citenamefont {Chen}\ and\ \citenamefont {Lidar}(2022)}]{HOQST2022}%
  \BibitemOpen
  \bibfield  {author} {\bibinfo {author} {\bibfnamefont {H.}~\bibnamefont {Chen}}\ and\ \bibinfo {author} {\bibfnamefont {D.~A.}\ \bibnamefont {Lidar}},\ }\bibfield  {title} {\bibinfo {title} {Hamiltonian open quantum system toolkit},\ }\href {https://doi.org/10.1038/s42005-022-00887-2} {\bibfield  {journal} {\bibinfo  {journal} {Commun. Phys.}\ }\textbf {\bibinfo {volume} {5}},\ \bibinfo {pages} {112} (\bibinfo {year} {2022})}\BibitemShut {NoStop}%
\bibitem [{\citenamefont {Kr{\"a}mer}\ \emph {et~al.}(2018)\citenamefont {Kr{\"a}mer}, \citenamefont {Plankensteiner}, \citenamefont {Ostermann},\ and\ \citenamefont {Ritsch}}]{QuantumOptics.jl-2018}%
  \BibitemOpen
  \bibfield  {author} {\bibinfo {author} {\bibfnamefont {S.}~\bibnamefont {Kr{\"a}mer}}, \bibinfo {author} {\bibfnamefont {D.}~\bibnamefont {Plankensteiner}}, \bibinfo {author} {\bibfnamefont {L.}~\bibnamefont {Ostermann}},\ and\ \bibinfo {author} {\bibfnamefont {H.}~\bibnamefont {Ritsch}},\ }\bibfield  {title} {\bibinfo {title} {Quantum{O}ptics. jl: {A} {J}ulia framework for simulating open quantum systems},\ }\href {https://doi.org/10.1016/j.cpc.2018.02.004} {\bibfield  {journal} {\bibinfo  {journal} {Comput. Phys. Commun.}\ }\textbf {\bibinfo {volume} {227}},\ \bibinfo {pages} {109} (\bibinfo {year} {2018})}\BibitemShut {NoStop}%
\bibitem [{\citenamefont {Luo}\ \emph {et~al.}(2020)\citenamefont {Luo}, \citenamefont {Liu}, \citenamefont {Zhang},\ and\ \citenamefont {Wang}}]{Yao2020}%
  \BibitemOpen
  \bibfield  {author} {\bibinfo {author} {\bibfnamefont {X.-Z.}\ \bibnamefont {Luo}}, \bibinfo {author} {\bibfnamefont {J.-G.}\ \bibnamefont {Liu}}, \bibinfo {author} {\bibfnamefont {P.}~\bibnamefont {Zhang}},\ and\ \bibinfo {author} {\bibfnamefont {L.}~\bibnamefont {Wang}},\ }\bibfield  {title} {\bibinfo {title} {Yao.jl: {E}xtensible, {E}fficient {F}ramework for {Q}uantum {A}lgorithm {D}esign},\ }\href {https://doi.org/10.22331/q-2020-10-11-341} {\bibfield  {journal} {\bibinfo  {journal} {{Quantum}}\ }\textbf {\bibinfo {volume} {4}},\ \bibinfo {pages} {341} (\bibinfo {year} {2020})}\BibitemShut {NoStop}%
\bibitem [{\citenamefont {Gawron}\ \emph {et~al.}(2018)\citenamefont {Gawron}, \citenamefont {Kurzyk},\ and\ \citenamefont {Pawela}}]{QuantumInformation2018}%
  \BibitemOpen
  \bibfield  {author} {\bibinfo {author} {\bibfnamefont {P.}~\bibnamefont {Gawron}}, \bibinfo {author} {\bibfnamefont {D.}~\bibnamefont {Kurzyk}},\ and\ \bibinfo {author} {\bibfnamefont {{\L}.}~\bibnamefont {Pawela}},\ }\bibfield  {title} {\bibinfo {title} {{QuantumInformation}.jl{\textemdash}{A} {J}ulia package for numerical computation in quantum information theory},\ }\href {https://doi.org/10.1371/journal.pone.0209358} {\bibfield  {journal} {\bibinfo  {journal} {{PLOS} {ONE}}\ }\textbf {\bibinfo {volume} {13}},\ \bibinfo {pages} {e0209358} (\bibinfo {year} {2018})}\BibitemShut {NoStop}%
\bibitem [{\citenamefont {Rackauckas}\ and\ \citenamefont {Nie}(2017)}]{DifferentialEquations.jl-2017}%
  \BibitemOpen
  \bibfield  {author} {\bibinfo {author} {\bibfnamefont {C.}~\bibnamefont {Rackauckas}}\ and\ \bibinfo {author} {\bibfnamefont {Q.}~\bibnamefont {Nie}},\ }\bibfield  {title} {\bibinfo {title} {{DifferentialEquations}.jl {\textendash} {A} performant and feature-rich ecosystem for solving differential equations in {J}ulia},\ }\href {https://doi.org/10.5334/jors.151} {\bibfield  {journal} {\bibinfo  {journal} {J. Open Res. Software}\ }\textbf {\bibinfo {volume} {5}},\ \bibinfo {pages} {15} (\bibinfo {year} {2017})}\BibitemShut {NoStop}%
\bibitem [{\citenamefont {Kimmerer}\ \emph {et~al.}()\citenamefont {Kimmerer}, \citenamefont {Puri},\ and\ \citenamefont {Rackauckas}}]{LinearSolve.jl}%
  \BibitemOpen
  \bibfield  {author} {\bibinfo {author} {\bibfnamefont {W.}~\bibnamefont {Kimmerer}}, \bibinfo {author} {\bibfnamefont {V.}~\bibnamefont {Puri}},\ and\ \bibinfo {author} {\bibfnamefont {C.}~\bibnamefont {Rackauckas}},\ }\href {https://github.com/SciML/LinearSolve.jl} {\bibinfo {title} {Linearsolve.jl}}\BibitemShut {NoStop}%
\bibitem [{\citenamefont {Hogben}\ \emph {et~al.}(2011)\citenamefont {Hogben}, \citenamefont {Krzystyniak}, \citenamefont {Charnock}, \citenamefont {Hore},\ and\ \citenamefont {Kuprov}}]{FastExpm.jl-2011}%
  \BibitemOpen
  \bibfield  {author} {\bibinfo {author} {\bibfnamefont {H.}~\bibnamefont {Hogben}}, \bibinfo {author} {\bibfnamefont {M.}~\bibnamefont {Krzystyniak}}, \bibinfo {author} {\bibfnamefont {G.}~\bibnamefont {Charnock}}, \bibinfo {author} {\bibfnamefont {P.}~\bibnamefont {Hore}},\ and\ \bibinfo {author} {\bibfnamefont {I.}~\bibnamefont {Kuprov}},\ }\bibfield  {title} {\bibinfo {title} {Spinach – {A} software library for simulation of spin dynamics in large spin systems},\ }\href {https://doi.org/10.1016/j.jmr.2010.11.008} {\bibfield  {journal} {\bibinfo  {journal} {J. Magn. Reson.}\ }\textbf {\bibinfo {volume} {208}},\ \bibinfo {pages} {179} (\bibinfo {year} {2011})}\BibitemShut {NoStop}%
\bibitem [{\citenamefont {Cirio}\ \emph {et~al.}(2016)\citenamefont {Cirio}, \citenamefont {De~Liberato}, \citenamefont {Lambert},\ and\ \citenamefont {Nori}}]{Mauro2016}%
  \BibitemOpen
  \bibfield  {author} {\bibinfo {author} {\bibfnamefont {M.}~\bibnamefont {Cirio}}, \bibinfo {author} {\bibfnamefont {S.}~\bibnamefont {De~Liberato}}, \bibinfo {author} {\bibfnamefont {N.}~\bibnamefont {Lambert}},\ and\ \bibinfo {author} {\bibfnamefont {F.}~\bibnamefont {Nori}},\ }\bibfield  {title} {\bibinfo {title} {Ground state electroluminescence},\ }\href {https://doi.org/10.1103/PhysRevLett.116.113601} {\bibfield  {journal} {\bibinfo  {journal} {Phys. Rev. Lett.}\ }\textbf {\bibinfo {volume} {116}},\ \bibinfo {pages} {113601} (\bibinfo {year} {2016})}\BibitemShut {NoStop}%
\bibitem [{\citenamefont {Stockklauser}\ \emph {et~al.}(2017)\citenamefont {Stockklauser}, \citenamefont {Scarlino}, \citenamefont {Koski}, \citenamefont {Gasparinetti}, \citenamefont {Andersen}, \citenamefont {Reichl}, \citenamefont {Wegscheider}, \citenamefont {Ihn}, \citenamefont {Ensslin},\ and\ \citenamefont {Wallraff}}]{Stockklauser2017}%
  \BibitemOpen
  \bibfield  {author} {\bibinfo {author} {\bibfnamefont {A.}~\bibnamefont {Stockklauser}}, \bibinfo {author} {\bibfnamefont {P.}~\bibnamefont {Scarlino}}, \bibinfo {author} {\bibfnamefont {J.~V.}\ \bibnamefont {Koski}}, \bibinfo {author} {\bibfnamefont {S.}~\bibnamefont {Gasparinetti}}, \bibinfo {author} {\bibfnamefont {C.~K.}\ \bibnamefont {Andersen}}, \bibinfo {author} {\bibfnamefont {C.}~\bibnamefont {Reichl}}, \bibinfo {author} {\bibfnamefont {W.}~\bibnamefont {Wegscheider}}, \bibinfo {author} {\bibfnamefont {T.}~\bibnamefont {Ihn}}, \bibinfo {author} {\bibfnamefont {K.}~\bibnamefont {Ensslin}},\ and\ \bibinfo {author} {\bibfnamefont {A.}~\bibnamefont {Wallraff}},\ }\bibfield  {title} {\bibinfo {title} {Strong coupling cavity {QED} with gate-defined double quantum dots enabled by a high impedance resonator},\ }\href {https://doi.org/10.1103/PhysRevX.7.011030} {\bibfield  {journal} {\bibinfo  {journal} {Phys. Rev. X}\ }\textbf {\bibinfo {volume} {7}},\ \bibinfo {pages} {011030} (\bibinfo {year}
  {2017})}\BibitemShut {NoStop}%
\bibitem [{\citenamefont {Gustafsson}\ \emph {et~al.}(2014)\citenamefont {Gustafsson}, \citenamefont {Aref}, \citenamefont {Kockum}, \citenamefont {Ekstr\"{o}m}, \citenamefont {Johansson},\ and\ \citenamefont {Delsing}}]{Gustafsson2014}%
  \BibitemOpen
  \bibfield  {author} {\bibinfo {author} {\bibfnamefont {M.~V.}\ \bibnamefont {Gustafsson}}, \bibinfo {author} {\bibfnamefont {T.}~\bibnamefont {Aref}}, \bibinfo {author} {\bibfnamefont {A.~F.}\ \bibnamefont {Kockum}}, \bibinfo {author} {\bibfnamefont {M.~K.}\ \bibnamefont {Ekstr\"{o}m}}, \bibinfo {author} {\bibfnamefont {G.}~\bibnamefont {Johansson}},\ and\ \bibinfo {author} {\bibfnamefont {P.}~\bibnamefont {Delsing}},\ }\bibfield  {title} {\bibinfo {title} {Propagating phonons coupled to an artificial atom},\ }\href {https://doi.org/10.1126/science.1257219} {\bibfield  {journal} {\bibinfo  {journal} {Science}\ }\textbf {\bibinfo {volume} {346}},\ \bibinfo {pages} {207} (\bibinfo {year} {2014})}\BibitemShut {NoStop}%
\bibitem [{\citenamefont {Manenti}\ \emph {et~al.}(2017)\citenamefont {Manenti}, \citenamefont {Kockum}, \citenamefont {Patterson}, \citenamefont {Behrle}, \citenamefont {Rahamim}, \citenamefont {Tancredi}, \citenamefont {Nori},\ and\ \citenamefont {Leek}}]{Manenti2017}%
  \BibitemOpen
  \bibfield  {author} {\bibinfo {author} {\bibfnamefont {R.}~\bibnamefont {Manenti}}, \bibinfo {author} {\bibfnamefont {A.~F.}\ \bibnamefont {Kockum}}, \bibinfo {author} {\bibfnamefont {A.}~\bibnamefont {Patterson}}, \bibinfo {author} {\bibfnamefont {T.}~\bibnamefont {Behrle}}, \bibinfo {author} {\bibfnamefont {J.}~\bibnamefont {Rahamim}}, \bibinfo {author} {\bibfnamefont {G.}~\bibnamefont {Tancredi}}, \bibinfo {author} {\bibfnamefont {F.}~\bibnamefont {Nori}},\ and\ \bibinfo {author} {\bibfnamefont {P.~J.}\ \bibnamefont {Leek}},\ }\bibfield  {title} {\bibinfo {title} {Circuit quantum acoustodynamics with surface acoustic waves},\ }\href {https://doi.org/10.1038/s41467-017-01063-9} {\bibfield  {journal} {\bibinfo  {journal} {Nat. Commun.}\ }\textbf {\bibinfo {volume} {8}},\ \bibinfo {pages} {975} (\bibinfo {year} {2017})}\BibitemShut {NoStop}%
\bibitem [{\citenamefont {Iorsh}\ \emph {et~al.}(2020)\citenamefont {Iorsh}, \citenamefont {Poshakinskiy},\ and\ \citenamefont {Poddubny}}]{Iorsh2020}%
  \BibitemOpen
  \bibfield  {author} {\bibinfo {author} {\bibfnamefont {I.}~\bibnamefont {Iorsh}}, \bibinfo {author} {\bibfnamefont {A.}~\bibnamefont {Poshakinskiy}},\ and\ \bibinfo {author} {\bibfnamefont {A.}~\bibnamefont {Poddubny}},\ }\bibfield  {title} {\bibinfo {title} {Waveguide quantum optomechanics: Parity-time phase transitions in ultrastrong coupling regime},\ }\href {https://doi.org/10.1103/PhysRevLett.125.183601} {\bibfield  {journal} {\bibinfo  {journal} {Phys. Rev. Lett.}\ }\textbf {\bibinfo {volume} {125}},\ \bibinfo {pages} {183601} (\bibinfo {year} {2020})}\BibitemShut {NoStop}%
\bibitem [{\citenamefont {Benz}\ \emph {et~al.}(2016)\citenamefont {Benz}, \citenamefont {Schmidt}, \citenamefont {Dreismann}, \citenamefont {Chikkaraddy}, \citenamefont {Zhang}, \citenamefont {Demetriadou}, \citenamefont {Carnegie}, \citenamefont {Ohadi}, \citenamefont {de~Nijs}, \citenamefont {Esteban}, \citenamefont {Aizpurua},\ and\ \citenamefont {Baumberg}}]{Benz2016}%
  \BibitemOpen
  \bibfield  {author} {\bibinfo {author} {\bibfnamefont {F.}~\bibnamefont {Benz}}, \bibinfo {author} {\bibfnamefont {M.~K.}\ \bibnamefont {Schmidt}}, \bibinfo {author} {\bibfnamefont {A.}~\bibnamefont {Dreismann}}, \bibinfo {author} {\bibfnamefont {R.}~\bibnamefont {Chikkaraddy}}, \bibinfo {author} {\bibfnamefont {Y.}~\bibnamefont {Zhang}}, \bibinfo {author} {\bibfnamefont {A.}~\bibnamefont {Demetriadou}}, \bibinfo {author} {\bibfnamefont {C.}~\bibnamefont {Carnegie}}, \bibinfo {author} {\bibfnamefont {H.}~\bibnamefont {Ohadi}}, \bibinfo {author} {\bibfnamefont {B.}~\bibnamefont {de~Nijs}}, \bibinfo {author} {\bibfnamefont {R.}~\bibnamefont {Esteban}}, \bibinfo {author} {\bibfnamefont {J.}~\bibnamefont {Aizpurua}},\ and\ \bibinfo {author} {\bibfnamefont {J.~J.}\ \bibnamefont {Baumberg}},\ }\bibfield  {title} {\bibinfo {title} {Single-molecule optomechanics in {\textquotedblleft}picocavities{\textquotedblright}},\ }\href {https://doi.org/10.1126/science.aah5243} {\bibfield  {journal} {\bibinfo  {journal}
  {Science}\ }\textbf {\bibinfo {volume} {354}},\ \bibinfo {pages} {726} (\bibinfo {year} {2016})}\BibitemShut {NoStop}%
\bibitem [{\citenamefont {Kuo}\ \emph {et~al.}(2020)\citenamefont {Kuo}, \citenamefont {Lambert}, \citenamefont {Miranowicz}, \citenamefont {Chen}, \citenamefont {Chen}, \citenamefont {Chen},\ and\ \citenamefont {Nori}}]{Po2020}%
  \BibitemOpen
  \bibfield  {author} {\bibinfo {author} {\bibfnamefont {P.~C.}\ \bibnamefont {Kuo}}, \bibinfo {author} {\bibfnamefont {N.}~\bibnamefont {Lambert}}, \bibinfo {author} {\bibfnamefont {A.}~\bibnamefont {Miranowicz}}, \bibinfo {author} {\bibfnamefont {H.~B.}\ \bibnamefont {Chen}}, \bibinfo {author} {\bibfnamefont {G.~Y.}\ \bibnamefont {Chen}}, \bibinfo {author} {\bibfnamefont {Y.~N.}\ \bibnamefont {Chen}},\ and\ \bibinfo {author} {\bibfnamefont {F.}~\bibnamefont {Nori}},\ }\bibfield  {title} {\bibinfo {title} {Collectively induced exceptional points of quantum emitters coupled to nanoparticle surface plasmons},\ }\href {https://doi.org/10.1103/PhysRevA.101.013814} {\bibfield  {journal} {\bibinfo  {journal} {Phys. Rev. A}\ }\textbf {\bibinfo {volume} {101}},\ \bibinfo {pages} {013814} (\bibinfo {year} {2020})}\BibitemShut {NoStop}%
\bibitem [{\citenamefont {Cirio}\ \emph {et~al.}(2022)\citenamefont {Cirio}, \citenamefont {Kuo}, \citenamefont {Chen}, \citenamefont {Nori},\ and\ \citenamefont {Lambert}}]{Mauro2022}%
  \BibitemOpen
  \bibfield  {author} {\bibinfo {author} {\bibfnamefont {M.}~\bibnamefont {Cirio}}, \bibinfo {author} {\bibfnamefont {P.~C.}\ \bibnamefont {Kuo}}, \bibinfo {author} {\bibfnamefont {Y.~N.}\ \bibnamefont {Chen}}, \bibinfo {author} {\bibfnamefont {F.}~\bibnamefont {Nori}},\ and\ \bibinfo {author} {\bibfnamefont {N.}~\bibnamefont {Lambert}},\ }\bibfield  {title} {\bibinfo {title} {Canonical derivation of the fermionic influence superoperator},\ }\href {https://doi.org/10.1103/PhysRevB.105.035121} {\bibfield  {journal} {\bibinfo  {journal} {Phys. Rev. B}\ }\textbf {\bibinfo {volume} {105}},\ \bibinfo {pages} {035121} (\bibinfo {year} {2022})}\BibitemShut {NoStop}%
\bibitem [{\citenamefont {Shi}\ \emph {et~al.}(2009)\citenamefont {Shi}, \citenamefont {Chen}, \citenamefont {Nan}, \citenamefont {Xu},\ and\ \citenamefont {Yan}}]{Shi2009}%
  \BibitemOpen
  \bibfield  {author} {\bibinfo {author} {\bibfnamefont {Q.}~\bibnamefont {Shi}}, \bibinfo {author} {\bibfnamefont {L.}~\bibnamefont {Chen}}, \bibinfo {author} {\bibfnamefont {G.}~\bibnamefont {Nan}}, \bibinfo {author} {\bibfnamefont {R.-X.}\ \bibnamefont {Xu}},\ and\ \bibinfo {author} {\bibfnamefont {Y.}~\bibnamefont {Yan}},\ }\bibfield  {title} {\bibinfo {title} {Efficient hierarchical {L}iouville space propagator to quantum dissipative dynamics},\ }\href {https://doi.org/10.1063/1.3077918} {\bibfield  {journal} {\bibinfo  {journal} {J. Chem. Phys.}\ }\textbf {\bibinfo {volume} {130}},\ \bibinfo {pages} {084105} (\bibinfo {year} {2009})}\BibitemShut {NoStop}%
\bibitem [{\citenamefont {Hu}\ \emph {et~al.}(2011)\citenamefont {Hu}, \citenamefont {Luo}, \citenamefont {Jiang}, \citenamefont {Xu},\ and\ \citenamefont {Yan}}]{Jie2011}%
  \BibitemOpen
  \bibfield  {author} {\bibinfo {author} {\bibfnamefont {J.}~\bibnamefont {Hu}}, \bibinfo {author} {\bibfnamefont {M.}~\bibnamefont {Luo}}, \bibinfo {author} {\bibfnamefont {F.}~\bibnamefont {Jiang}}, \bibinfo {author} {\bibfnamefont {R.-X.}\ \bibnamefont {Xu}},\ and\ \bibinfo {author} {\bibfnamefont {Y.}~\bibnamefont {Yan}},\ }\bibfield  {title} {\bibinfo {title} {Pad{\'e} spectrum decompositions of quantum distribution functions and optimal hierarchical equations of motion construction for quantum open systems},\ }\href {https://doi.org/10.1063/1.3602466} {\bibfield  {journal} {\bibinfo  {journal} {J. Chem. Phys.}\ }\textbf {\bibinfo {volume} {134}},\ \bibinfo {pages} {244106} (\bibinfo {year} {2011})}\BibitemShut {NoStop}%
\bibitem [{\citenamefont {Kuo}\ \emph {et~al.}(2023)\citenamefont {Kuo}, \citenamefont {Lambert}, \citenamefont {Cirio}, \citenamefont {Huang}, \citenamefont {Nori},\ and\ \citenamefont {Chen}}]{Pochen2023}%
  \BibitemOpen
  \bibfield  {author} {\bibinfo {author} {\bibfnamefont {P.-C.}\ \bibnamefont {Kuo}}, \bibinfo {author} {\bibfnamefont {N.}~\bibnamefont {Lambert}}, \bibinfo {author} {\bibfnamefont {M.}~\bibnamefont {Cirio}}, \bibinfo {author} {\bibfnamefont {Y.-T.}\ \bibnamefont {Huang}}, \bibinfo {author} {\bibfnamefont {F.}~\bibnamefont {Nori}},\ and\ \bibinfo {author} {\bibfnamefont {Y.-N.}\ \bibnamefont {Chen}},\ }\bibfield  {title} {\bibinfo {title} {{Kondo QED}: The {K}ondo effect and photon trapping in a two-impurity {A}nderson model ultra-strongly coupled to light},\ }\href {https://doi.org/10.48550/arXiv.2302.01044} {\bibfield  {journal} {\bibinfo  {journal} {arXiv preprint arXiv:2302.01044}\ } (\bibinfo {year} {2023})}\BibitemShut {NoStop}%
\bibitem [{\citenamefont {Wenderoth}\ \emph {et~al.}(2016)\citenamefont {Wenderoth}, \citenamefont {B\"atge},\ and\ \citenamefont {H\"artle}}]{Wenderoth2016}%
  \BibitemOpen
  \bibfield  {author} {\bibinfo {author} {\bibfnamefont {S.}~\bibnamefont {Wenderoth}}, \bibinfo {author} {\bibfnamefont {J.}~\bibnamefont {B\"atge}},\ and\ \bibinfo {author} {\bibfnamefont {R.}~\bibnamefont {H\"artle}},\ }\bibfield  {title} {\bibinfo {title} {Sharp peaks in the conductance of a double quantum dot and a quantum-dot spin valve at high temperatures: A hierarchical quantum master equation approach},\ }\href {https://doi.org/10.1103/PhysRevB.94.121303} {\bibfield  {journal} {\bibinfo  {journal} {Phys. Rev. B}\ }\textbf {\bibinfo {volume} {94}},\ \bibinfo {pages} {121303} (\bibinfo {year} {2016})}\BibitemShut {NoStop}%
\bibitem [{\citenamefont {H\"artle}\ \emph {et~al.}(2015)\citenamefont {H\"artle}, \citenamefont {Cohen}, \citenamefont {Reichman},\ and\ \citenamefont {Millis}}]{Hartle2015}%
  \BibitemOpen
  \bibfield  {author} {\bibinfo {author} {\bibfnamefont {R.}~\bibnamefont {H\"artle}}, \bibinfo {author} {\bibfnamefont {G.}~\bibnamefont {Cohen}}, \bibinfo {author} {\bibfnamefont {D.~R.}\ \bibnamefont {Reichman}},\ and\ \bibinfo {author} {\bibfnamefont {A.~J.}\ \bibnamefont {Millis}},\ }\bibfield  {title} {\bibinfo {title} {Transport through an {A}nderson impurity: Current ringing, nonlinear magnetization, and a direct comparison of continuous-time quantum monte carlo and hierarchical quantum master equations},\ }\href {https://doi.org/10.1103/PhysRevB.92.085430} {\bibfield  {journal} {\bibinfo  {journal} {Phys. Rev. B}\ }\textbf {\bibinfo {volume} {92}},\ \bibinfo {pages} {085430} (\bibinfo {year} {2015})}\BibitemShut {NoStop}%
\bibitem [{\citenamefont {Str\"{u}mpfer}\ and\ \citenamefont {Schulten}(2012)}]{PHI}%
  \BibitemOpen
  \bibfield  {author} {\bibinfo {author} {\bibfnamefont {J.}~\bibnamefont {Str\"{u}mpfer}}\ and\ \bibinfo {author} {\bibfnamefont {K.}~\bibnamefont {Schulten}},\ }\bibfield  {title} {\bibinfo {title} {Open quantum dynamics calculations with the hierarchy equations of motion on parallel computers},\ }\href {https://doi.org/10.1021/ct3003833} {\bibfield  {journal} {\bibinfo  {journal} {J. Chem. Theory Comput.}\ }\textbf {\bibinfo {volume} {8}},\ \bibinfo {pages} {2808} (\bibinfo {year} {2012})}\BibitemShut {NoStop}%
\bibitem [{\citenamefont {Kramer}\ \emph {et~al.}(2018)\citenamefont {Kramer}, \citenamefont {Noack}, \citenamefont {Reinefeld}, \citenamefont {Rodr{\'{\i}}guez},\ and\ \citenamefont {Zelinskyy}}]{DM-HEOM}%
  \BibitemOpen
  \bibfield  {author} {\bibinfo {author} {\bibfnamefont {T.}~\bibnamefont {Kramer}}, \bibinfo {author} {\bibfnamefont {M.}~\bibnamefont {Noack}}, \bibinfo {author} {\bibfnamefont {A.}~\bibnamefont {Reinefeld}}, \bibinfo {author} {\bibfnamefont {M.}~\bibnamefont {Rodr{\'{\i}}guez}},\ and\ \bibinfo {author} {\bibfnamefont {Y.}~\bibnamefont {Zelinskyy}},\ }\bibfield  {title} {\bibinfo {title} {Efficient calculation of open quantum system dynamics and time-resolved spectroscopy with distributed memory {HEOM} ({DM}-{HEOM})},\ }\href {https://doi.org/10.1002/jcc.25354} {\bibfield  {journal} {\bibinfo  {journal} {J. Comput. Chem.}\ }\textbf {\bibinfo {volume} {39}},\ \bibinfo {pages} {1779} (\bibinfo {year} {2018})}\BibitemShut {NoStop}%
\bibitem [{\citenamefont {Ikeda}\ and\ \citenamefont {Scholes}(2020)}]{PyHEOM}%
  \BibitemOpen
  \bibfield  {author} {\bibinfo {author} {\bibfnamefont {T.}~\bibnamefont {Ikeda}}\ and\ \bibinfo {author} {\bibfnamefont {G.~D.}\ \bibnamefont {Scholes}},\ }\bibfield  {title} {\bibinfo {title} {Generalization of the hierarchical equations of motion theory for efficient calculations with arbitrary correlation functions},\ }\href {https://doi.org/10.1063/5.0007327} {\bibfield  {journal} {\bibinfo  {journal} {J. Chem. Phys.}\ }\textbf {\bibinfo {volume} {152}},\ \bibinfo {pages} {204101} (\bibinfo {year} {2020})}\BibitemShut {NoStop}%
\bibitem [{\citenamefont {Velizhanin}\ \emph {et~al.}(2008)\citenamefont {Velizhanin}, \citenamefont {Wang},\ and\ \citenamefont {Thoss}}]{Velizhanin2008}%
  \BibitemOpen
  \bibfield  {author} {\bibinfo {author} {\bibfnamefont {K.~A.}\ \bibnamefont {Velizhanin}}, \bibinfo {author} {\bibfnamefont {H.}~\bibnamefont {Wang}},\ and\ \bibinfo {author} {\bibfnamefont {M.}~\bibnamefont {Thoss}},\ }\bibfield  {title} {\bibinfo {title} {Heat transport through model molecular junctions: A multilayer multiconfiguration time-dependent hartree approach},\ }\href {https://doi.org/10.1016/j.cplett.2008.05.065} {\bibfield  {journal} {\bibinfo  {journal} {Chem. Phys. Lett.}\ }\textbf {\bibinfo {volume} {460}},\ \bibinfo {pages} {325} (\bibinfo {year} {2008})}\BibitemShut {NoStop}%
\bibitem [{\citenamefont {Kato}\ and\ \citenamefont {Tanimura}(2015)}]{Kato2015}%
  \BibitemOpen
  \bibfield  {author} {\bibinfo {author} {\bibfnamefont {A.}~\bibnamefont {Kato}}\ and\ \bibinfo {author} {\bibfnamefont {Y.}~\bibnamefont {Tanimura}},\ }\bibfield  {title} {\bibinfo {title} {Quantum heat transport of a two-qubit system: Interplay between system-bath coherence and qubit-qubit coherence},\ }\href {https://doi.org/10.1063/1.4928192} {\bibfield  {journal} {\bibinfo  {journal} {J. Chem. Phys.}\ }\textbf {\bibinfo {volume} {143}},\ \bibinfo {pages} {064107} (\bibinfo {year} {2015})}\BibitemShut {NoStop}%
\bibitem [{\citenamefont {Song}\ and\ \citenamefont {Shi}(2017)}]{Song2017}%
  \BibitemOpen
  \bibfield  {author} {\bibinfo {author} {\bibfnamefont {L.}~\bibnamefont {Song}}\ and\ \bibinfo {author} {\bibfnamefont {Q.}~\bibnamefont {Shi}},\ }\bibfield  {title} {\bibinfo {title} {Hierarchical equations of motion method applied to nonequilibrium heat transport in model molecular junctions: Transient heat current and high-order moments of the current operator},\ }\href {https://doi.org/10.1103/PhysRevB.95.064308} {\bibfield  {journal} {\bibinfo  {journal} {Phys. Rev. B}\ }\textbf {\bibinfo {volume} {95}},\ \bibinfo {pages} {064308} (\bibinfo {year} {2017})}\BibitemShut {NoStop}%
\bibitem [{\citenamefont {Sprinzak}\ \emph {et~al.}(2002)\citenamefont {Sprinzak}, \citenamefont {Ji}, \citenamefont {Heiblum}, \citenamefont {Mahalu},\ and\ \citenamefont {Shtrikman}}]{Sprinzak2002}%
  \BibitemOpen
  \bibfield  {author} {\bibinfo {author} {\bibfnamefont {D.}~\bibnamefont {Sprinzak}}, \bibinfo {author} {\bibfnamefont {Y.}~\bibnamefont {Ji}}, \bibinfo {author} {\bibfnamefont {M.}~\bibnamefont {Heiblum}}, \bibinfo {author} {\bibfnamefont {D.}~\bibnamefont {Mahalu}},\ and\ \bibinfo {author} {\bibfnamefont {H.}~\bibnamefont {Shtrikman}},\ }\bibfield  {title} {\bibinfo {title} {Charge distribution in a {K}ondo-correlated quantum dot},\ }\href {https://doi.org/10.1103/PhysRevLett.88.176805} {\bibfield  {journal} {\bibinfo  {journal} {Phys. Rev. Lett.}\ }\textbf {\bibinfo {volume} {88}},\ \bibinfo {pages} {176805} (\bibinfo {year} {2002})}\BibitemShut {NoStop}%
\bibitem [{\citenamefont {Keller}\ \emph {et~al.}(2013)\citenamefont {Keller}, \citenamefont {Amasha}, \citenamefont {Weymann}, \citenamefont {Moca}, \citenamefont {Rau}, \citenamefont {Katine}, \citenamefont {Shtrikman}, \citenamefont {Zar{\'{a}}nd},\ and\ \citenamefont {Goldhaber-Gordon}}]{Keller2013}%
  \BibitemOpen
  \bibfield  {author} {\bibinfo {author} {\bibfnamefont {A.~J.}\ \bibnamefont {Keller}}, \bibinfo {author} {\bibfnamefont {S.}~\bibnamefont {Amasha}}, \bibinfo {author} {\bibfnamefont {I.}~\bibnamefont {Weymann}}, \bibinfo {author} {\bibfnamefont {C.~P.}\ \bibnamefont {Moca}}, \bibinfo {author} {\bibfnamefont {I.~G.}\ \bibnamefont {Rau}}, \bibinfo {author} {\bibfnamefont {J.~A.}\ \bibnamefont {Katine}}, \bibinfo {author} {\bibfnamefont {H.}~\bibnamefont {Shtrikman}}, \bibinfo {author} {\bibfnamefont {G.}~\bibnamefont {Zar{\'{a}}nd}},\ and\ \bibinfo {author} {\bibfnamefont {D.}~\bibnamefont {Goldhaber-Gordon}},\ }\bibfield  {title} {\bibinfo {title} {Emergent {SU}(4) {K}ondo physics in a spin{\textendash}charge-entangled double quantum dot},\ }\href {https://doi.org/10.1038/nphys2844} {\bibfield  {journal} {\bibinfo  {journal} {Nat. Phys.}\ }\textbf {\bibinfo {volume} {10}},\ \bibinfo {pages} {145} (\bibinfo {year} {2013})}\BibitemShut {NoStop}%
\bibitem [{\citenamefont {Le~Hur}(2015)}]{Hur2015}%
  \BibitemOpen
  \bibfield  {author} {\bibinfo {author} {\bibfnamefont {K.}~\bibnamefont {Le~Hur}},\ }\bibfield  {title} {\bibinfo {title} {Quantum dots and the {K}ondo effect},\ }\href {https://doi.org/10.1038/526203a} {\bibfield  {journal} {\bibinfo  {journal} {Nature}\ }\textbf {\bibinfo {volume} {526}},\ \bibinfo {pages} {203} (\bibinfo {year} {2015})}\BibitemShut {NoStop}%
\bibitem [{\citenamefont {Park}\ \emph {et~al.}(2002)\citenamefont {Park}, \citenamefont {Pasupathy}, \citenamefont {Goldsmith}, \citenamefont {Chang}, \citenamefont {Yaish}, \citenamefont {Petta}, \citenamefont {Rinkoski}, \citenamefont {Sethna}, \citenamefont {Abru{\~{n}}a}, \citenamefont {McEuen},\ and\ \citenamefont {Ralph}}]{Park2002}%
  \BibitemOpen
  \bibfield  {author} {\bibinfo {author} {\bibfnamefont {J.}~\bibnamefont {Park}}, \bibinfo {author} {\bibfnamefont {A.~N.}\ \bibnamefont {Pasupathy}}, \bibinfo {author} {\bibfnamefont {J.~I.}\ \bibnamefont {Goldsmith}}, \bibinfo {author} {\bibfnamefont {C.}~\bibnamefont {Chang}}, \bibinfo {author} {\bibfnamefont {Y.}~\bibnamefont {Yaish}}, \bibinfo {author} {\bibfnamefont {J.~R.}\ \bibnamefont {Petta}}, \bibinfo {author} {\bibfnamefont {M.}~\bibnamefont {Rinkoski}}, \bibinfo {author} {\bibfnamefont {J.~P.}\ \bibnamefont {Sethna}}, \bibinfo {author} {\bibfnamefont {H.~D.}\ \bibnamefont {Abru{\~{n}}a}}, \bibinfo {author} {\bibfnamefont {P.~L.}\ \bibnamefont {McEuen}},\ and\ \bibinfo {author} {\bibfnamefont {D.~C.}\ \bibnamefont {Ralph}},\ }\bibfield  {title} {\bibinfo {title} {Coulomb blockade and the {K}ondo effect in single-atom transistors},\ }\href {https://doi.org/10.1038/nature00791} {\bibfield  {journal} {\bibinfo  {journal} {Nature}\ }\textbf {\bibinfo {volume} {417}},\ \bibinfo {pages} {722} (\bibinfo
  {year} {2002})}\BibitemShut {NoStop}%
\bibitem [{\citenamefont {Wingreen}(2004)}]{Wingreen2004}%
  \BibitemOpen
  \bibfield  {author} {\bibinfo {author} {\bibfnamefont {N.~S.}\ \bibnamefont {Wingreen}},\ }\bibfield  {title} {\bibinfo {title} {Quantum many-body effects in a single-electron transistor},\ }\href {https://doi.org/10.1126/science.1098302} {\bibfield  {journal} {\bibinfo  {journal} {Science}\ }\textbf {\bibinfo {volume} {304}},\ \bibinfo {pages} {1258} (\bibinfo {year} {2004})}\BibitemShut {NoStop}%
\bibitem [{\citenamefont {Yu}\ \emph {et~al.}(2004)\citenamefont {Yu}, \citenamefont {Keane}, \citenamefont {Ciszek}, \citenamefont {Cheng}, \citenamefont {Stewart}, \citenamefont {Tour},\ and\ \citenamefont {Natelson}}]{Natelson2004}%
  \BibitemOpen
  \bibfield  {author} {\bibinfo {author} {\bibfnamefont {L.~H.}\ \bibnamefont {Yu}}, \bibinfo {author} {\bibfnamefont {Z.~K.}\ \bibnamefont {Keane}}, \bibinfo {author} {\bibfnamefont {J.~W.}\ \bibnamefont {Ciszek}}, \bibinfo {author} {\bibfnamefont {L.}~\bibnamefont {Cheng}}, \bibinfo {author} {\bibfnamefont {M.~P.}\ \bibnamefont {Stewart}}, \bibinfo {author} {\bibfnamefont {J.~M.}\ \bibnamefont {Tour}},\ and\ \bibinfo {author} {\bibfnamefont {D.}~\bibnamefont {Natelson}},\ }\bibfield  {title} {\bibinfo {title} {Inelastic electron tunneling via molecular vibrations in single-molecule transistors},\ }\href {https://doi.org/10.1103/PhysRevLett.93.266802} {\bibfield  {journal} {\bibinfo  {journal} {Phys. Rev. Lett.}\ }\textbf {\bibinfo {volume} {93}},\ \bibinfo {pages} {266802} (\bibinfo {year} {2004})}\BibitemShut {NoStop}%
\bibitem [{\citenamefont {Smith}\ \emph {et~al.}(2019)\citenamefont {Smith}, \citenamefont {Kim}, \citenamefont {Pollmann},\ and\ \citenamefont {Knolle}}]{Smith2019}%
  \BibitemOpen
  \bibfield  {author} {\bibinfo {author} {\bibfnamefont {A.}~\bibnamefont {Smith}}, \bibinfo {author} {\bibfnamefont {M.~S.}\ \bibnamefont {Kim}}, \bibinfo {author} {\bibfnamefont {F.}~\bibnamefont {Pollmann}},\ and\ \bibinfo {author} {\bibfnamefont {J.}~\bibnamefont {Knolle}},\ }\bibfield  {title} {\bibinfo {title} {Simulating quantum many-body dynamics on a current digital quantum computer},\ }\href {https://doi.org/10.1038/s41534-019-0217-0} {\bibfield  {journal} {\bibinfo  {journal} {npj Quantum Inf.}\ }\textbf {\bibinfo {volume} {5}},\ \bibinfo {pages} {106} (\bibinfo {year} {2019})}\BibitemShut {NoStop}%
\bibitem [{\citenamefont {Wang}\ \emph {et~al.}(2013)\citenamefont {Wang}, \citenamefont {Zheng}, \citenamefont {Jin},\ and\ \citenamefont {Yan}}]{Yan_7_2013}%
  \BibitemOpen
  \bibfield  {author} {\bibinfo {author} {\bibfnamefont {S.}~\bibnamefont {Wang}}, \bibinfo {author} {\bibfnamefont {X.}~\bibnamefont {Zheng}}, \bibinfo {author} {\bibfnamefont {J.}~\bibnamefont {Jin}},\ and\ \bibinfo {author} {\bibfnamefont {Y.}~\bibnamefont {Yan}},\ }\bibfield  {title} {\bibinfo {title} {Hierarchical {L}iouville-space approach to nonequilibrium dynamical properties of quantum impurity systems},\ }\href {https://doi.org/10.1103/PhysRevB.88.035129} {\bibfield  {journal} {\bibinfo  {journal} {Phys. Rev. B}\ }\textbf {\bibinfo {volume} {88}},\ \bibinfo {pages} {035129} (\bibinfo {year} {2013})}\BibitemShut {NoStop}%
\bibitem [{\citenamefont {Kouwenhoven}\ and\ \citenamefont {Glazman}(2001)}]{Kouwenhoven2001}%
  \BibitemOpen
  \bibfield  {author} {\bibinfo {author} {\bibfnamefont {L.}~\bibnamefont {Kouwenhoven}}\ and\ \bibinfo {author} {\bibfnamefont {L.}~\bibnamefont {Glazman}},\ }\bibfield  {title} {\bibinfo {title} {Revival of the {K}ondo effect},\ }\href {https://doi.org/10.1088/2058-7058/14/1/28} {\bibfield  {journal} {\bibinfo  {journal} {Phys. World}\ }\textbf {\bibinfo {volume} {14}},\ \bibinfo {pages} {33} (\bibinfo {year} {2001})}\BibitemShut {NoStop}%
\bibitem [{\citenamefont {Borzenets}\ \emph {et~al.}(2020)\citenamefont {Borzenets}, \citenamefont {Shim}, \citenamefont {Chen}, \citenamefont {Ludwig}, \citenamefont {Wieck}, \citenamefont {Tarucha}, \citenamefont {Sim},\ and\ \citenamefont {Yamamoto}}]{VBorzenets2020}%
  \BibitemOpen
  \bibfield  {author} {\bibinfo {author} {\bibfnamefont {I.~V.}\ \bibnamefont {Borzenets}}, \bibinfo {author} {\bibfnamefont {J.}~\bibnamefont {Shim}}, \bibinfo {author} {\bibfnamefont {J.~C.~H.}\ \bibnamefont {Chen}}, \bibinfo {author} {\bibfnamefont {A.}~\bibnamefont {Ludwig}}, \bibinfo {author} {\bibfnamefont {A.~D.}\ \bibnamefont {Wieck}}, \bibinfo {author} {\bibfnamefont {S.}~\bibnamefont {Tarucha}}, \bibinfo {author} {\bibfnamefont {H.-S.}\ \bibnamefont {Sim}},\ and\ \bibinfo {author} {\bibfnamefont {M.}~\bibnamefont {Yamamoto}},\ }\bibfield  {title} {\bibinfo {title} {Observation of the {K}ondo screening cloud},\ }\href {https://doi.org/10.1038/s41586-020-2058-6} {\bibfield  {journal} {\bibinfo  {journal} {Nature}\ }\textbf {\bibinfo {volume} {579}},\ \bibinfo {pages} {210} (\bibinfo {year} {2020})}\BibitemShut {NoStop}%
\bibitem [{\citenamefont {Smith}\ \emph {et~al.}(2022)\citenamefont {Smith}, \citenamefont {Chen}, \citenamefont {Chang}, \citenamefont {Wu}, \citenamefont {Lo}, \citenamefont {Chao}, \citenamefont {Farrer}, \citenamefont {Beere}, \citenamefont {Griffiths}, \citenamefont {Jones}, \citenamefont {Ritchie}, \citenamefont {Chen},\ and\ \citenamefont {Chen}}]{Luke2022}%
  \BibitemOpen
  \bibfield  {author} {\bibinfo {author} {\bibfnamefont {L.~W.}\ \bibnamefont {Smith}}, \bibinfo {author} {\bibfnamefont {H.-B.}\ \bibnamefont {Chen}}, \bibinfo {author} {\bibfnamefont {C.-W.}\ \bibnamefont {Chang}}, \bibinfo {author} {\bibfnamefont {C.-W.}\ \bibnamefont {Wu}}, \bibinfo {author} {\bibfnamefont {S.-T.}\ \bibnamefont {Lo}}, \bibinfo {author} {\bibfnamefont {S.-H.}\ \bibnamefont {Chao}}, \bibinfo {author} {\bibfnamefont {I.}~\bibnamefont {Farrer}}, \bibinfo {author} {\bibfnamefont {H.~E.}\ \bibnamefont {Beere}}, \bibinfo {author} {\bibfnamefont {J.~P.}\ \bibnamefont {Griffiths}}, \bibinfo {author} {\bibfnamefont {G.~A.~C.}\ \bibnamefont {Jones}}, \bibinfo {author} {\bibfnamefont {D.~A.}\ \bibnamefont {Ritchie}}, \bibinfo {author} {\bibfnamefont {Y.-N.}\ \bibnamefont {Chen}},\ and\ \bibinfo {author} {\bibfnamefont {T.-M.}\ \bibnamefont {Chen}},\ }\bibfield  {title} {\bibinfo {title} {Electrically controllable {K}ondo correlation in spin-orbit-coupled quantum point contacts},\ }\href
  {https://doi.org/10.1103/PhysRevLett.128.027701} {\bibfield  {journal} {\bibinfo  {journal} {Phys. Rev. Lett.}\ }\textbf {\bibinfo {volume} {128}},\ \bibinfo {pages} {027701} (\bibinfo {year} {2022})}\BibitemShut {NoStop}%
\bibitem [{\citenamefont {van~der Wiel}\ \emph {et~al.}(2002)\citenamefont {van~der Wiel}, \citenamefont {De~Franceschi}, \citenamefont {Elzerman}, \citenamefont {Fujisawa}, \citenamefont {Tarucha},\ and\ \citenamefont {Kouwenhoven}}]{vanderWiel2002}%
  \BibitemOpen
  \bibfield  {author} {\bibinfo {author} {\bibfnamefont {W.~G.}\ \bibnamefont {van~der Wiel}}, \bibinfo {author} {\bibfnamefont {S.}~\bibnamefont {De~Franceschi}}, \bibinfo {author} {\bibfnamefont {J.~M.}\ \bibnamefont {Elzerman}}, \bibinfo {author} {\bibfnamefont {T.}~\bibnamefont {Fujisawa}}, \bibinfo {author} {\bibfnamefont {S.}~\bibnamefont {Tarucha}},\ and\ \bibinfo {author} {\bibfnamefont {L.~P.}\ \bibnamefont {Kouwenhoven}},\ }\bibfield  {title} {\bibinfo {title} {Electron transport through double quantum dots},\ }\href {https://doi.org/10.1103/RevModPhys.75.1} {\bibfield  {journal} {\bibinfo  {journal} {Rev. Mod. Phys.}\ }\textbf {\bibinfo {volume} {75}},\ \bibinfo {pages} {1} (\bibinfo {year} {2002})}\BibitemShut {NoStop}%
\bibitem [{\citenamefont {Bruhat}\ \emph {et~al.}(2018)\citenamefont {Bruhat}, \citenamefont {Cubaynes}, \citenamefont {Viennot}, \citenamefont {Dartiailh}, \citenamefont {Desjardins}, \citenamefont {Cottet},\ and\ \citenamefont {Kontos}}]{Kontos2018}%
  \BibitemOpen
  \bibfield  {author} {\bibinfo {author} {\bibfnamefont {L.~E.}\ \bibnamefont {Bruhat}}, \bibinfo {author} {\bibfnamefont {T.}~\bibnamefont {Cubaynes}}, \bibinfo {author} {\bibfnamefont {J.~J.}\ \bibnamefont {Viennot}}, \bibinfo {author} {\bibfnamefont {M.~C.}\ \bibnamefont {Dartiailh}}, \bibinfo {author} {\bibfnamefont {M.~M.}\ \bibnamefont {Desjardins}}, \bibinfo {author} {\bibfnamefont {A.}~\bibnamefont {Cottet}},\ and\ \bibinfo {author} {\bibfnamefont {T.}~\bibnamefont {Kontos}},\ }\bibfield  {title} {\bibinfo {title} {Circuit {QED} with a quantum-dot charge qubit dressed by {C}ooper pairs},\ }\href {https://doi.org/10.1103/PhysRevB.98.155313} {\bibfield  {journal} {\bibinfo  {journal} {Phys. Rev. B}\ }\textbf {\bibinfo {volume} {98}},\ \bibinfo {pages} {155313} (\bibinfo {year} {2018})}\BibitemShut {NoStop}%
\bibitem [{\citenamefont {van Woerkom}\ \emph {et~al.}(2018)\citenamefont {van Woerkom}, \citenamefont {Scarlino}, \citenamefont {Ungerer}, \citenamefont {M\"uller}, \citenamefont {Koski}, \citenamefont {Landig}, \citenamefont {Reichl}, \citenamefont {Wegscheider}, \citenamefont {Ihn}, \citenamefont {Ensslin},\ and\ \citenamefont {Wallraff}}]{Wallraff2018}%
  \BibitemOpen
  \bibfield  {author} {\bibinfo {author} {\bibfnamefont {D.~J.}\ \bibnamefont {van Woerkom}}, \bibinfo {author} {\bibfnamefont {P.}~\bibnamefont {Scarlino}}, \bibinfo {author} {\bibfnamefont {J.~H.}\ \bibnamefont {Ungerer}}, \bibinfo {author} {\bibfnamefont {C.}~\bibnamefont {M\"uller}}, \bibinfo {author} {\bibfnamefont {J.~V.}\ \bibnamefont {Koski}}, \bibinfo {author} {\bibfnamefont {A.~J.}\ \bibnamefont {Landig}}, \bibinfo {author} {\bibfnamefont {C.}~\bibnamefont {Reichl}}, \bibinfo {author} {\bibfnamefont {W.}~\bibnamefont {Wegscheider}}, \bibinfo {author} {\bibfnamefont {T.}~\bibnamefont {Ihn}}, \bibinfo {author} {\bibfnamefont {K.}~\bibnamefont {Ensslin}},\ and\ \bibinfo {author} {\bibfnamefont {A.}~\bibnamefont {Wallraff}},\ }\bibfield  {title} {\bibinfo {title} {Microwave photon-mediated interactions between semiconductor qubits},\ }\href {https://doi.org/10.1103/PhysRevX.8.041018} {\bibfield  {journal} {\bibinfo  {journal} {Phys. Rev. X}\ }\textbf {\bibinfo {volume} {8}},\ \bibinfo {pages} {041018}
  (\bibinfo {year} {2018})}\BibitemShut {NoStop}%
\bibitem [{\citenamefont {Scarlino}\ \emph {et~al.}(2019)\citenamefont {Scarlino}, \citenamefont {van Woerkom}, \citenamefont {Mendes}, \citenamefont {Koski}, \citenamefont {Landig}, \citenamefont {Andersen}, \citenamefont {Gasparinetti}, \citenamefont {Reichl}, \citenamefont {Wegscheider}, \citenamefont {Ensslin}, \citenamefont {Ihn}, \citenamefont {Blais},\ and\ \citenamefont {Wallraff}}]{Scarlino2019}%
  \BibitemOpen
  \bibfield  {author} {\bibinfo {author} {\bibfnamefont {P.}~\bibnamefont {Scarlino}}, \bibinfo {author} {\bibfnamefont {D.~J.}\ \bibnamefont {van Woerkom}}, \bibinfo {author} {\bibfnamefont {U.~C.}\ \bibnamefont {Mendes}}, \bibinfo {author} {\bibfnamefont {J.~V.}\ \bibnamefont {Koski}}, \bibinfo {author} {\bibfnamefont {A.~J.}\ \bibnamefont {Landig}}, \bibinfo {author} {\bibfnamefont {C.~K.}\ \bibnamefont {Andersen}}, \bibinfo {author} {\bibfnamefont {S.}~\bibnamefont {Gasparinetti}}, \bibinfo {author} {\bibfnamefont {C.}~\bibnamefont {Reichl}}, \bibinfo {author} {\bibfnamefont {W.}~\bibnamefont {Wegscheider}}, \bibinfo {author} {\bibfnamefont {K.}~\bibnamefont {Ensslin}}, \bibinfo {author} {\bibfnamefont {T.}~\bibnamefont {Ihn}}, \bibinfo {author} {\bibfnamefont {A.}~\bibnamefont {Blais}},\ and\ \bibinfo {author} {\bibfnamefont {A.}~\bibnamefont {Wallraff}},\ }\bibfield  {title} {\bibinfo {title} {Coherent microwave-photon-mediated coupling between a semiconductor and a superconducting qubit},\ }\href
  {https://doi.org/10.1038/s41467-019-10798-6} {\bibfield  {journal} {\bibinfo  {journal} {Nat. Commun.}\ }\textbf {\bibinfo {volume} {10}},\ \bibinfo {pages} {3011} (\bibinfo {year} {2019})}\BibitemShut {NoStop}%
\bibitem [{\citenamefont {Viennot}\ \emph {et~al.}(2014)\citenamefont {Viennot}, \citenamefont {Delbecq}, \citenamefont {Dartiailh}, \citenamefont {Cottet},\ and\ \citenamefont {Kontos}}]{Kontos2014}%
  \BibitemOpen
  \bibfield  {author} {\bibinfo {author} {\bibfnamefont {J.~J.}\ \bibnamefont {Viennot}}, \bibinfo {author} {\bibfnamefont {M.~R.}\ \bibnamefont {Delbecq}}, \bibinfo {author} {\bibfnamefont {M.~C.}\ \bibnamefont {Dartiailh}}, \bibinfo {author} {\bibfnamefont {A.}~\bibnamefont {Cottet}},\ and\ \bibinfo {author} {\bibfnamefont {T.}~\bibnamefont {Kontos}},\ }\bibfield  {title} {\bibinfo {title} {Out-of-equilibrium charge dynamics in a hybrid circuit quantum electrodynamics architecture},\ }\href {https://doi.org/10.1103/PhysRevB.89.165404} {\bibfield  {journal} {\bibinfo  {journal} {Phys. Rev. B}\ }\textbf {\bibinfo {volume} {89}},\ \bibinfo {pages} {165404} (\bibinfo {year} {2014})}\BibitemShut {NoStop}%
\bibitem [{\citenamefont {Bruhat}\ \emph {et~al.}(2016)\citenamefont {Bruhat}, \citenamefont {Viennot}, \citenamefont {Dartiailh}, \citenamefont {Desjardins}, \citenamefont {Kontos},\ and\ \citenamefont {Cottet}}]{Cottet2016}%
  \BibitemOpen
  \bibfield  {author} {\bibinfo {author} {\bibfnamefont {L.~E.}\ \bibnamefont {Bruhat}}, \bibinfo {author} {\bibfnamefont {J.~J.}\ \bibnamefont {Viennot}}, \bibinfo {author} {\bibfnamefont {M.~C.}\ \bibnamefont {Dartiailh}}, \bibinfo {author} {\bibfnamefont {M.~M.}\ \bibnamefont {Desjardins}}, \bibinfo {author} {\bibfnamefont {T.}~\bibnamefont {Kontos}},\ and\ \bibinfo {author} {\bibfnamefont {A.}~\bibnamefont {Cottet}},\ }\bibfield  {title} {\bibinfo {title} {Cavity photons as a probe for charge relaxation resistance and photon emission in a quantum dot coupled to normal and superconducting continua},\ }\href {https://doi.org/10.1103/PhysRevX.6.021014} {\bibfield  {journal} {\bibinfo  {journal} {Phys. Rev. X}\ }\textbf {\bibinfo {volume} {6}},\ \bibinfo {pages} {021014} (\bibinfo {year} {2016})}\BibitemShut {NoStop}%
\bibitem [{\citenamefont {Souquet}\ \emph {et~al.}(2014)\citenamefont {Souquet}, \citenamefont {Woolley}, \citenamefont {Gabelli}, \citenamefont {Simon},\ and\ \citenamefont {Clerk}}]{Souquet2014}%
  \BibitemOpen
  \bibfield  {author} {\bibinfo {author} {\bibfnamefont {J.~R.}\ \bibnamefont {Souquet}}, \bibinfo {author} {\bibfnamefont {M.~J.}\ \bibnamefont {Woolley}}, \bibinfo {author} {\bibfnamefont {J.}~\bibnamefont {Gabelli}}, \bibinfo {author} {\bibfnamefont {P.}~\bibnamefont {Simon}},\ and\ \bibinfo {author} {\bibfnamefont {A.~A.}\ \bibnamefont {Clerk}},\ }\bibfield  {title} {\bibinfo {title} {Photon-assisted tunnelling with nonclassical light},\ }\href {https://doi.org/10.1038/ncomms6562} {\bibfield  {journal} {\bibinfo  {journal} {Nat. Commun.}\ }\textbf {\bibinfo {volume} {5}},\ \bibinfo {pages} {5562} (\bibinfo {year} {2014})}\BibitemShut {NoStop}%
\bibitem [{\citenamefont {Kockum}\ \emph {et~al.}(2019)\citenamefont {Kockum}, \citenamefont {Miranowicz}, \citenamefont {Liberato}, \citenamefont {Savasta},\ and\ \citenamefont {Nori}}]{Kockum2019}%
  \BibitemOpen
  \bibfield  {author} {\bibinfo {author} {\bibfnamefont {A.~F.}\ \bibnamefont {Kockum}}, \bibinfo {author} {\bibfnamefont {A.}~\bibnamefont {Miranowicz}}, \bibinfo {author} {\bibfnamefont {S.~D.}\ \bibnamefont {Liberato}}, \bibinfo {author} {\bibfnamefont {S.}~\bibnamefont {Savasta}},\ and\ \bibinfo {author} {\bibfnamefont {F.}~\bibnamefont {Nori}},\ }\bibfield  {title} {\bibinfo {title} {Ultrastrong coupling between light and matter},\ }\href {https://doi.org/10.1038/s42254-019-0046-2} {\bibfield  {journal} {\bibinfo  {journal} {Nat. Rev. Phys.}\ }\textbf {\bibinfo {volume} {1}},\ \bibinfo {pages} {295} (\bibinfo {year} {2019})}\BibitemShut {NoStop}%
\bibitem [{\citenamefont {Breuer}\ and\ \citenamefont {Petruccione}(2007)}]{Breuer2007}%
  \BibitemOpen
  \bibfield  {author} {\bibinfo {author} {\bibfnamefont {H.-P.}\ \bibnamefont {Breuer}}\ and\ \bibinfo {author} {\bibfnamefont {F.}~\bibnamefont {Petruccione}},\ }\href {https://doi.org/10.1093/acprof:oso/9780199213900.001.0001} {\emph {\bibinfo {title} {{The Theory of Open Quantum Systems}}}}\ (\bibinfo  {publisher} {Oxford University Press},\ \bibinfo {year} {2007})\BibitemShut {NoStop}%
\bibitem [{\citenamefont {Chen}\ \emph {et~al.}(2009)\citenamefont {Chen}, \citenamefont {Chen}, \citenamefont {Liao}, \citenamefont {Lambert},\ and\ \citenamefont {Nori}}]{Yueh-Nan2009}%
  \BibitemOpen
  \bibfield  {author} {\bibinfo {author} {\bibfnamefont {Y.-N.}\ \bibnamefont {Chen}}, \bibinfo {author} {\bibfnamefont {G.-Y.}\ \bibnamefont {Chen}}, \bibinfo {author} {\bibfnamefont {Y.-Y.}\ \bibnamefont {Liao}}, \bibinfo {author} {\bibfnamefont {N.}~\bibnamefont {Lambert}},\ and\ \bibinfo {author} {\bibfnamefont {F.}~\bibnamefont {Nori}},\ }\bibfield  {title} {\bibinfo {title} {Detecting non-{M}arkovian plasmonic band gaps in quantum dots using electron transport},\ }\href {https://doi.org/10.1103/PhysRevB.79.245312} {\bibfield  {journal} {\bibinfo  {journal} {Phys. Rev. B}\ }\textbf {\bibinfo {volume} {79}},\ \bibinfo {pages} {245312} (\bibinfo {year} {2009})}\BibitemShut {NoStop}%
\bibitem [{\citenamefont {Xiong}\ \emph {et~al.}(2015)\citenamefont {Xiong}, \citenamefont {Lo}, \citenamefont {Zhang}, \citenamefont {Feng},\ and\ \citenamefont {Nori}}]{Xiong2015}%
  \BibitemOpen
  \bibfield  {author} {\bibinfo {author} {\bibfnamefont {H.-N.}\ \bibnamefont {Xiong}}, \bibinfo {author} {\bibfnamefont {P.-Y.}\ \bibnamefont {Lo}}, \bibinfo {author} {\bibfnamefont {W.-M.}\ \bibnamefont {Zhang}}, \bibinfo {author} {\bibfnamefont {D.~H.}\ \bibnamefont {Feng}},\ and\ \bibinfo {author} {\bibfnamefont {F.}~\bibnamefont {Nori}},\ }\bibfield  {title} {\bibinfo {title} {Non-{M}arkovian complexity in the quantum-to-classical transition},\ }\href {https://doi.org/10.1038/srep13353} {\bibfield  {journal} {\bibinfo  {journal} {Sci. Rep.}\ }\textbf {\bibinfo {volume} {5}},\ \bibinfo {pages} {13353} (\bibinfo {year} {2015})}\BibitemShut {NoStop}%
\bibitem [{\citenamefont {Johansson}\ \emph {et~al.}(2012)\citenamefont {Johansson}, \citenamefont {Nation},\ and\ \citenamefont {Nori}}]{QuTiP}%
  \BibitemOpen
  \bibfield  {author} {\bibinfo {author} {\bibfnamefont {J.}~\bibnamefont {Johansson}}, \bibinfo {author} {\bibfnamefont {P.}~\bibnamefont {Nation}},\ and\ \bibinfo {author} {\bibfnamefont {F.}~\bibnamefont {Nori}},\ }\bibfield  {title} {\bibinfo {title} {{QuTiP}: An open-source {P}ython framework for the dynamics of open quantum systems},\ }\href {https://doi.org/10.1016/j.cpc.2012.02.021} {\bibfield  {journal} {\bibinfo  {journal} {Comput. Phys. Commun.}\ }\textbf {\bibinfo {volume} {183}},\ \bibinfo {pages} {1760} (\bibinfo {year} {2012})}\BibitemShut {NoStop}%
\bibitem [{\citenamefont {Johansson}\ \emph {et~al.}(2013)\citenamefont {Johansson}, \citenamefont {Nation},\ and\ \citenamefont {Nori}}]{QuTiP2}%
  \BibitemOpen
  \bibfield  {author} {\bibinfo {author} {\bibfnamefont {J.}~\bibnamefont {Johansson}}, \bibinfo {author} {\bibfnamefont {P.}~\bibnamefont {Nation}},\ and\ \bibinfo {author} {\bibfnamefont {F.}~\bibnamefont {Nori}},\ }\bibfield  {title} {\bibinfo {title} {{QuTiP} 2: A {P}ython framework for the dynamics of open quantum systems},\ }\href {https://doi.org/10.1016/j.cpc.2012.11.019} {\bibfield  {journal} {\bibinfo  {journal} {Comput. Phys. Commun.}\ }\textbf {\bibinfo {volume} {184}},\ \bibinfo {pages} {1234} (\bibinfo {year} {2013})}\BibitemShut {NoStop}%
\end{thebibliography}
%

\end{document}